\newcommand*{\ATLASLATEXPATH}{./}

\def\llgg{$\ell^{+}\ell^{-}\gamma\gamma$\xspace}
\def\llg{\ensuremath{\ell^{+}\ell^{-}\gamma}\xspace}
\def\Zg{\ensuremath{\Zzero\gamma}\xspace}
\def\Zgg{\ensuremath{\Zzero\gamma\gamma}\xspace}
\def\eeg{\ensuremath{e^{+}e^{-}\gamma}\xspace}
\def\mumug{\ensuremath{\mu^{+}\mu^{-}\gamma}\xspace}
\def\eegg{\ensuremath{e^{+}e^{-}\gamma\gamma}\xspace}
\def\mumugg{\ensuremath{\mu^{+}\mu^{-}\gamma\gamma}\xspace}

\def\rrr{\rightarrow \xspace}
\documentclass[cernpreprint,texlive=2011,txfonts,UKenglish]{\ATLASLATEXPATH atlasdoc}
\pdfoutput=1

\usepackage[biblatex=true,subfigure]{\ATLASLATEXPATH atlaspackage}
\usepackage{\ATLASLATEXPATH atlasbiblatex}

\usepackage{\ATLASLATEXPATH atlasphysics}

\graphicspath{{./}{./}}

\usepackage{multirow}
\usepackage{xspace}
\usepackage{textcomp}
\usepackage{setspace}

\AtlasTitle{Measurements of $Z\gamma$ and $Z\gamma\gamma$ production in $pp$ collisions
       at $\sqrt{s}=8$ \TeV~with the ATLAS detector}

\author{The ATLAS Collaboration}

\AtlasRefCode{STDM-2014-01}

\PreprintIdNumber{CERN-EP-2016-049}

 \AtlasJournal{Physical Review D}

\AtlasAbstract{%
The production of $\Zboson$ bosons with one or two isolated high-energy photons is studied using $pp$ collisions at $\sqrt{s}$ = 8 \TeV. The analyses use a data sample with an integrated luminosity of 20.3 $\ifb$ collected by the ATLAS detector during the 2012 LHC data taking. 
The $Z\gamma$ and $Z\gamma\gamma$ production cross sections are measured with leptonic ($\ee$, $\mumu$, $\nnbar$) decays of the $Z$ boson, in extended fiducial regions defined in terms of the lepton and photon acceptance.
They are then compared to cross-section predictions from the Standard Model, where the sources of the photons are radiation off initial-state quarks and radiative $Z$-boson decay to charged leptons, and from fragmentation of final-state quarks and gluons into photons.
The yields of events with photon transverse energy $\ET > 250$ \GeV~from $\leplep\gamma$ events and with $\ET > 400$ \GeV~from $\nnbar\gamma$ events are used to search for anomalous triple gauge-boson couplings $ZZ\gamma$ and $Z\gamma\gamma$.
The yields of events with diphoton invariant mass $m_{\gamma\gamma} > 200$ \GeV~from $\leplep\gamma\gamma$ events and with $m_{\gamma\gamma} > 300$ \GeV~from $\nnbar\gamma\gamma$ events are used to search for anomalous quartic gauge-boson couplings $ZZ\gamma\gamma$ and $Z\gamma\gamma\gamma$.  
No deviations from Standard Model predictions are observed and limits are placed on parameters used to describe anomalous triple and quartic gauge-boson couplings.
}

 \AtlasCoverSupportingNote{~~~~~~~~~~~~~~~~~~~~~~ATL-COM-PHYS-2013-1573}{https://cds.cern.ch/record/1631102}
 \AtlasCoverCommentsDeadline{2 Mar 2016}

 \AtlasCoverAnalysisTeam{Benjamin Auerbach, Neil Delwadia, Al Goshaw, Ashutosh Kotwal, Dimitrii Krasnopevtsev, Anastasia Kurova, Tom LeCompte, Shu Li, S. Lona, Jackson Van Horn Matteucci,
Duong Nguyen, Nadezda Proklova, James Proudfoot, Anatoli Romaniouk, Evgeny Soldatov, Ryszard Stroynowski, Hulin Wang, Jingbo Ye}

 \AtlasCoverEdBoardMember{Giovanni Marchiori (chair), Bruce Schumm, Isabel Marian Trigger, Marc-Andre Pleier}

 \AtlasCoverEgroupEditors{atlas-stdm-2014-01-editors}

 \AtlasCoverEgroupEdBoard{atlas-stdm-2014-01-editorial-board}

\hypersetup{pdftitle={ATLAS draft},pdfauthor={The ATLAS Collaboration}}

\begin{document}
\nolinenumbers
\maketitle



\section {Introduction}

The production of $\Zboson$ bosons has been used in many experiments to test the electroweak sector of the Standard Model (SM). 
Precision measurements made at LEP and at the SLAC Linear Collider established $\Zboson$ boson properties that are consistent with the SM assumption of a gauge boson without internal structure.  
Studies of the $\Zboson$ boson in hadroproduction experiments at the Tevatron and Large Hadron Collider (LHC) are in agreement with the production dynamics predicted by the 
$SU(2)_L\times U(1)_Y$ gauge group of the SM's electroweak sector.  
The couplings of the $\Zboson$ boson to $\Wpm$ bosons have been observed and agree with SM predictions. 
 No experimental evidence has been reported for
couplings of $\Zboson$ bosons to photons.
Anomalous properties of the $\Zboson$ boson are often constrained in terms of limits on the triple ($\Zboson\Zboson\gamma$ and  $\Zboson\gamma\gamma$) and quartic ($ZZ\gamma\gamma$ and $Z\gamma\gamma\gamma$) gauge-boson couplings. 
Such limits have been reported by many experiments at LEP~\cite{Achard:2004ds,Abdallah:2007ae,Abbiendi:2000cu,Abbiendi:2004bf}, the Tevatron~\cite{Abazov:2011qp,Abazov:2009cj,Aaltonen:2011zc}, and the LHC~\cite{Aad:2013izg,Chatrchyan:2013nda,CMS:2013zg,Khachatryan:2015kea}. 
In addition, searches for new gauge bosons decaying to $\Zboson\gamma$ have been used to further constrain physics beyond the SM~\cite{EXOT-2013-09,Aad:2013izg}.
 
\begin{figure}[hbtp]
\begin{center}
\centering
\mbox{
\begin{subfigure}[]{}
\includegraphics[scale=0.45,angle=0]{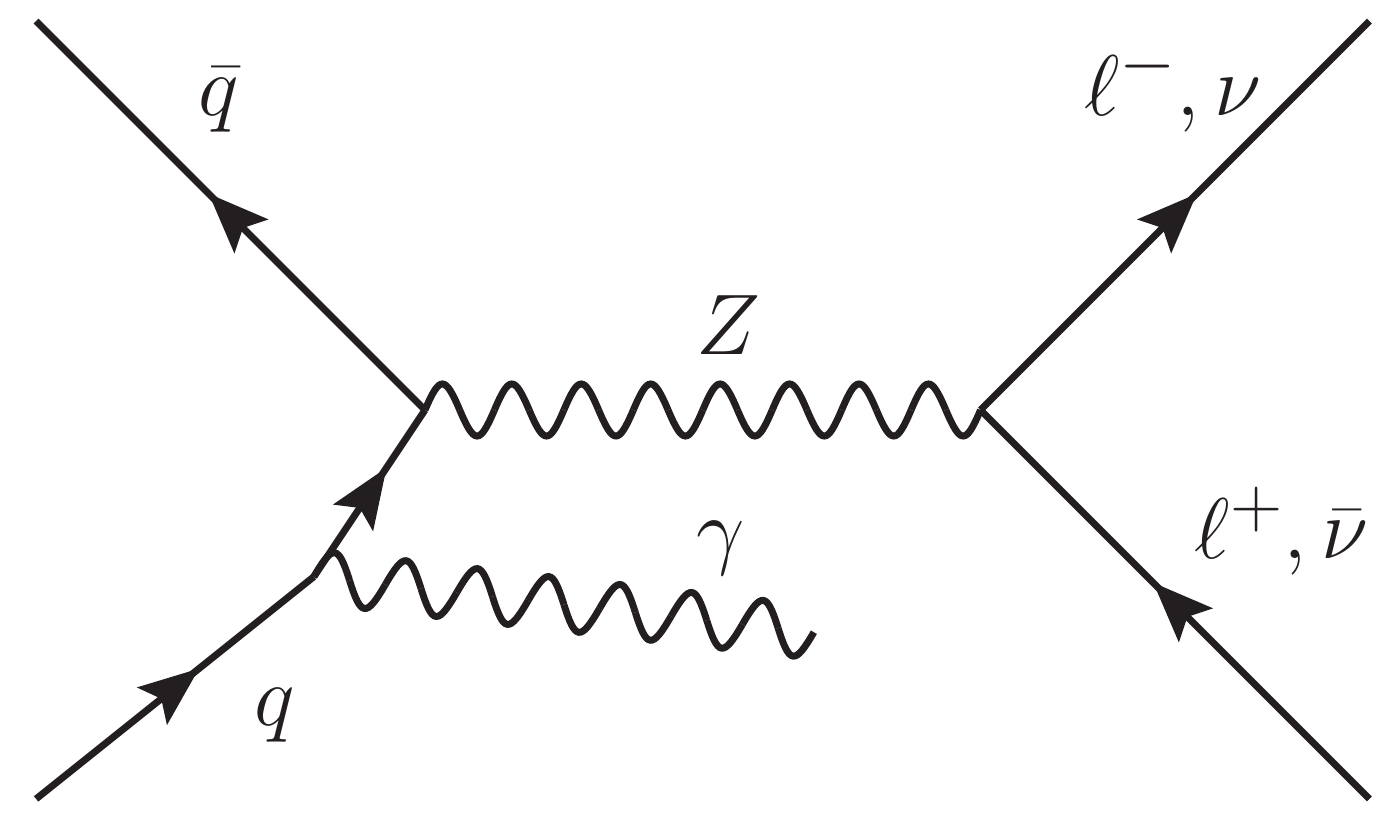}
\end{subfigure}

\begin{subfigure}[]{}
\includegraphics[scale=0.45,angle=0]{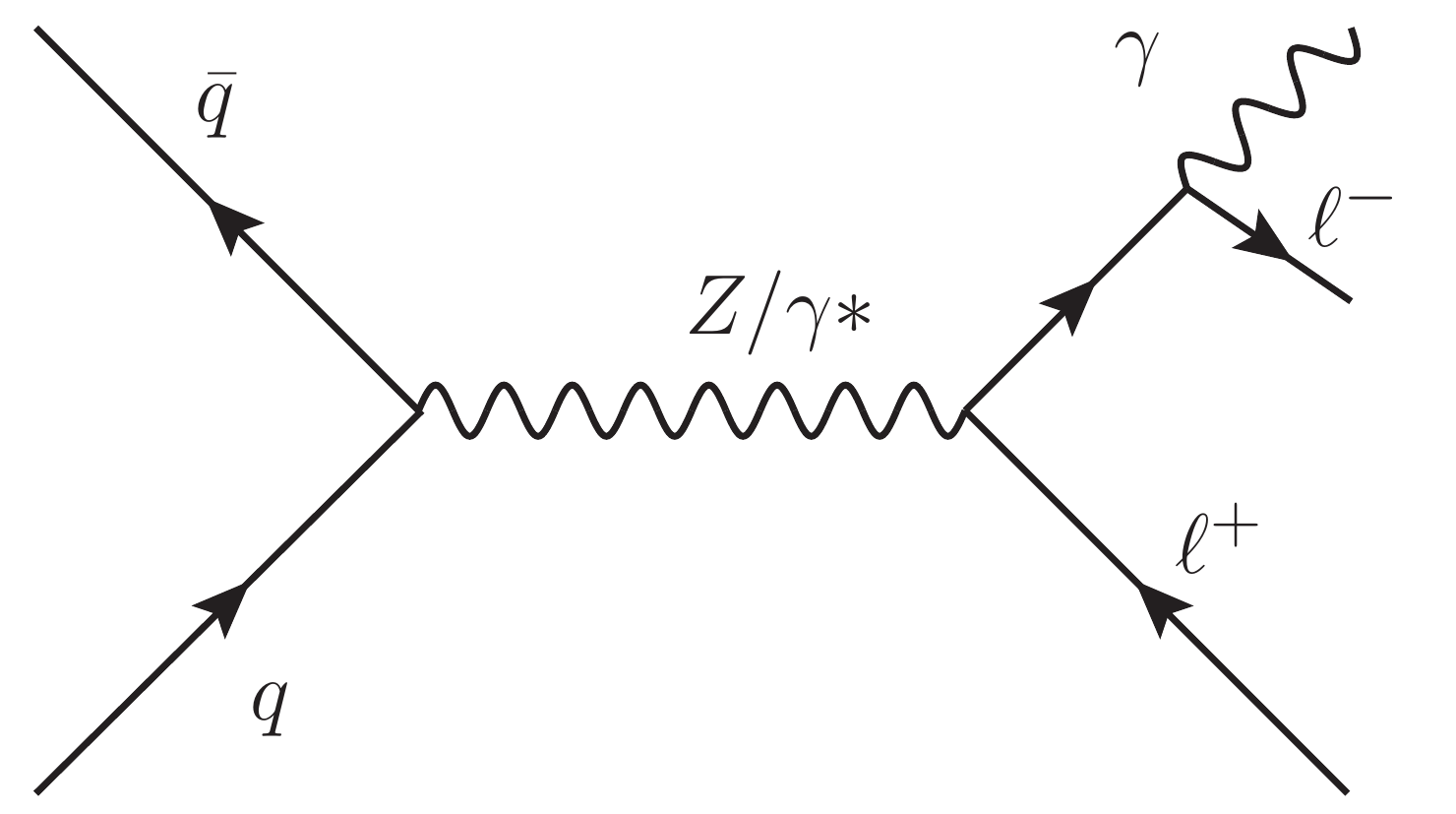}
\end{subfigure}
}
\mbox{
\begin{subfigure}[]{}
\includegraphics[scale=0.45,angle=0]{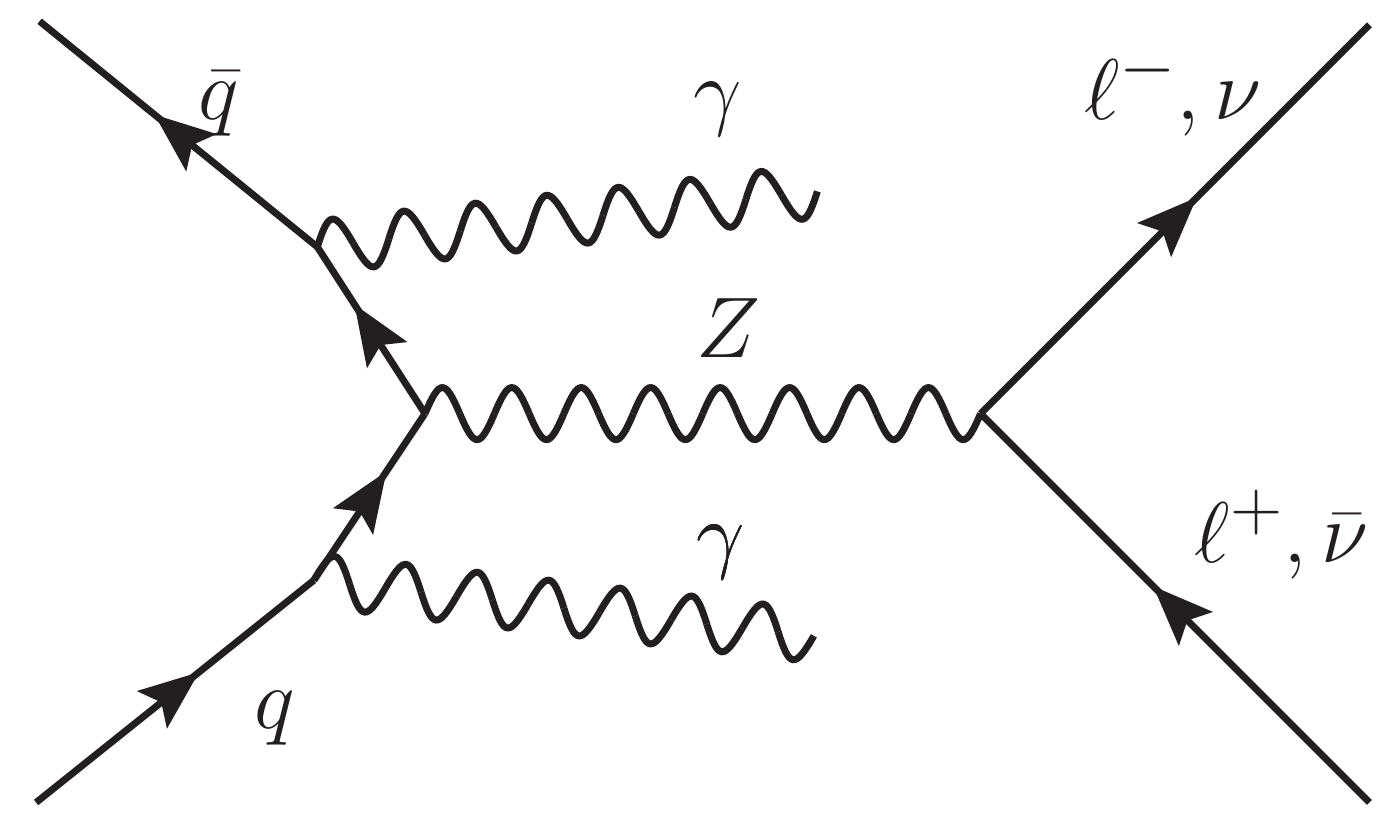}
\end{subfigure}
\begin{subfigure}[]{}
\includegraphics[scale=0.45,angle=0]{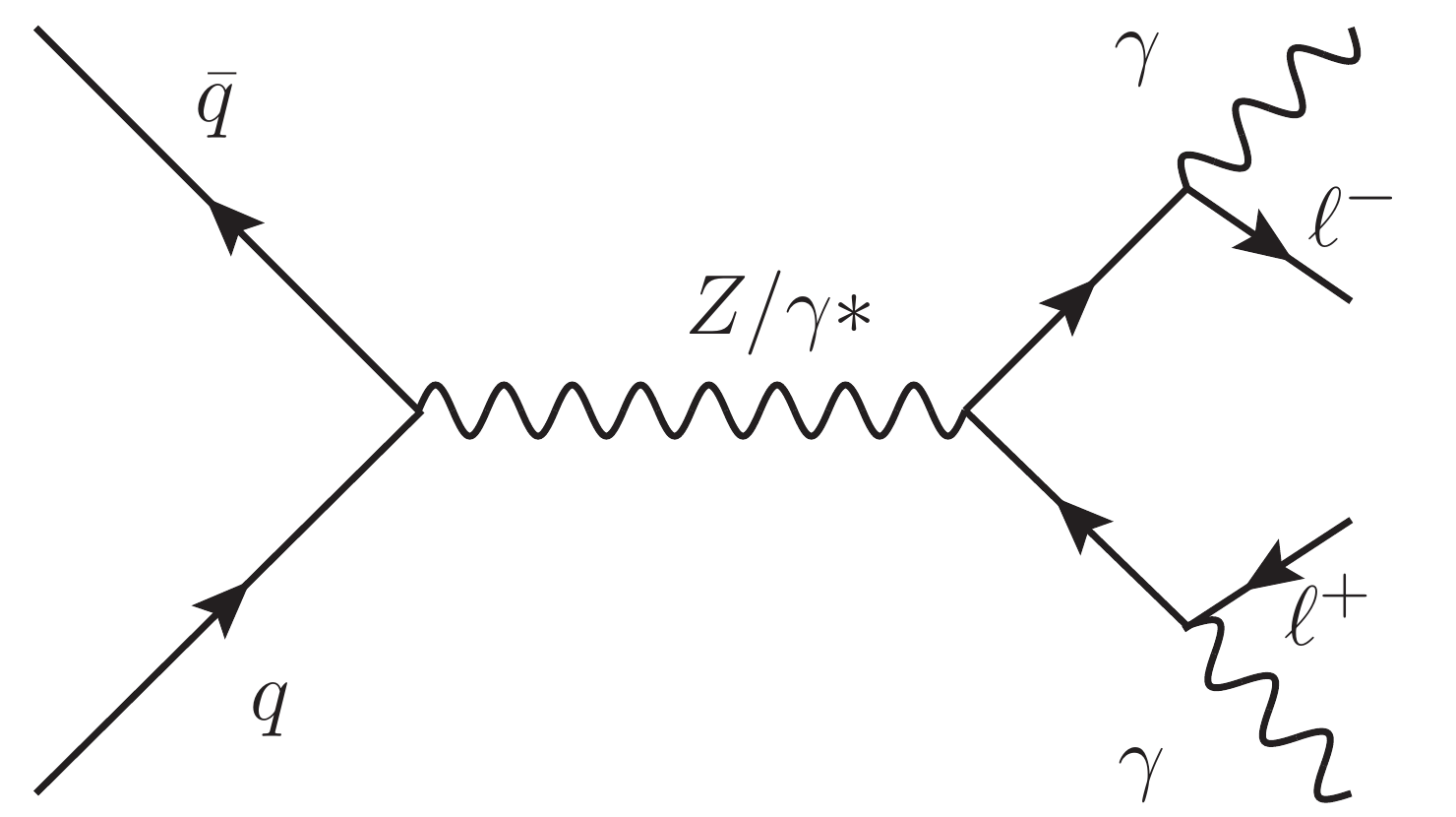}
\end{subfigure}
}
\mbox{
\begin{subfigure}[]{}
\includegraphics[scale=0.45,angle=0]{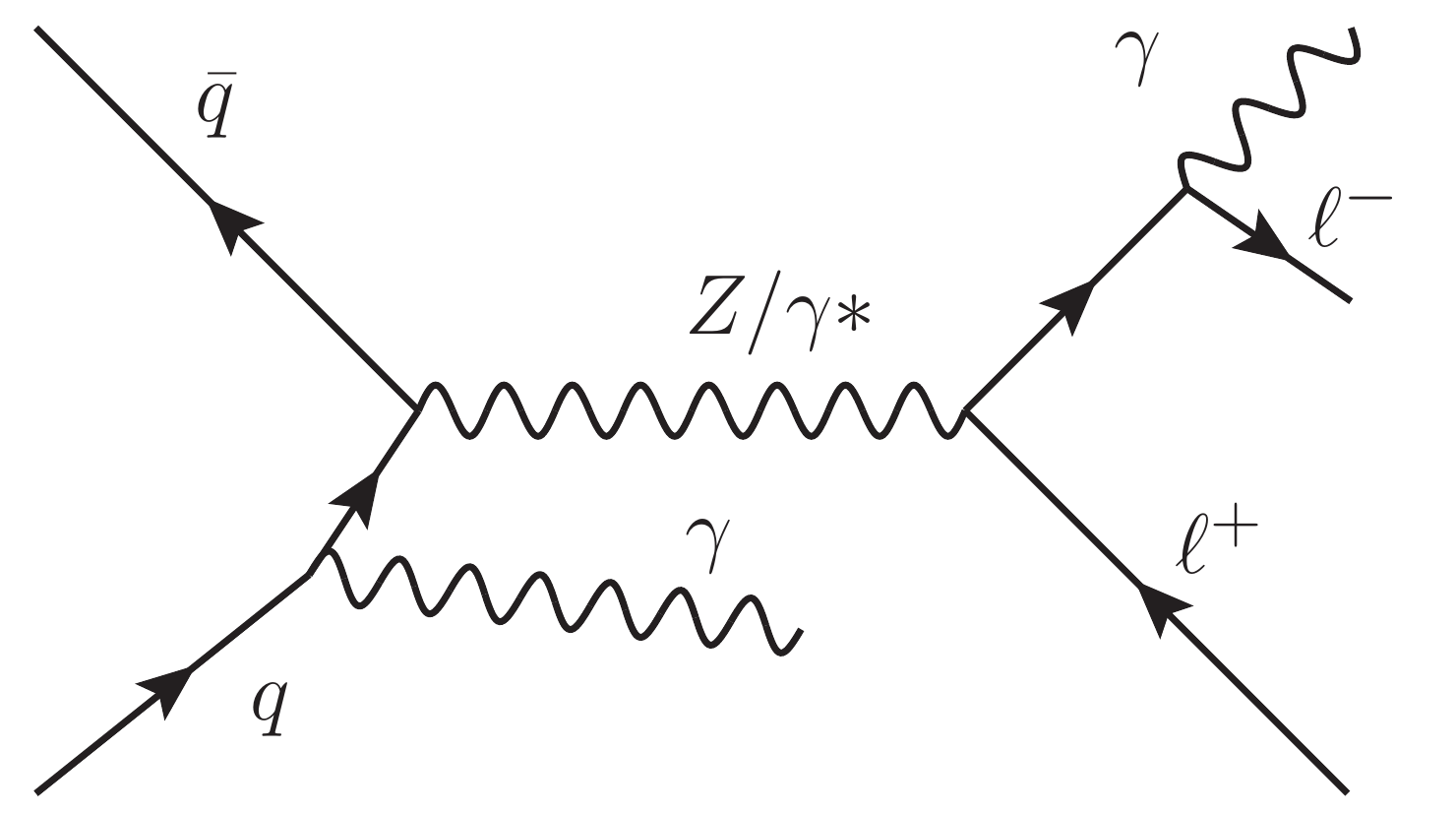}
\end{subfigure}
}
\mbox{
\begin{subfigure}[]{}
\includegraphics[scale=0.45,angle=0]{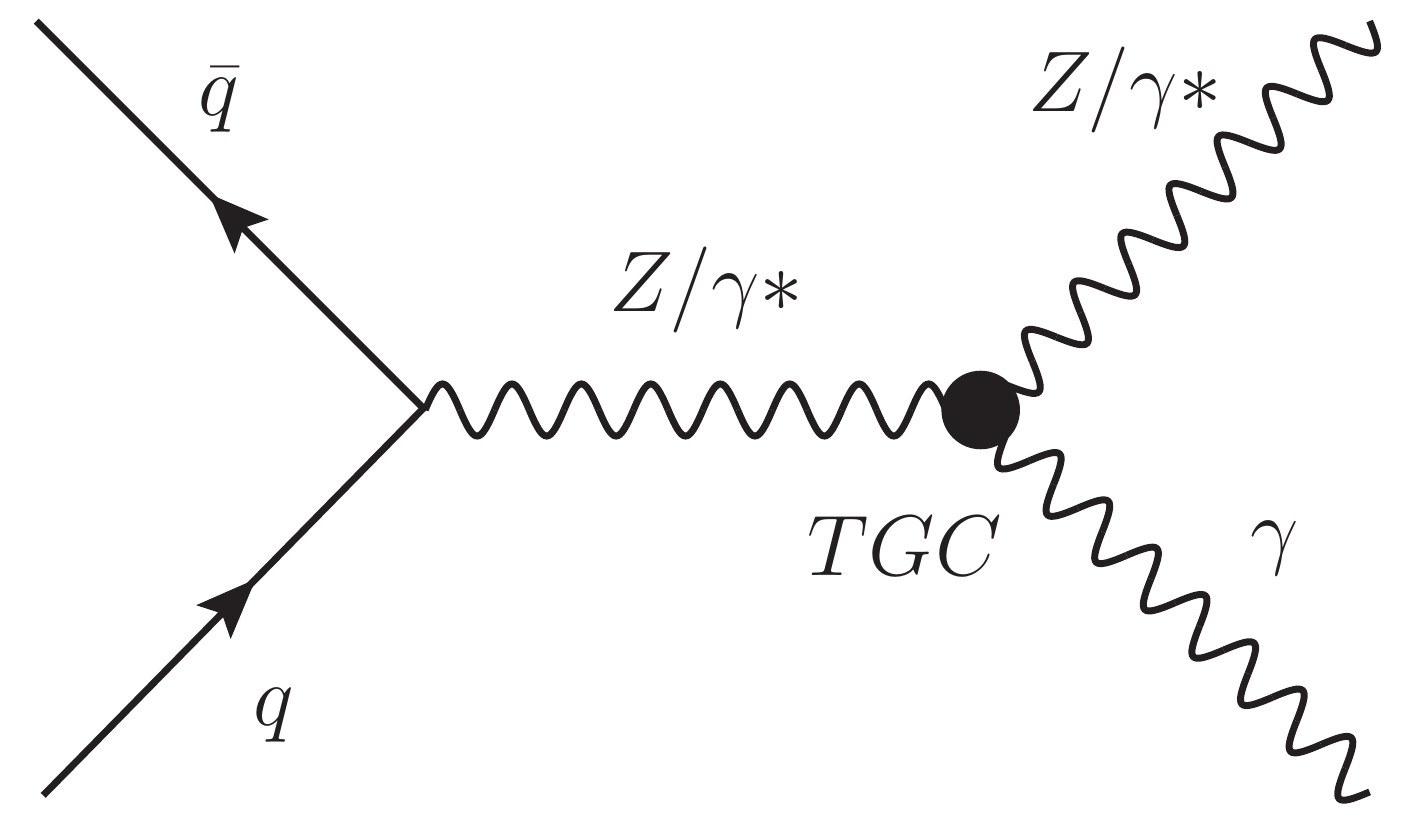}
\end{subfigure}
\begin{subfigure}[]{}
\includegraphics[scale=0.45,angle=0]{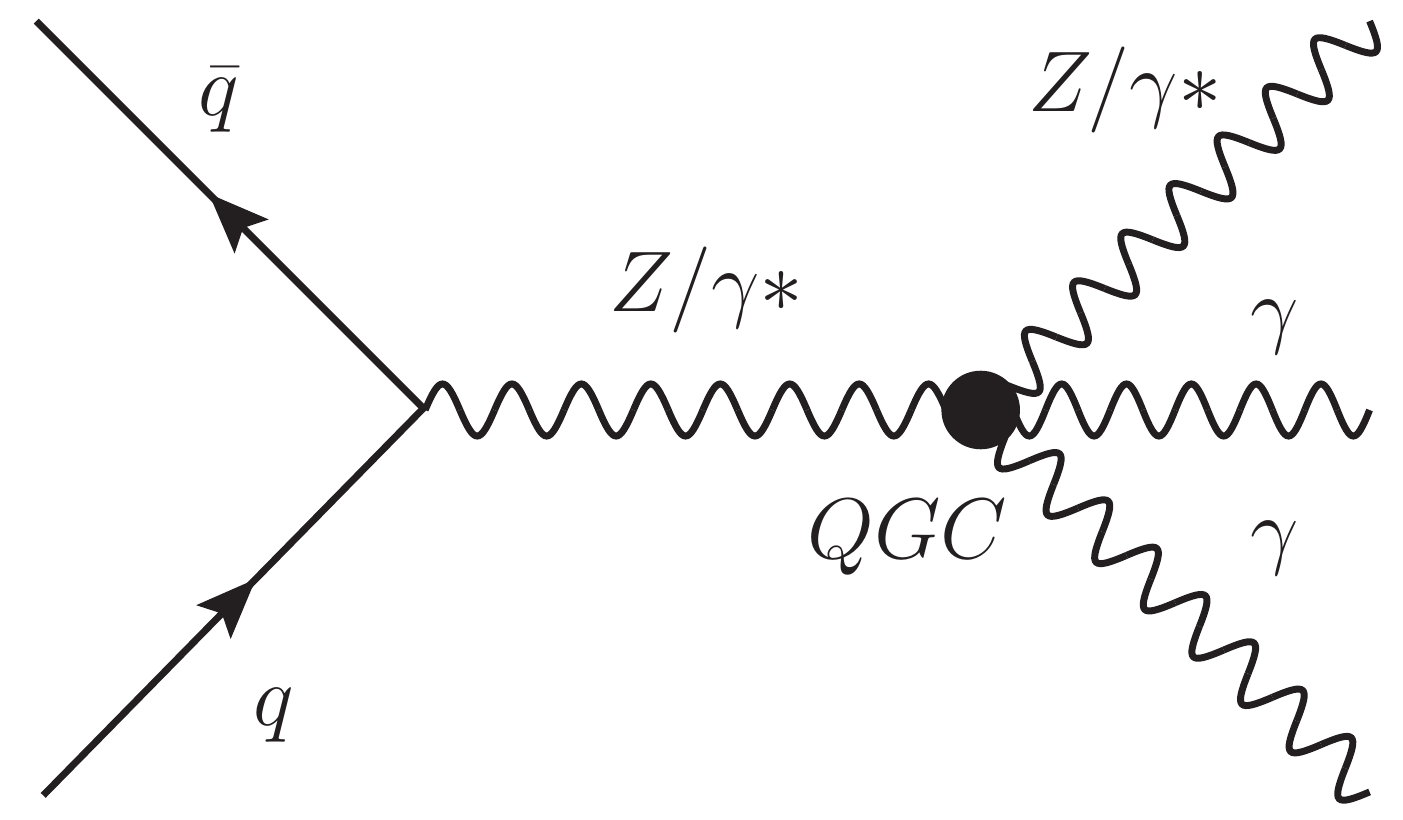}
\end{subfigure}
}

\caption{Feynman diagrams of $\Zboson\gamma(\gamma)$ production: (a,c) initial-state photon radiation (ISR), (b,d) final-state photon radiation (FSR), (e) mixed channel (FSR+ISR), (f) triple gauge-boson coupling (TGC) vertex, and (g) quartic gauge-boson coupling (QGC) vertex.}
\label{fig:feynmanZgg}
\end{center}
\end{figure}

Some of the elementary processes resulting in the production of a $Z$ boson in association with one or two photons are illustrated
by the leading-order Feynman diagrams shown in Figures~\ref{fig:feynmanZgg}(a)--\ref{fig:feynmanZgg}(e). Examples of triple and quartic gauge-boson couplings involving $\Zboson$ bosons and photons are shown in Figures~\ref{fig:feynmanZgg}(f) and~\ref{fig:feynmanZgg}(g). These couplings are forbidden at tree level in the SM, 
but can arise in theories predicting anomalous couplings.

This paper presents measurements of the hadroproduction of $\Zboson$ bosons associated with one or two isolated photons.  
The measurements use 20.3 fb$^{-1}$ of proton--proton ($pp$) collisions collected with the ATLAS detector at the CERN LHC operating at a center of mass energy of 8 \TeV.
The analyses use
the decays $\Zboson/\gamma^*\to\leplep$ (where $\ell$ =$e$ or $\mu$), with
the invariant mass of the dilepton pair above 40 \GeV, and $\Zboson\to\nnbar$. The $\Zboson/\gamma^*$ decays to charged leptons are selected using triggers on high transverse momentum\footnote{
ATLAS uses a right-handed coordinate system with its origin at the nominal
interaction point (IP) in the center of the detector and the $z$-axis
along the beam pipe.  The $x$-axis points from the IP to the center of the
LHC ring, and the $y$-axis points upward.  Cylindrical coordinates ($r,\phi$)
are used in the transverse ($x$,$y$) plane, with $\phi$ being the azimuthal
angle around the beam pipe.  The pseudorapidity is defined in terms of the
polar angle $\theta$ as $\eta=-\ln\tan(\theta/2)$.  The distance $\Delta R$
in the $\eta$--$\phi$ space is defined as $\Delta R$ = $\sqrt{(\Delta \eta)^2+(\Delta \phi)^2}$.
The transverse energy \ET~is defined as $\ET = E\times\sin\theta$.
} ($\pT$) electrons or muons. 
The production channels studied are $pp\to\ell^{+}\ell^{-}\gamma + X$ 
and $pp\to\ell^{+}\ell^{-}\gamma\gamma  + X$ where the photons are required to have transverse energy $\ET > 15$ \GeV. 
The events with $\Zboson$ boson decays to neutrinos are selected using high $\ET$ photon triggers.   
Measurements are made of the processes $pp\to\nu\bar{\nu}\gamma + X$ with photon $\ET > 130$ \GeV~and $pp\to\nu\bar{\nu}\gamma\gamma + X$ where both the photons have $\ET>$ 22 \GeV. 
In all the production channels, the measurements are made with no restriction on the recoil system $X$ (inclusive events) and by requiring that the system $X$ has no central jet ($|\eta|<4.5$) with $\pT$ > 30 \GeV~(exclusive events).
The SM sources of the direct photons are radiation off initial-state quarks and radiative $Z$-boson decay to charged leptons, and from fragmentation of final-state quarks and gluons into photons. 

The measurements are compared to SM predictions obtained with a parton-shower Monte Carlo (MC) simulation and with two higher-order perturbative parton-level calculations at next-to-leading order (NLO) and next-to-next-to-leading order (NNLO) in the strong coupling constant $\alpha_{\mathrm{s}}$.
The measured $Z\gamma$ production cross section at high values of
the photon $\ET$ is used to search for anomalous triple gauge-boson ($ZZ\gamma$ and $Z\gamma\gamma$)
couplings (aTGC). The measured $Z\gamma\gamma$ production cross section at high values of the diphoton mass
$m_{\gamma\gamma}$ is used to search for anomalous quartic gauge-boson ($ZZ\gamma\gamma$ and $Z\gamma\gamma\gamma$) couplings (aQGC).
Deviations from the SM Lagrangian are parameterized by adding higher-order operators that introduce couplings of photons to the $Z$ bosons.

This paper is organized as follows. 
The ATLAS detector is briefly described in Section~\ref{sec:atlasdet}. 
The signal and background simulation is presented in Section~\ref{sec:simulation}.
The object and event selections and the background estimation are described in Section~\ref{sec:selection} and Section~\ref{sec:backgrounds}, respectively.
The results of cross-section measurements and their comparison with the Standard Model predictions are presented in Section~\ref{sec:crossection} and Section~\ref{sec:comparison}, respectively.
The limits on the anomalous triple and quartic gauge-boson couplings are presented in Section~\ref{sec:agc}.
Section~\ref{sec:summary} provides the conclusions.  

\section{The ATLAS detector and LHC data sample}
\label{sec:atlasdet}

The ATLAS detector has been described in detail elsewhere~\cite{ATLAS1}.
A short overview is presented here with an emphasis on the electromagnetic calorimeter needed for precision measurement of the high-energy photons. The major components of the ATLAS detector are an
inner tracking detector (ID) surrounded by a thin
superconducting solenoid providing a 2~T axial magnetic field,
electromagnetic (EM) and hadronic calorimeters, and a muon spectrometer (MS).  
The ID is composed of three subsystems. The pixel and silicon
microstrip detectors cover the pseudorapidity range $|\eta|<2.5$, while
the transition radiation tracker (TRT) has an acceptance range of
$|\eta|<2.0$. The TRT provides identification information for
electrons by the detection of transition radiation. The MS is composed of three large superconducting air-core toroid magnets, a system of three stations of chambers for tracking
measurements with high precision in the range $|\eta|<2.7$, and a muon trigger system effective over the range $|\eta|<$ 2.4.

The electromagnetic calorimeter is a lead/liquid-argon detector
composed of a barrel ($|\eta|<1.475$) and two endcaps ($1.375<|\eta|<3.2$).  
For |$\eta$| $<$ 2.5 the calorimeter has three layers, longitudinal in shower depth, with the first layer having the highest
granularity in the $\eta$ direction, and the second layer collecting most
of the electromagnetic shower energy for high-$p_{\mathrm{T}}$ objects.  A thin presampler layer
covering the range $|\eta|<1.8$ is used to correct for the energy
lost by EM particles upstream of the calorimeter.
The hadronic calorimeter system, which
surrounds the electromagnetic calorimeter, is based on two different
detector technologies, with scintillator tiles or liquid-argon as the active
medium, and with either steel, copper, or tungsten as the absorber
material. Photons are identified as narrow, isolated showers 
in the EM calorimeter with no penetration into the hadronic calorimeter. 
The fine segmentation of the ATLAS calorimeter allows efficient rejection of jets fragmenting to high-energy $\pi^0$ or $\eta$ mesons that could be misidentified as isolated direct photons.

Collision events are selected using a three-level trigger system.
The first-level trigger is based on custom-built
electronics that use a subset of the total detector
information to reduce the data rate to below the design value of
75 kHz. The subsequent two trigger levels run on a processor farm
and analyze detector information with greater precision. The resulting
recorded event rate from LHC $pp$ collisions at $\sqrt{s}$ = 8 \TeV~during the data-taking period in 2012 was approximately 400 Hz.
After applying criteria to ensure nominal
ATLAS detector operation, the total integrated luminosity useful for data 
analysis is 20.3 fb$^{-1}$.
The uncertainty in the integrated luminosity is determined to be 1.9$\%$.  
It is derived, following the same methodology as that detailed in Ref.~\cite{Aad:2013lumi}, from a calibration of the luminosity scale obtained from beam-separation scans.

Online triggers based on high-energy electrons, muons, and photons are used
to select events with final states consistent with one of the four following processes:
\begin{itemize}
\item $pp\to e^+e^{-}\gamma(\gamma) + X$,
\item $pp\to\mu^+\mu^-\gamma(\gamma) + X$,
\item $pp\to\nu\bar{\nu}\gamma + X$,
\item $pp\to\nu\bar{\nu}\gamma\gamma + X$.
\end{itemize}

The $\leplep\gamma$ and $\leplep\gamma\gamma$ events are selected using single-lepton or dilepton triggers.
The $\pT$ thresholds are 24 \GeV~for single-lepton triggers, and 12 \GeV~(13
\GeV) for dielectron (dimuon) triggers.
A dimuon trigger with asymmetric muon $\pT$ thresholds of 8 \GeV~and 18 \GeV~is also used.
The $\nnbar\gamma$ and $\nnbar\gamma\gamma$ events are selected using a single-photon trigger with a threshold of $\ET>$ 120 \GeV~and a diphoton trigger with a threshold of $\ET>$ 20 \GeV, respectively.
For the events falling within the acceptance of the measurement, the trigger efficiency is close to $100\%$ for $e^+e^{-}\gamma(\gamma)$ and $\nnbar\gamma$ final states, about $99\%$ for $\nnbar\gamma\gamma$ final states, and about $95\%$ for $\mu^+\mu^-\gamma(\gamma)$ final states.

\section {Simulation of signals and backgrounds}
\label{sec:simulation}

Simulated signal and background events are produced with various Monte Carlo event generators, processed through a full ATLAS detector simulation~\cite{SOFT-2010-01} using \textsc{Geant4}~\cite{bib-geant4}, and then reconstructed with the same procedure as for data.
Additional $pp$ interactions (pileup), in the same and neighboring bunch crossings, are overlaid on the hard scattering process in MC simulation.
The MC events are then reweighted to reproduce the distribution of the number of interactions per bunch crossing in data. 
The mean number of interactions per bunch crossing in the dataset considered is 20.7.

\subsection{Monte Carlo generation of SM $pp \to \Zboson \gamma (\gamma) + X$ and anomalous gauge-boson couplings processes}
\label{sec:signalMC}

The efficiency of the event selection is studied using a MC simulation of the $\Zboson\gamma$ and $\Zboson\gamma\gamma$ signals using the $\SHERPA$ 1.4
generator \cite{Gleisberg:2008ta} with the CT10 parton distribution function (PDF) set~\cite{Lai:2010vv}, and leading-order (LO) matrix elements with up to three additional final-state partons 
for $\Zboson\gamma$ and up to one additional final-state parton for $\Zboson\gamma\gamma$.
$\SHERPA$ uses the CKKW scheme~\cite{Catani:2001cc,Krauss:2002up} to merge matrix elements and parton showers.
This "multileg" approach ensures that the first few hardest emissions are modeled by the real-emission matrix elements. 
$\SHERPA$ was found to adequately characterize the distributions of selected $\Zboson\gamma$ candidates in a previous publication~\cite{Aad:2013izg}. 
Theoretical uncertainties in the $\SHERPA$ predictions in Figures~\ref{fig:photonET_Zllg}--\ref{fig:ZnunuGGPlots} are taken to be the
same as those estimated with MCFM in Section~\ref{sec:theory_calc}.

Signal samples with anomalous triple and quartic gauge-boson couplings are generated using $\SHERPA$ for aTGC and {\textsc{Vbfnlo}} 2.7.0~\cite{bib:vbfnlo1, bib:vbfnlo2, bib:vbfnlo3} interfaced to $\PYTHIA$ 8.175~\cite{Sjostrand:2007gs} for parton showering, hadronization, and the underlying event for aQGC.
More details are given in Section~\ref{sec:agc}.

\subsection{Monte Carlo generation of background processes}

In the measurements of the $\ee\gamma(\gamma)$, $\mumu\gamma(\gamma)$, and $\nnbar\gamma(\gamma)$ production cross sections,
backgrounds are estimated either from simulation or from data.
The main backgrounds arise from object misidentification and are obtained using data-driven techniques, as described in Section~\ref{sec:backgrounds}. MC simulated backgrounds are used for validation in this case.
Smaller backgrounds are estimated directly from simulation.

The $W\Zboson$ and $\Zboson\Zboson$ backgrounds are generated with {\textsc{Powheg-box}} \cite{Frixione:2007vw, powheg-diboson} and the CT10 PDF set,
with parton showering, hadronization, and the underlying event modeled by $\PYTHIA$ 8.165 with the AU2 set of tuned parameters~\cite{ATL-PHYS-PUB-2012-003}.
The background arising from $t\bar{t}\gamma$ is generated with {\textsc{MadGraph5\_aMC@NLO}} 5.2.1.0~\cite{madgraph} and the CTEQ6L1~\cite{cteq} PDF set, 
with parton showering, hadronization, and the underlying event modeled by $\PYTHIA$ 8.183.
$\SHERPA$ 1.4 with the CT10 PDF set is used to simulate $\tautau\gamma(\gamma)$, $\gamma$+jets, and
 $W\gamma(\gamma)$ events. 
An alternative MC sample of simulated $\gamma+\mathrm{jet}$ events is generated using $\PYTHIA$ 8.165 with the CTEQ6L1 PDF set.
An alternative MC sample of simulated $W\gamma$ events is generated using $\ALPGEN$ 2.14~\cite{Mangano:2002ea} with the CTEQ6L1 PDF set, interfaced to $\HERWIG$ 6.520~\cite{herwig} with {\textsc{Jimmy}} 4.30~\cite{Butterworth:1996zw} and the AUET2 set of tuned parameters~\cite{ATL-PHYS-PUB-2011-008} for parton showering, hadronization, and the underlying event. 
The $t\bar{t}\gamma$, $W\Zboson$, and $\Zboson\Zboson$ backgrounds are normalized using the NLO cross sections~\cite{powheg-diboson,Melnikov:2011ta}; the $\tautau\gamma$ and $\tautau\gamma\gamma$ backgrounds are normalized using the cross sections predicted by $\SHERPA$.

\section {Selection of $\Zboson\gamma$ and $\Zboson\gamma\gamma$ signal events}
\label{sec:selection}

The event selection criteria are chosen to provide precise cross section measurements of $Z\gamma$ and $Z\gamma\gamma$ production, and to provide good sensitivities to anomalous gauge-boson couplings between photons and the $\Zboson$ bosons.
The selections are optimized
for each of these measurements to obtain
high signal efficiency together with good background rejection.

\subsection{Physics object reconstruction and identification}
\label{subsec:sel_id_and_reco}

Collision events are selected by requiring at least one reconstructed primary vertex candidate with at least three charged-particle tracks with $\pT > 0.4$ 
\GeV. The vertex candidate with the highest sum of the $\pT^2$ of the associated 
tracks is chosen as the event primary vertex. 
This criterion may choose the wrong primary vertex in $\nnbar\gamma(\gamma)$ events. 
The effect of such a wrong choice was studied in simulation and found to have negligible impact on the photon transverse energy resolution for this analysis.

Electron candidates are reconstructed within the fiducial acceptance region $|\eta| <$ 2.47 from an energy cluster in the EM calorimeter 
associated with a reconstructed track in the ID~\cite{Aad:2014elid}. Photon candidates are reconstructed from energy clusters with $|\eta| <$ 2.37~\cite{ATLAS-CONF-2012-123}.   
The EM cluster of the electron/photon candidate must lie outside the transition region between the barrel and endcap EM calorimeters, 
thus electrons and photons with 1.37 $< |\eta| <$ 1.52 are rejected. The cluster energies are corrected using an in situ calibration based on the known $\Zboson$ boson mass~\cite{Aad:2012elid}.
Clusters without matching tracks are classified as unconverted photon candidates, whereas clusters that are matched to one or two tracks that originate from a conversion vertex are
considered as converted photon candidates. Both the unconverted and converted candidates are used in the analysis.
Electron tracks are required to be matched to the event primary vertex. 
The electron $d_{0}$ significance, defined as the ratio of the absolute value of the transverse impact parameter, $d_{0}$, with respect to the primary vertex, to its measured uncertainty, must be less than 6.0, and
the weighted electron longitudinal impact parameter with respect to the primary vertex $|z_{0}\times\sin\theta|$ must be less than 0.5 mm. 
Reconstructed electrons are required to have $\pT > 25$ \GeV. 
The photon $\ET$  threshold depends on the analysis channel.

Muon candidates are identified, within pseudorapidity $|\eta| <$ 2.5, by matching complete tracks or track segments in the MS to tracks in the ID~\cite{PERF-2014-05}. 
Similarly to electrons, the muon candidates are required to be matched to the
primary vertex with a transverse impact parameter significance of less
than 3.0, and a weighted longitudinal impact parameter $|z_{0}\times\sin\theta|$ of
less than 0.5 mm.
Reconstructed muons are required to have $\pT > 25$ \GeV.

Photons and electrons are required to meet identification criteria
based on shower shapes in the EM calorimeter,
leakage into the hadronic calorimeter, and ID tracking information. The
resulting selected photons are classified as
"loose" or "tight" and the electrons as "medium" as defined 
in Refs.~\cite{ATLAS-CONF-2012-123, ATLAS-CONF-2014-032, Aad:2014elid}. The "tight" identification
criterion for photons is used to suppress the 
background from multiple showers produced in meson (e.g., $\pi^0, \eta $) decays~\cite{ATLAS-CONF-2012-123}. The electron
identification criteria are used to suppress background electrons (primarily from photon conversions and Dalitz decays) and jets faking electrons~\cite{Aad:2012elid}.

Photons, electrons, and muons are required to be isolated from nearby hadronic activity.
Photons are considered isolated if the sum of transverse energy calculated from clusters of calorimeter energy deposits~\cite{Lampl:1099735} in an "isolation" cone of size $\Delta R=0.4$ around the candidate, $\ET^\mathrm{iso}$,
is smaller than 4 \GeV~after subtracting the contribution from the photon itself, and corrected for the leakage of the photon energy and the effects of underlying event and pileup~\cite{Aad:2010sp,Cacciari:2008gn}.
For electrons to be isolated, the calorimeter transverse energy deposits and the sum of the transverse momenta of tracks in a cone of size $\Delta R = 
0.2$ around the candidate after subtracting the contribution from the electron itself must be below $0.14 \times \pT^e$ and $0.13 \times \pT^e$, respectively, where $\pT^e$ is the electron transverse momentum. 
Muons are considered isolated if the sum of the transverse momenta of ID tracks excluding the track associated with the muon in a cone of size $\Delta R = 0.2$ is below $0.1 \times \pT^{\mu}$, where $\pT^{\mu}$ is the muon transverse momentum.

All lepton and photon efficiencies of the trigger, reconstruction, and identification are corrected in the simulation with data-derived correction factors.

Jets are reconstructed from clustered energy deposits in the calorimeter using the anti-$k_t$ algorithm~\cite{Cacciari:2008gp} with radius parameter $R = 0.4$ and are required
to have $\pT >$ 30 \GeV~and $|\eta| <$ 4.5. 
Reconstructed calorimeter jets are corrected for effects of
noncompensating response, energy losses in the dead material, shower leakage,
as well as inefficiencies in energy clustering and jet reconstruction
by applying a simulation-based correction derived in bins of $\eta$ and $E$. An in situ calibration corrects for differences between data and
simulation in the jet response. This jet energy scale
calibration is thoroughly discussed in Ref.~\cite{ATLAS-CONF-2015-037}.
In order to reduce pileup effects, 
for jets with $\pT <$ 50 \GeV~and $|\eta| <$ 2.4 the jet vertex fraction (JVF), 
defined as the ratio of the summed scalar $\pT$ of tracks associated with both the $R = 0.4$ jet and the primary vertex to that of all tracks associated with the jet,
must be greater than 0.5.

To reject electrons reconstructed from a bremsstrahlung photon emitted by a muon traversing the calorimeter, any electron candidate within a $\Delta R = 0.1$ cone around a selected muon is removed. 
Jets are removed if they are found within a $\Delta R = 0.3$ cone around a selected lepton or photon.

The missing transverse momentum vector $\vec{p}_{\mathrm{T}}^{\mathrm{\;miss}}$ is the vector of momentum imbalance in the transverse plane. The reconstruction of the direction and magnitude
of the missing transverse momentum vector is described in Ref.~\cite{ATLAS-CONF-2013-082}.
The $\vec{p}_{\mathrm{T}}^{\mathrm{\;miss}}$ is calculated from the vector sum of the calibrated transverse momenta of all jets with
$\pT$ > 20 \GeV~and $|\eta|$ < 4.5, the transverse momenta of electron and muon candidates, and all calorimeter energy clusters not belonging to a reconstructed object (soft-term).
Selection criteria based on $\vec{p}_{\mathrm{T}}^{\mathrm{\;miss}}$ or its magnitude $\met$ are used only in the neutrino channels, as described in Section~\ref{subsec:sel_Znunu}.

\subsection{Selection of $\leplep\gamma$ and $\leplep\gamma\gamma$ event candidates}
\label{subsec:selllgg}

Selected $\leplep\gamma$ or $\leplep\gamma\gamma$ event candidates must contain exactly one pair of same-flavor, opposite-charge isolated leptons (electrons or muons) and at least one or two isolated photons with $\ET^{\gamma} > $ 15 \GeV, respectively. 
In the case of additional photon candidates, those with the highest $\ET^{\gamma}$ are selected.
The dilepton invariant mass $m_{\leplep}$ is required to be greater than 40 \GeV.
The reconstructed photons are removed if they are found within a $\Delta R =$ 0.7 (0.4) cone around a selected lepton for $\leplep\gamma$ ($\leplep\gamma\gamma$) events.
A further requirement on the photon--photon separation of $\Delta R(\gamma,\gamma)$ > 0.4 is applied in \llgg events.
The selected events are categorized as inclusive events, referring to those with no requirement on the jets, and exclusive events, which are defined to be those with no selected jet with $\pT > 30$ \GeV~and $|\eta|<4.5$.

\subsection{Selection of $\nnbar\gamma$ and $\nnbar\gamma\gamma$ event candidates}
\label{subsec:sel_Znunu}

The $\nnbar\gamma$ event candidates are selected by considering events with $\met > $ 100 \GeV~and at least one isolated photon with $\ET^{\gamma} > $ 130 \GeV.
The separation between the reconstructed photon direction and $\vec{p}_{\mathrm{T}}^{\mathrm{\;miss}}$ in the transverse plane is required to be 
$\Delta \phi(\vec{p}_{\mathrm{T}}^{\mathrm{\;miss}},\gamma) > \pi/2$, 
since in signal events the $\Zboson$ boson should recoil against the photon.
The $\nnbar\gamma\gamma$ event candidates are selected by considering events with $\met>110$ \GeV~and at least two isolated photons with $\ET > 22$ \GeV~and $\Delta R(\gamma,\gamma)>0.4$.
The directions of the di-photon system and the $\vec{p}_{\mathrm{T}}^{\mathrm{\;miss}}$ are required to be separated in the transverse plane by
$\Delta \phi(\vec{p}_{\mathrm{T}}^{\mathrm{\;miss}},\gamma\gamma) > 5\pi/6$.
In the case of additional photon candidates in $\nnbar\gamma$/$\nnbar\gamma\gamma$ events, one/two photons with the highest $\ET^{\gamma}$ are selected.
To suppress $W(\gamma)$+jets and $W\gamma(\gamma)$ backgrounds, 
events containing an identified muon or electron (as defined in Section~\ref{subsec:sel_id_and_reco} without isolation requirement) are rejected.
The selected events are categorized as inclusive events and exclusive events, as described in Section~\ref{subsec:selllgg}.

\section {Estimation of backgrounds}
\label{sec:backgrounds}

This section describes the background estimation in each of the final states. 
The backgrounds in the $\leplep\gamma$ and $\leplep\gamma\gamma$ final states are discussed in Section~\ref{sec:backgrounds_Zllg_Zllgg}. 
The dominant backgrounds in these final states are $Z$+jets and $Z\gamma$+jets with jets misidentified as photons. 
The backgrounds in the $\nnbar\gamma$  and $\nnbar\gamma\gamma$  final states are discussed in Section~\ref{sec:bkgnngnngg}.
The dominant backgrounds in these final states are those with jets misidentified as photons, those with electrons misidentified as photons, as well as $W(\ell\nu)\gamma$ and $W(\ell\nu)\gamma\gamma$ where the lepton from the $W$ decay is not detected.

\subsection{Backgrounds to $\leplep\gamma$ and $\leplep\gamma\gamma$}
\label{sec:backgrounds_Zllg_Zllgg}

Backgrounds in the selected $\leplep\gamma$ and $\leplep\gamma\gamma$ samples are dominated by events in which hadronic jets, which contain photons from $\pi^{0}$ or $\eta$ decays, are misidentified as prompt photons. 
In the $\leplep\gamma$ measurement, the background from jets misidentified as photons originates from the production of $Z$ bosons in association with jets ($Z$+jets),
while in the $\leplep\gamma\gamma$ measurement this background originates from both $\Zboson\gamma$ in association with jets ($\Zboson\gamma$+jets) and $\Zboson$+jets events with one or two jets misidentified as photons, respectively. 
The backgrounds from jets misidentified as photons are estimated using data-driven methods as described in Sections~\ref{sec:dd_bkg_Zllg} and~\ref{sec:dd_bkg_Zllgg}. 
Smaller backgrounds originate from $t\bar{t}\gamma$, $WZ$, and $\tautau\gamma$ for $\leplep\gamma$, and from $WZ$, $ZZ$, and $\tautau\gamma\gamma$ for $\leplep\gamma\gamma$.
The backgrounds from $t\bar{t}\gamma$ and $\tautau\gamma(\gamma)$ yield the same final states as the signals, while the backgrounds from $WZ$ and $ZZ$ meet the selection criteria when the electrons from the $W$ or $Z$ decay are misidentified as photons or when final-state photons are radiated.
These are expected to contribute in total less than $1.5\%$ of the selected event yield in both the $\leplep\gamma$ and $\leplep\gamma\gamma$ final states, and are derived from simulation as described in Section~\ref{sec:bkgresults}.

\subsubsection{Estimation of the background from jets misidentified as photons in $\leplep\gamma$ measurements}
\label{sec:dd_bkg_Zllg}

For the $\leplep\gamma$ measurement, a two-dimensional sideband method is used to measure the background from jets misidentified as photons, as described in Refs.~\cite{Aad:2013izg,Aad:2010sp}. 
In this method, a looser photon selection is considered, in which the isolation and some identification requirements on the photon are discarded.
After this selection, the $\leplep\gamma$ events are separated into one signal and three control regions, defined by varying the photon identification and isolation requirements.
Photon candidates failing a subset of requirements on the photon shower-shape variables but satisfying all other requirements in the "tight" photon identification are considered as "nontight".
Events in the signal region (A) have the photon satisfying the nominal photon isolation and "tight" identification requirements as described in Section~\ref{subsec:sel_id_and_reco}. 
The three control regions are defined as:

\begin{description}
\item[i.] Control region B: the photon candidate meets the "tight" identification criteria and is not isolated ($\ET^\mathrm{iso} > 4$ \GeV);
\item[ii.] Control region C: the photon candidate meets the "nontight" identification criteria and is isolated ($\ET^\mathrm{iso} < 4$ \GeV);
\item[iii.] Control region D: the photon candidate meets the "nontight" identification criteria and is not isolated ($\ET^\mathrm{iso} > 4$ \GeV).
\end{description}

The shower-shape requirements that the "nontight" photons are required to fail are chosen to enhance the $\Zboson$+jets background events in the control regions while minimizing the correlation with the photon isolation. 
The number of $Z+$jets events in the signal region, $N^{j\rightarrow\gamma}_{\mathrm{A}}$, can be derived from the number of observed events in the control regions $N_{i}$ ($i$=B,C,D) :
\begin{linenomath}
\begin{align}
N^{j\rightarrow\gamma}_{\mathrm{A}} & = \left( (N_{\mathrm{B}} - N_{\mathrm{B}}^{\mathrm{Other~BKG}} - c_{\mathrm{B}} N_{\mathrm{A}}^{Z\gamma})\frac{N_{\mathrm{C}} - N_{\mathrm{C}}^{\mathrm{Other~BKG}} - c_{\mathrm{C}} N_{\mathrm{A}}^{Z\gamma}}{N_{\mathrm{D}} - N_{\mathrm{D}}^{\mathrm{Other~BKG}} - c_{\mathrm{D}} N_{\mathrm{A}}^{\Zboson\gamma}} \right) R, \label{eq:ABCD1} \\
N_{\mathrm{A}}^{Z\gamma} & = N_{\mathrm{A}} - N_{\mathrm{A}}^{\mathrm{Other~BKG}} - N^{j\rightarrow\gamma}_{\mathrm{A}}. \label{eq:ABCD2}
\end{align}
\end{linenomath}
The coefficients $c_i$ ($i$=B,C,D) are equal to the ratio of the \llg yields in the control regions to the signal region, and are estimated from simulation. 
The $R$ factor accounts for a potential correlation between the photon identification and isolation variables for the $Z$+jets background.
The central value of $R$ is taken to be one, as would be the case for no correlation. 
Its uncertainty of $20\%$ is determined by the deviation of the $R$ value from one as determined from simulation studies of the $Z$+jets background. 
The yields $N_{i}^{\mathrm{Other~BKG}}$ ($i$=A,B,C,D) are the contributions from other electroweak backgrounds in each region taken from simulation.
Equations~(\ref{eq:ABCD1}) and~(\ref{eq:ABCD2}) yield a quadratic expression in the unknown variable $N^{j\rightarrow\gamma}_{\mathrm{A}}$. 
The solution with physical meaning is retained.

The uncertainty in the value of $R$ represents the dominant systematic uncertainty of $24\%$ in the estimate of the $Z$+jets background.
The second largest systematic uncertainty of $10\%$ arises from the inaccuracy in modeling of the coefficients $c_i$, mainly due to the uncertainties in photon identification and isolation efficiencies. 
An additional $Z$+jets background uncertainty of $5\%$ arises from uncertainties in the estimates of the $N_{i}^{\mathrm{Other~BKG}}$ in each of the control regions.

\subsubsection{Estimation of the background from jets misidentified as photons in \llgg measurements}
\label{sec:dd_bkg_Zllgg}

A matrix method as described in Ref.~\cite{STDM-2011-05} is used to estimate the background from jets misidentified as photons in $\leplep\gamma\gamma$ events from $Z(\leplep)\gamma$+jets and $Z(\leplep)$+jets events with one or two jets misidentified as photons. 
The method uses as inputs the jet-to-photon misidentification rate (fake rate), $f$, which is the probability for a jet satisfying "loose" photon identification criteria~\cite{ATLAS-CONF-2012-123} to be identified as a "tight" and isolated photon, 
and the real photon identification efficiency, $\epsilon$, which is the probability for "loose" prompt photons to be identified as "tight" and isolated photons.
The fake rate and the real photon identification efficiency are estimated from data and from MC simulation, respectively.
A 4$\times$4 matrix is constructed from the fake rate and the real photon identification efficiency, relating the observed number of events, $N_{\mathrm{TT}}$, $N_{\mathrm{TL}}$, $N_{\mathrm{LT}}$, $N_{\mathrm{LL}}$, 
to the unknown number of each type of event, $N_{\gamma\gamma}$, $N_{\gamma\mathrm{jet}}$, $N_{\mathrm{jet} \gamma}$, $N_{\mathrm{jet}\mathrm{jet}}$, by a set of linear equations: 
\begin{linenomath}
\begin{equation}
\left( \begin{array}{c}
N_{\mathrm{TT}}\\
N_{\mathrm{TL}}\\
N_{\mathrm{LT}}\\
N_{\mathrm{LL}}\end{array}\right) =
 \left( \begin{array}{cccc}
\epsilon_1\epsilon_2 & \epsilon_1 f_2 & f_1 \epsilon_2 & f_1 f_2\\
\epsilon_1(1-\epsilon_2) & \epsilon_1(1-f_2) & f_1(1-\epsilon_2) & f_1(1-f_2)\\
(1-\epsilon_1)\epsilon_2 & (1-\epsilon_1)f_2 & (1-f_1)\epsilon_2 & (1-f_1)f_2\\
(1-\epsilon_1)(1-\epsilon_2) & (1-\epsilon_1)(1-f_2) & (1-f_1)(1-\epsilon_2) & (1-f_1)(1-f_2)\end{array} \right)  \left( \begin{array}{c}
N_{\gamma\gamma} \\
N_{\gamma\mathrm{jet}}\\
N_{\mathrm{jet} \gamma}\\
N_{\mathrm{jet}\mathrm{jet}}\end{array}\right).
\label{equ:matrix}
\end{equation}
\end{linenomath}
In the subscripts TT, TL, LT, LL, the first (second) subscript refers to the leading (subleading) reconstructed photon candidate; T means that it is "tight" and isolated while L corresponds to a "loose", not "tight" or not isolated candidate.
Similarly, the subscripts $\gamma\gamma$, $\gamma$jet, jet$\gamma$, and jetjet correspond to the cases of two photons, leading photon and subleading jet, leading jet and subleading photon, and two jets, respectively. 
The subscripts 1 and 2 refer to the leading and subleading photon candidates, respectively.
The number of each type of event, $N_{\gamma\gamma}$, $N_{\gamma\mathrm{jet}}$, $N_{\mathrm{jet} \gamma}$, $N_{\mathrm{jet}\mathrm{jet}}$, is obtained by solving Equation~(\ref{equ:matrix}), from which
the number of background events with jets misidentified as photons in the signal region, $N_{\mathrm{TT}}^{j\rightarrow\gamma}$, is then obtained: $N_{\mathrm{TT}}^{j\rightarrow\gamma} = \epsilon_1f_2\times N_{\gamma\mathrm{jet}} + f_1\epsilon_2\times N_{\mathrm{jet} \gamma} + f_1f_2\times N_{\mathrm{jet}\mathrm{jet}}$.

The fake rate is estimated from data using a sample enriched in $\Zboson(\leplep)+$jets with one jet misidentified as a photon.
To suppress the contribution from $\Zboson\rightarrow \leplep\gamma$, the invariant mass of opposite-charge dilepton pairs in the events is required to be within 8 \GeV~of the $\Zboson$ boson mass. 
A two-dimensional sideband method similar to that described in Section~\ref{sec:dd_bkg_Zllg} is used to estimate the number of $\leplep+$jets events in which the "loose" jets satisfy the "tight" identification and isolation requirements.  
As the fake rate depends on the photon $\ET$, a fake rate as a function of the photon $\ET$ is used in the matrix method.
The real photon identification efficiency, which is also a function of the photon $\ET$, is estimated from MC simulation.

The systematic uncertainty related to the background from jets misidentified as photons is dominated by the potential bias of the two-dimensional sideband method to estimate the fake rate.
It is evaluated from $Z$+jets MC simulation to be about $23\%$, by comparing the fake rate calculated by the two-dimensional sideband method to the fake rate calculated using the generator-level information in the MC simulation. 
Other systematic uncertainties, arising from possible inaccuracy in modeling of the real photon identification efficiency, other electroweak backgrounds, as well as the dependence of $\epsilon$ and $f$ on photon $\eta$, sum to about $10\%$.

\subsubsection{Results of the background estimation for $\leplep\gamma$ and $\leplep\gamma\gamma$}
\label{sec:bkgresults}
The backgrounds other than those from jets misidentified as photons are estimated using MC simulation.
The systematic uncertainties in these backgrounds consist of the experimental uncertainties described in Section~\ref{sec:intXsec} and the cross-section uncertainties,
which are $22\%$ ($t\bar{t}\gamma$~\cite{Melnikov:2011ta}), 
$10\%$ ($W\Zboson$~\cite{Campbell:2011bn,Campanario:2012fk}) and $15\%$ ($\Zboson\Zboson$~\cite{Campbell:2011bn,Cascioli:2014yka}). 
The cross-section uncertainties in the $\tautau\gamma$ and $\tautau\gamma\gamma$ backgrounds are evaluated to be 7\% using MCFM, as described in Section~\ref{sec:theory_calc}. 
An additional uncertainty of $30\%$ ($60\%$) is assigned to the $WZ$ ($ZZ$) background to account for the mismodeling of the electron-to-photon fake rate.
This uncertainty is estimated by comparing the fake rate predicted by simulation to that estimated in data, using the method described in Section~\ref{sec:bkg-Znunug-Wenu}.

The number of events observed in data, $N^{\mathrm{obs}}_{\Zg}$, as well as the estimated background yields in the \llg and \llgg measurements, are summarized in Tables~\ref{table:results_Zllg} and~\ref{table:results_Zllgg}, respectively.

\begin{figure}[hbtp]
\begin{center}
        \includegraphics[width=0.49\textwidth]{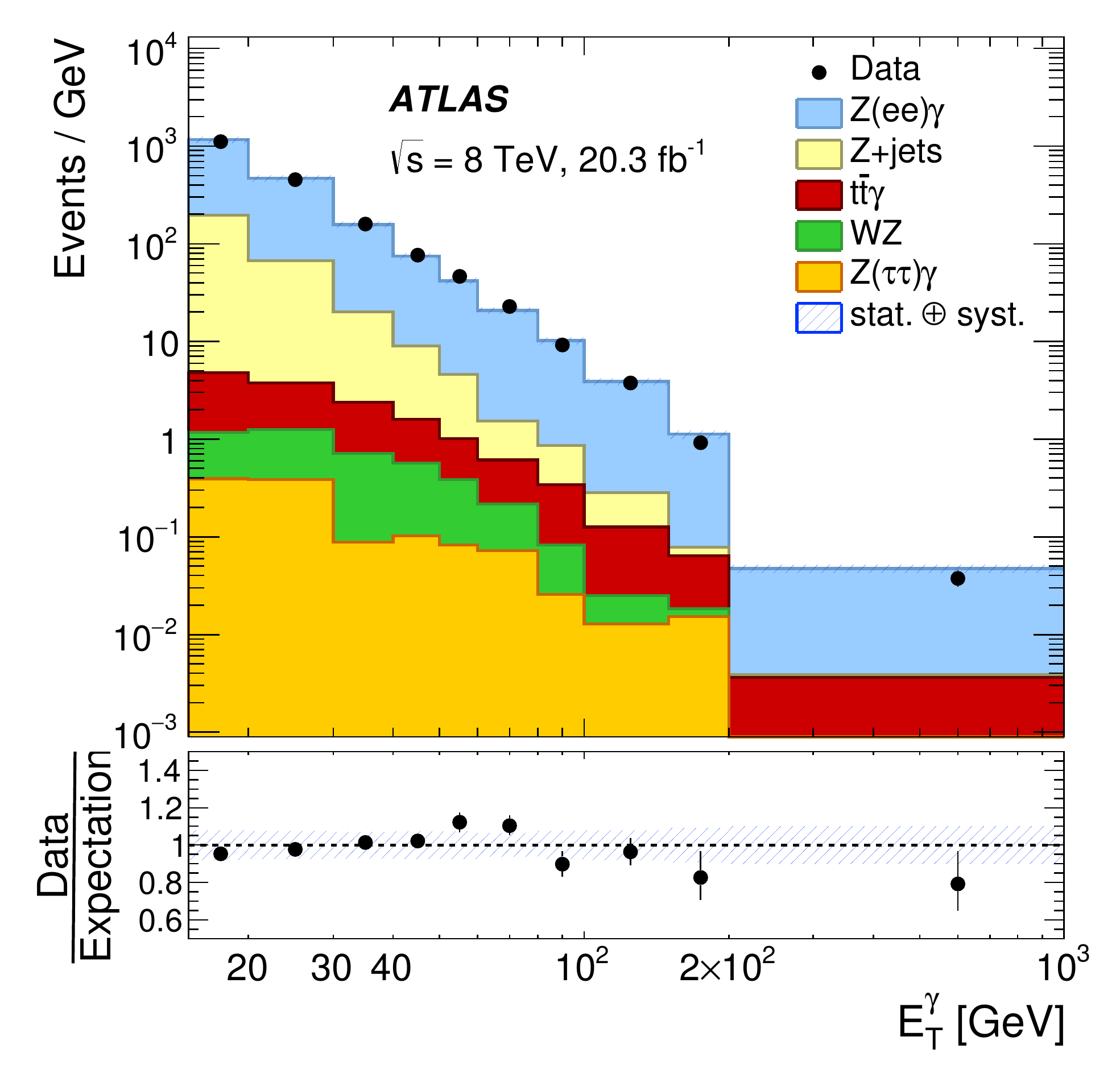}
        \includegraphics[width=0.49\textwidth]{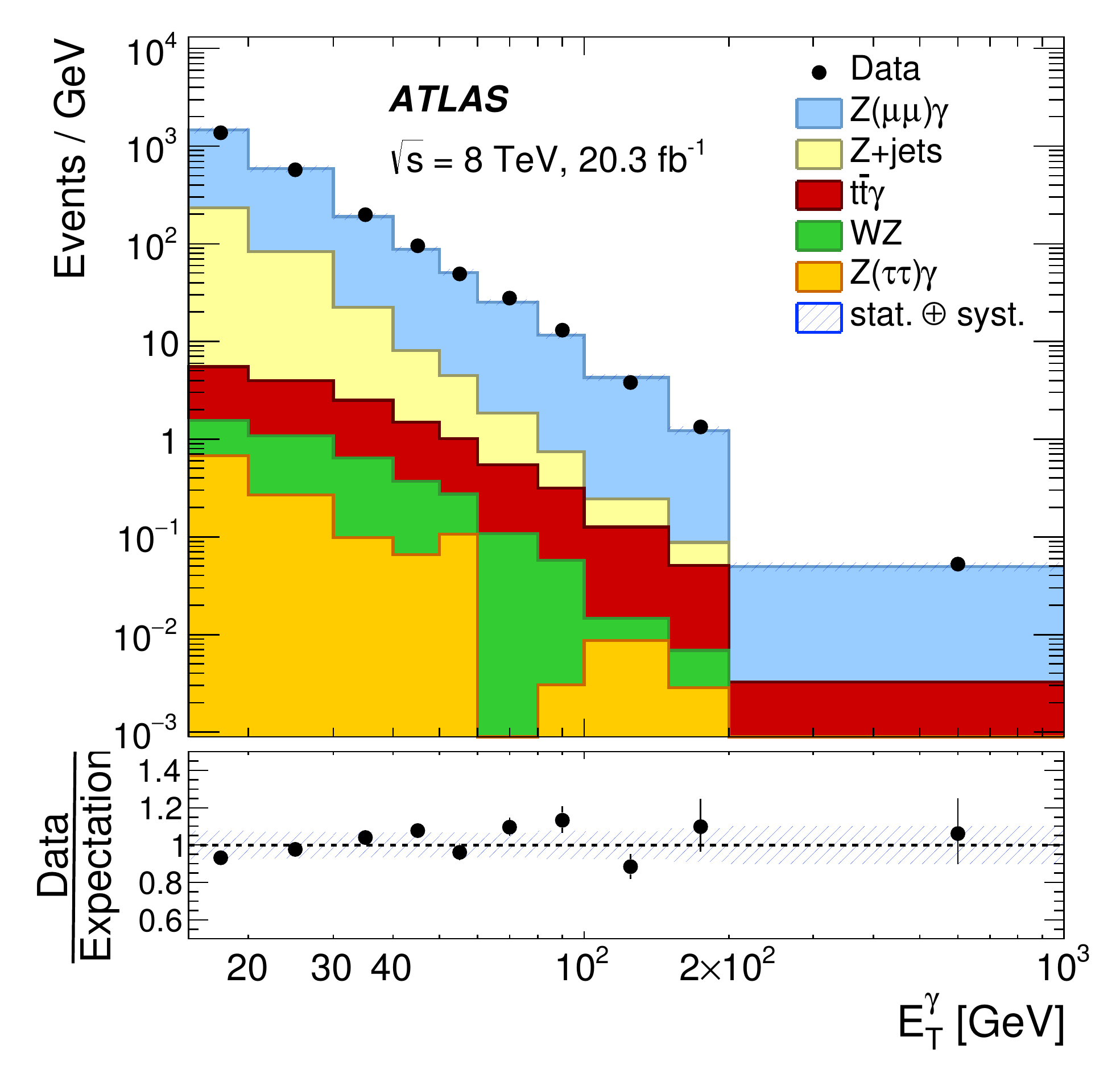}
\end{center}
\caption{The photon transverse energy ($\ET^{\gamma}$) distributions from inclusive ($N_{\mathrm{jet}}\geq 0$) $\leplep\gamma$ events for the electron (left) and muon (right) channels. 
The numbers of candidates observed in data (points with error bars) are compared to the sum of the SM signal predicted from $\SHERPA$ and the various backgrounds discussed in Section~\ref{sec:backgrounds_Zllg_Zllgg}.
The uncertainty band on the sum of expected signal and backgrounds includes both the statistical and systematic uncertainties in the MC simulations and the data-driven background estimate added in quadrature.
The signal is normalized using the cross sections predicted by $\SHERPA$.
The theoretical uncertainties in the signal cross sections are evaluated bin-by-bin using MCFM, as described in Section~\ref{sec:theory_calc}.
The ratio of the numbers of candidates observed in data to the sum of expected signal and backgrounds is also shown.}
\label{fig:photonET_Zllg}
\end{figure}

\begin{figure}[hbtp]
\begin{center}
        \includegraphics[width=0.49\textwidth]{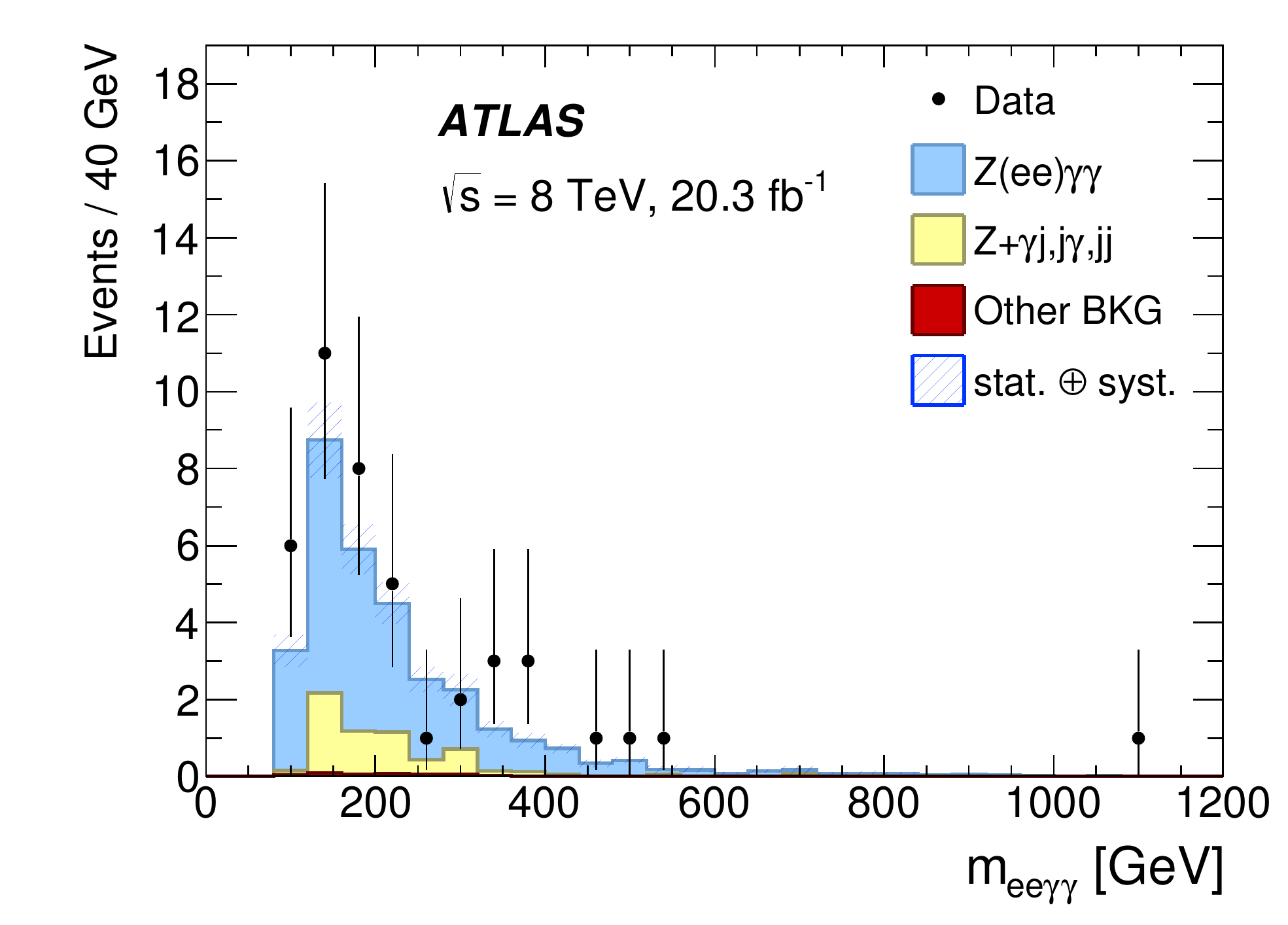}
        \includegraphics[width=0.49\textwidth]{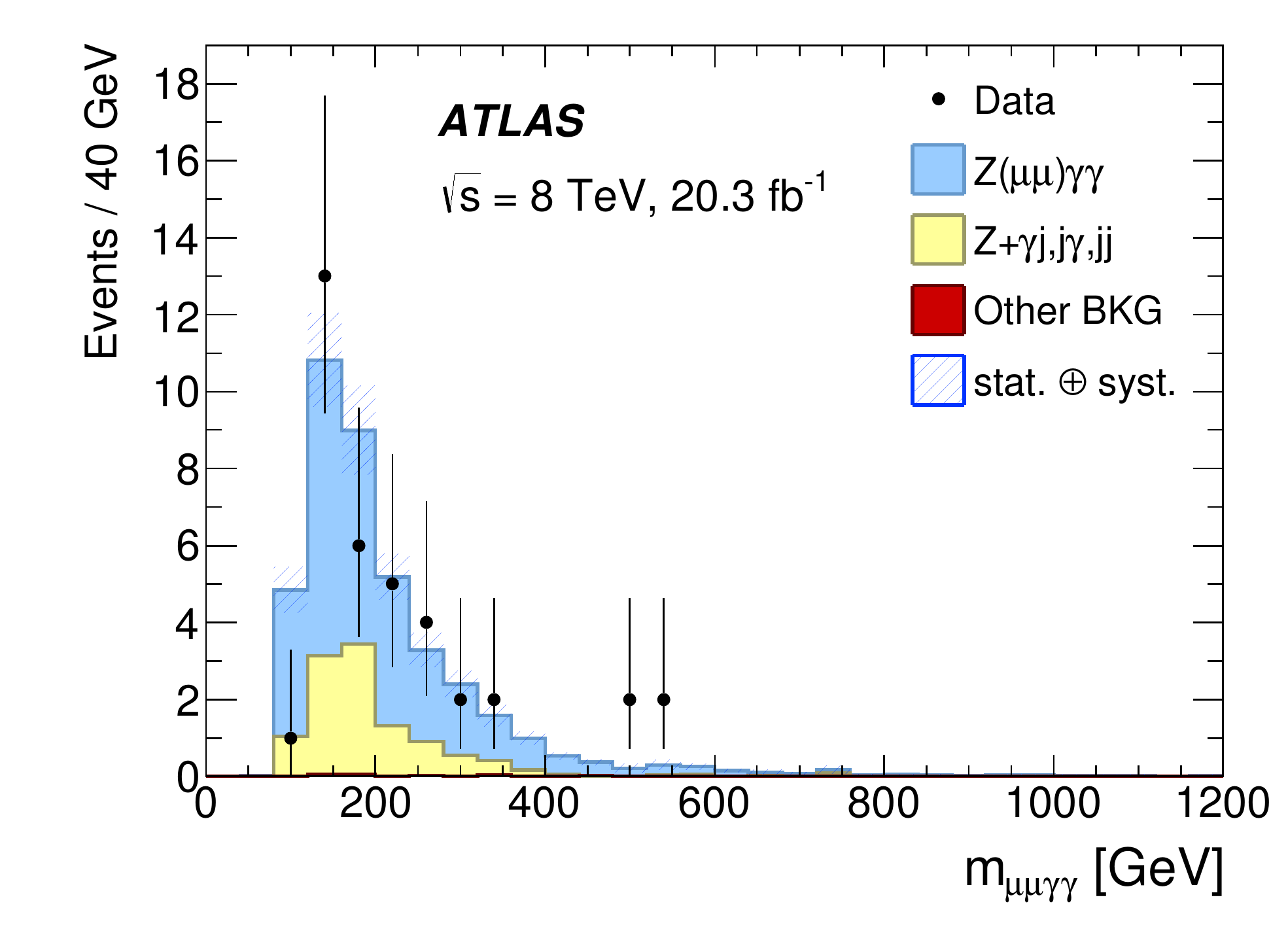}
\end{center}
\caption{The four-body invariant mass ($m_{\leplep\gamma\gamma}$) distributions from inclusive ($N_{\mathrm{jet}}\geq 0$) $\leplep\gamma\gamma$ events for the electron (left) and muon (right) channels.
The numbers of candidates observed in data (points with error bars) are compared to the sum of the SM signal predicted from $\SHERPA$ and the various backgrounds discussed in Section~\ref{sec:backgrounds_Zllg_Zllgg}.
The uncertainty band on the sum of expected signal and backgrounds includes both the statistical and systematic uncertainties in the MC simulations and the data-driven background estimate added in quadrature.
The signal is normalized using the cross sections predicted by $\SHERPA$.
The theoretical uncertainties in the signal cross sections are evaluated using MCFM, as described in Section~\ref{sec:theory_calc}.
}
\label{fig:mllgg_Zllgg}
\end{figure}

\begin{figure}[hbtp]
\begin{center}
        \includegraphics[width=0.49\textwidth]{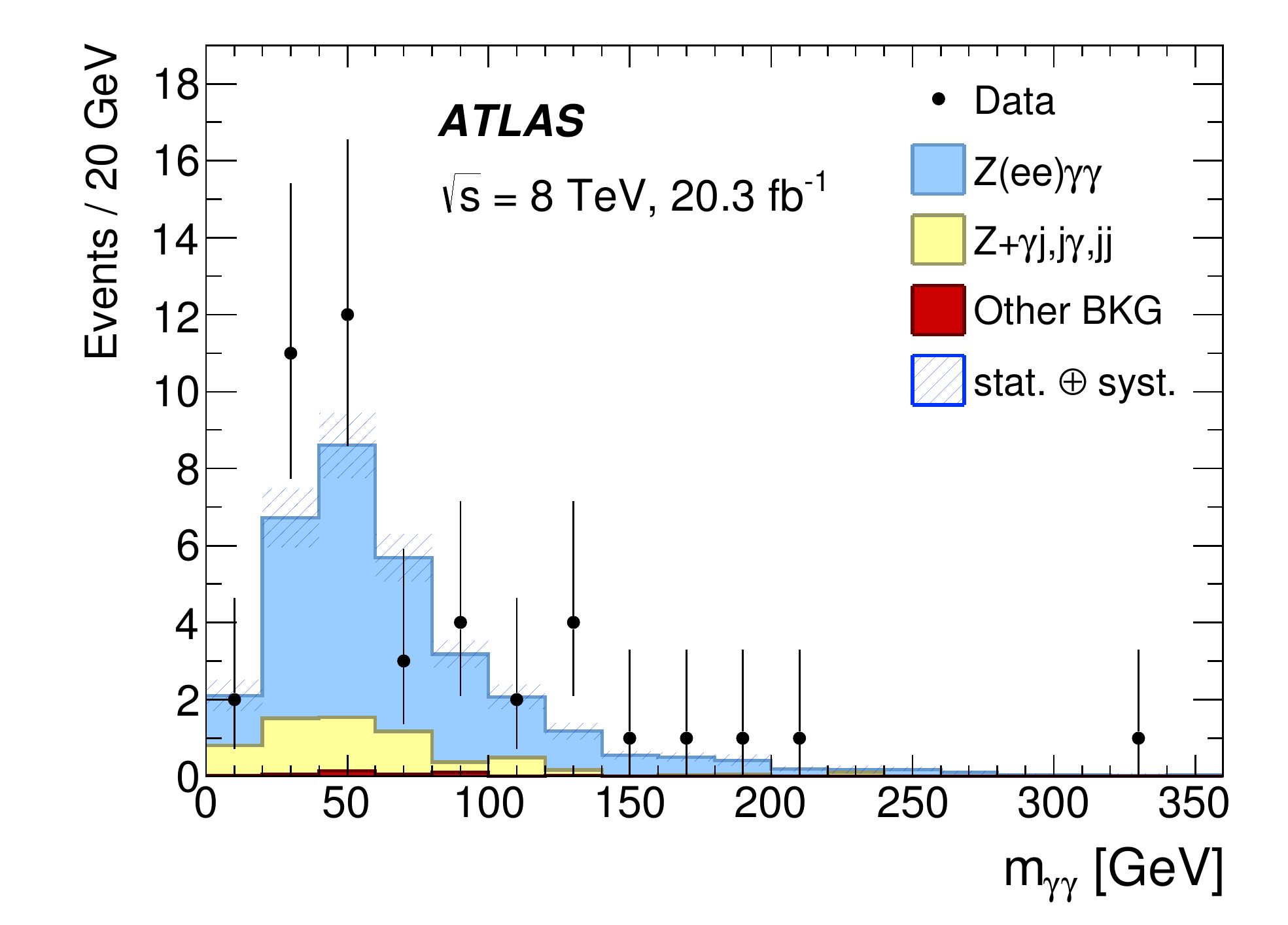}
        \includegraphics[width=0.49\textwidth]{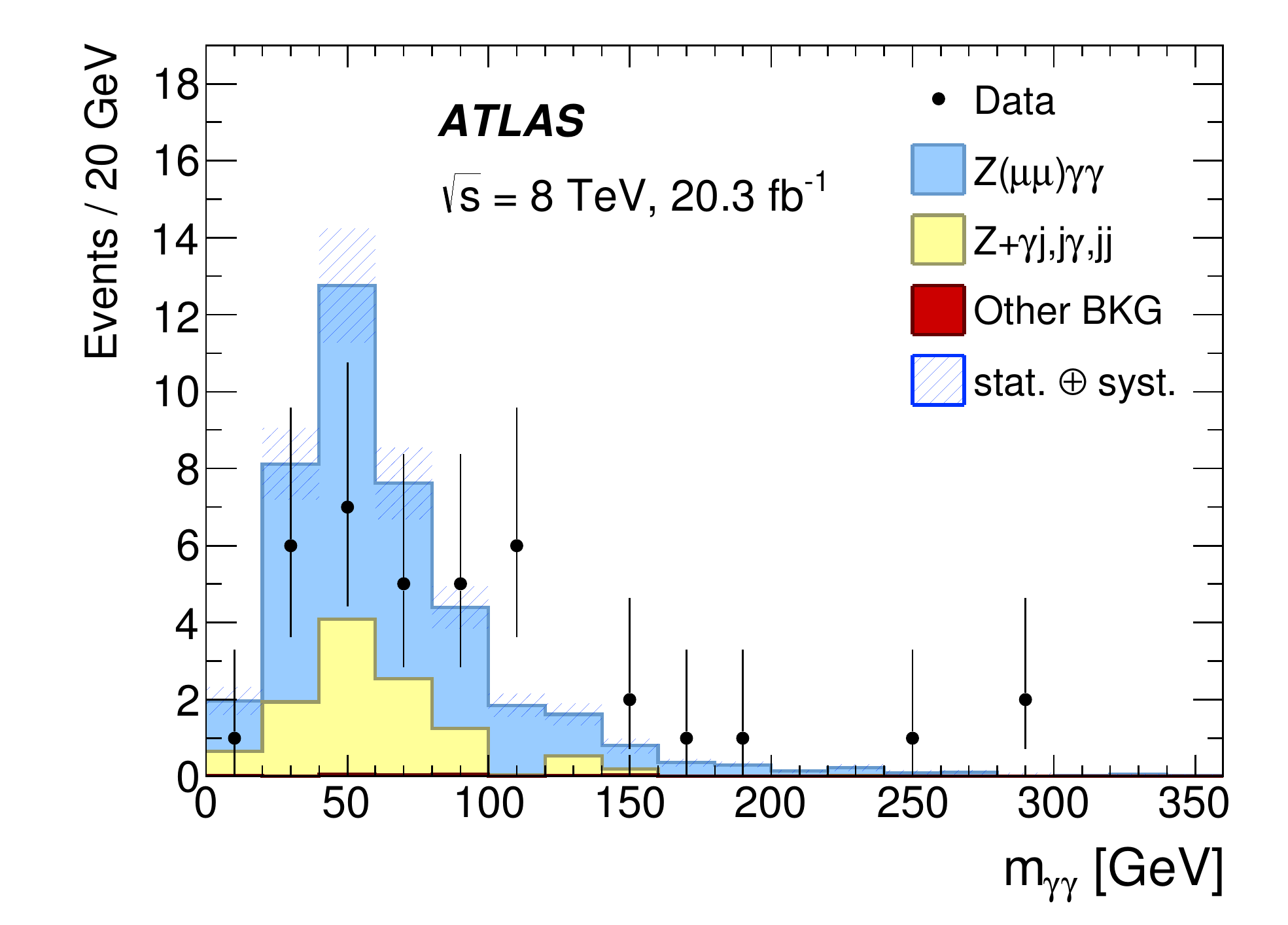}
\end{center}
\caption{The diphoton invariant mass ($m_{\gamma\gamma}$) distributions from inclusive ($N_{\mathrm{jet}}\geq 0$) $\leplep\gamma\gamma$ events for the electron (left) and muon (right) channels. 
The numbers of candidates observed in data (points with error bars) are compared to the sum of the SM signal predicted from $\SHERPA$ and the various backgrounds discussed in Section~\ref{sec:backgrounds_Zllg_Zllgg}.
The uncertainty band on the sum of expected signal and backgrounds includes both the statistical and systematic uncertainties in the MC simulations and the data-driven background estimate added in quadrature.
The signal is normalized using the cross sections predicted by $\SHERPA$.
The theoretical uncertainties in the signal cross sections are evaluated using MCFM, as described in Section~\ref{sec:theory_calc}.
}
\label{fig:mgg_Zllgg}
\end{figure}

The $\ET$ distributions of photons selected in the \eeg and \mumug inclusive measurements are shown in Figure~\ref{fig:photonET_Zllg}.
The highest-$\ET$ photon is measured as $\ET^{\gamma}=585~(570)$ \GeV~in the \eeg (\mumug) final state.
The background from jets misidentified as photons ($Z$+jets) in each $\ET$ bin results from the data-driven estimation for that bin. The distributions of other backgrounds are taken from MC simulation normalized to the integrated luminosity with the cross sections of the background processes.
Similarly, Figures~\ref{fig:mllgg_Zllgg} and~\ref{fig:mgg_Zllgg} present the distributions of the invariant mass of the $\leplep\gamma\gamma$ four-body system and the diphoton invariant mass distributions, respectively, in the \eegg and \mumugg inclusive measurements. 

\begin{table}[hbtp]
\begin{center}
\begin{tabular}{lcccc}
\hline
\hline
 & \eeg & \mumug & \eeg & \mumug \\
$ $ & \multicolumn{2}{c}{$N_{\mathrm{jets}}$ $\geq$ 0} & \multicolumn{2}{c}{$N_{\mathrm{jets}}$ = 0}\\
$N^{\mathrm{obs}}_{\Zg}$ & 13807 & 17054 & 10268 & 12738\\
\hline
$N^{j\rrr\gamma}_{\Zg}$ & $1840 \pm 90 \pm 480$ & $ 2120 \pm 90 \pm 560 $ & $1260 \pm 80 \pm 330$ & $ 1510 \pm 80 \pm 400 $      \\
$N^{\mathrm{Other~BKG}}_{\Zg}$ & $143 \pm 3 \pm 28$ & $ 146 \pm 2 \pm 29 $    & $30.8 \pm 1.6 \pm 6.7$ & $ 26.9 \pm 1.5 \pm 5.8 $    \\ 
\hline
$N^{\mathrm{sig}}_{\Zg}$ ($\SHERPA$) & $12040 \pm 40 \pm 820$ & $15070 \pm 40 \pm 960$ & $9160 \pm 30 \pm 750$ & $11570 \pm 40 \pm 910$ \\
\hline
\hline
\end{tabular}
\caption{Total number of events satisfying the $\leplep\gamma$ selection requirements in data $(N^{\mathrm{obs}}_{\Zg})$, predicted number of signal events from $\SHERPA$ ($N^{\mathrm{sig}}_{\Zg}$), and the estimated number of background events ($N^{j\rrr\gamma}_{\Zg}$ and $N^{\mathrm{Other~BKG}}_{\Zg}$) in the $\eeg$ and $\mumug$ channels with the inclusive ($N_{\mathrm{jets}}$ $\geq$ 0) and exclusive ($N_{\mathrm{jets}}$ = 0) selections.  
The first uncertainty is statistical and the second is the sum of all contributions to the systematic uncertainty. 
The statistical uncertainties arise from the numbers of events in the control regions and the simulation.
The systematic uncertainties in the signal include both the experimental uncertainties described in Section~\ref{sec:intXsec} and the theoretical uncertainties in the cross sections evaluated using MCFM, as described in Section~\ref{sec:theory_calc}.
}
\label{table:results_Zllg}
\end{center}
\end{table}

\begin{table}[hbtp]
\begin{center}
\resizebox{\textwidth}{!}{
\begin{tabular}{lcccc}
\hline
\hline
 & \eegg & \mumugg & \eegg & \mumugg \\
$ $ & \multicolumn{2}{c}{$N_{\mathrm{jets}}$ $\geq$ 0} & \multicolumn{2}{c}{$N_{\mathrm{jets}}$ = 0}\\
$N^{\mathrm{obs}}_{\Zgg}$ & 43 & 37 & 29 & 22 \\
\hline
$N^{j\rrr\gamma}_{\Zgg}$ & $ 5.8 \pm 1.0 \pm 1.4 $      & $ 10.9 \pm 1.1 \pm 2.8 $  & $ 3.08 \pm 0.73 \pm 0.75 $          & $ 6.4 \pm 0.9 \pm 1.8 $       \\
$N^{\mathrm{Other~BKG}}_{\Zgg}$      & $ 0.42 \pm 0.08 \pm 0.18 $          & $ 0.194 \pm 0.047 \pm 0.097 $   & $ 0.24 \pm 0.05 \pm 0.11 $          & $ 0.105 \pm 0.028 \pm 0.055 $    \\
\hline
$N^{\mathrm{sig}}_{\Zgg}$ ($\SHERPA$) &  $ 25.7 \pm 0.5 \pm 1.6 $      & $ 29.5 \pm 0.6 \pm 1.7 $        & $ 18.9 \pm 0.5 \pm 1.5 $        & $ 21.8 \pm 0.5 \pm 1.7 $  \\
\hline
\hline
\end{tabular}
}
\caption{Total number of events satisfying the $\leplep\gamma\gamma$ selection requirements in data $(N^{\mathrm{obs}}_{\Zgg})$, predicted number of signal events from $\SHERPA$ ($N^{\mathrm{sig}}_{\Zgg}$), and the estimated number of background events ($N^{j\rrr\gamma}_{\Zgg}$ and $N^{\mathrm{Other~BKG}}_{\Zgg}$) in the $\eegg$ and $\mumugg$ channels with the inclusive ($N_{\mathrm{jets}}$ $\geq$ 0) and exclusive ($N_{\mathrm{jets}}$ = 0) selections.
The first uncertainty is statistical and the second is the sum of all contributions to the systematic uncertainty. 
The statistical uncertainties arise from the numbers of events in the control regions and the simulation. 
The systematic uncertainties in the signal include both the experimental uncertainties described in Section~\ref{sec:intXsec} and the theoretical uncertainties in the cross sections evaluated using MCFM, as described in Section~\ref{sec:theory_calc}.
}
\label{table:results_Zllgg}
\end{center}
\end{table}

\subsection{Backgrounds to $\nnbar\gamma$  and $\nnbar\gamma\gamma$}
\label{sec:bkgnngnngg}

Backgrounds to the $\nnbar\gamma$ and $\nnbar\gamma\gamma$ signals originate from several sources (listed in decreasing order of significance): events with prompt photons and mismeasured jet momenta causing missing transverse momentum (dominant for the inclusive measurement); nonsignal electroweak processes, such as $W(\ell\nu)\gamma$, with partial event detection; events with real $\met$ from neutrinos (such as $\Zboson(\nnbar)$ or $W(e\nu)$) and misidentified photons from electrons or jets.
The largest contributions are determined using data-driven techniques. The procedures used to
estimate these backgrounds follow closely those in a previous ATLAS measurement~\cite{Aad:2013izg}. 
Smaller backgrounds originate from $\tautau\gamma$ for $\nnbar\gamma$ and $\tautau\gamma\gamma$ for $\nnbar\gamma\gamma$.
These are expected to contribute less than $1.5\%$ of the selected event yield and are derived from MC simulation. 
The backgrounds from multijet and $\leplep\gamma$ processes are negligible.
Each source of background is discussed in detail together with the method used for its estimation in the following subsections.

\subsubsection{$\gamma+$jets background to $\nnbar\gamma$} 
\label{sec:bkg-Znunug-gj}

An imprecise measurement of jet activity in the calorimeter can cause the appearance of fake $\met$ in the event. 
Photon+jets events are one of the dominant background contributions to the $\nnbar\gamma$ channel. 
Although the high-$\met$ requirement reduces the $\gamma$+jets background, a residual contamination from this background remains for the inclusive measurement and is estimated with the following data-driven method.

In order to measure this background from data, a control sample enriched in $\gamma$+jets events is selected by applying all the signal 
region (SR) selection criteria, but inverting the angular separation requirement such that $\Delta\phi(\vec{p}_{\mathrm{T}}^{\mathrm{\;miss}},\gamma)<\pi/2$.  
The data yield in this control region (CR), after subtraction of
signal and other backgrounds obtained using the MC simulation, is then extrapolated to the signal region with a
transfer factor determined from a $\gamma$+jets simulation. The transfer factor equals the ratio of the numbers of $\gamma$+jets events in the SR to the CR. The nominal
transfer factor is determined to be 1.1 from $\SHERPA$ and a 30\% uncertainty is
estimated using an alternative prediction from $\PYTHIA$.

\subsubsection{$W(\ell\nu)\gamma$ background to $\nnbar\gamma$ }
\label{sec:bkg-Znunug-Wg}

Misidentified events from $W(\ell\nu)\gamma$ production are one of the dominant background contributions to the $\nnbar\gamma$ signal. 
A large fraction (about 60\%) of this contamination originates from $W(\tau\nu)\gamma$ events. 
A scale factor is defined to correct the yield of $W\gamma$ events estimated by MC simulation to match the $W\gamma$ event yield measured 
in a control data region constructed by requiring exactly one identified electron or muon instead of the charged-lepton veto. 
Since the control region contains some amount of signal leakage and other background contaminations, these contributions are estimated using the methods described in Sections~\ref{sec:bkg-Znunug-gj},~\ref{sec:bkg-Znunug-Wenu} as well as with MC simulation and then subtracted.
With equal branching fraction of the $W$ boson leptonic decays, the MC scale factor for the dominant $W(\tau\nu)\gamma$ events in the signal region and its uncertainty are taken from the measurement of $W(\ell\nu)\gamma$ events in the control region.
The main uncertainty of 34\% in this background prediction is due to the extrapolation transfer factor from the control region to the signal region. This is estimated by comparing transfer factors between two MC samples generated with $\SHERPA$ and $\ALPGEN$, respectively. The transfer factor between the control and the signal regions is taken from $\SHERPA$ as the baseline and equals $2.2\pm0.7$ for the inclusive selection and $1.8\pm0.7$ for the exclusive selection.

\subsubsection{$W(e\nu)$ background to $\nnbar\gamma$ } 
\label{sec:bkg-Znunug-Wenu}

Misidentification of electrons as photons also contributes to the background yield in the signal region. 
The estimation of this background is made in two steps. The first is the determination of the probability for an electron to
be misidentified as a photon using $\Zboson(\ee)$ decays reconstructed as $e+\gamma$, as described in Ref.~\cite{Aad:2015exgm}. 
The probability of observing an $e+\gamma$ pair with invariant mass near the $\Zboson$ boson mass is used to determine an electron-to-photon fake factor $f_{e\rightarrow\gamma}$.
This increases from 2\% to 6\% as $|\eta|$ increases from 0 to 2.37.
The second step is the construction of a control region with nominal $\nnbar\gamma$ selection criteria, except that an electron is required instead 
of the photon in the final state. This control region contains $W(e\nu)$+jets as the dominant process and some fractions of other processes containing genuine electrons and jets.
The estimated $W(e\nu)$ background is then the product of the electron-to-photon fake factor by the number of events in the chosen control sample.
The total uncertainty in this background varies from 10\% to 30\% as a function of photon $\ET$ and $\eta$ and is dominated by the number of events in the $e+\gamma$ control sample used to measure the electron misidentification probability.

\subsubsection{$\Zboson(\nnbar)$+jets backgrounds to $\nnbar\gamma$} 

Misidentification of jets as photons gives a non-negligible background contribution to the $\nnbar\gamma$ signal. 
A data-driven method similar to the one described for $\Zboson(\leplep)$+jets in Section~\ref{sec:dd_bkg_Zllg} is 
used to determine the background contribution from $\Zboson(\nnbar)$+jets events. 
A systematic uncertainty of 25\% in this background is assigned, dominated by the uncertainty in the correlation factor between identification and isolation of jets reconstructed as photons.

\subsubsection{$\gamma$+jets and $\gamma\gamma$+jets backgrounds to $\nnbar\gamma\gamma$}
\label{sec:bkg-Znunugg-QCD}

The estimation of $\gamma$+jets and $\gamma\gamma$+jets backgrounds to the $Z(\nnbar)\gamma\gamma$ signal uses a two-dimensional sideband method.
Four regions are constructed using two orthogonal selections: different $\ET^\mathrm{miss}$ requirements ($\ET^\mathrm{miss}$ $<$ 20 \GeV~or $\ET^\mathrm{miss}$ $>$ 110 \GeV)
and different identification requirements for photons (two "tight" photons or one "tight" photon and one photon meeting the looser criteria but not the "tight" ones). 
Since the correlations between these regions are small, the number of background events in the signal region can be estimated by scaling the number of events 
in the high-$\ET^\mathrm{miss}$ control region by the ratio of the events from control samples in the low $\ET^\mathrm{miss}$ region. 
Corrections are applied for the $Z(\nnbar)\gamma\gamma$ signal and other backgrounds leaking into the control samples.
The largest uncertainty in this procedure is due to the number of events in the control regions. Systematic uncertainties for this background are evaluated with 
alternative low $\ET^\mathrm{miss}$ control regions (5 $<\ET^\mathrm{miss}<$ 25 \GeV) and from the uncertainty in the correlation between control regions (15\%).

\subsubsection{$W(\ell\nu)\gamma\gamma$ background to $\nnbar\gamma\gamma$}

The background from $W(\ell\nu)\gamma\gamma$ events is dominated by the
$\tau\nu\gamma\gamma$ contribution and is estimated using 
techniques similar to those described above in Section~\ref{sec:bkg-Znunug-Wg}. A control region is
defined by requiring exactly one identified electron or muon
instead of the charged-lepton veto. After accounting for signal leakage and
other background contributions, the control region yield is compared
to the $W\gamma\gamma$ simulation. Good agreement is found, as in the
recent measurement of the $W\gamma\gamma$ cross section~\cite{STDM-2013-05}, although in the high-$\ET^\mathrm{miss}$ region considered here the size of the control sample leads to a
100\% uncertainty in the transfer factor.

\subsubsection{$W(e\nu)\gamma$ background to $\nnbar\gamma\gamma$}

One of the dominant backgrounds in the $\nu\bar{\nu}\gamma\gamma$ channel originates from the misidentification of electrons as photons.
This background is estimated by selecting a control sample in which an electron is required instead of one of the photons in the $\nu\bar{\nu}\gamma\gamma$ final state.
The electron fake rate is estimated as described in Section~\ref{sec:bkg-Znunug-Wenu}. 
The estimated background in the signal region is then obtained by rescaling the yield in the control sample by the electron-to-photon fake rate.
The largest uncertainty in this background is 20\% and is derived from MC events in a closure test of the method.

\subsubsection{$\Zboson(\nnbar)\gamma$+jets background to $\nnbar\gamma\gamma$}

The $\Zboson(\nnbar)\gamma$+jets background falls into the signal region when one jet is misidentified as a photon.
This background contributes less than 5\% of the total event yield and is estimated from the MC simulation.
The systematic uncertainty arises from the mismodeling of the jet-to-photon misidentification rate in the MC simulation.
It is evaluated to be 127\% (106\%) in the inclusive (exclusive) channel, based on $\Zboson(\leplep)\gamma$+jets events with one jet misidentified as a photon, by comparing its estimate from data (as described in Section~\ref{sec:dd_bkg_Zllgg}) with the prediction from MC simulation.

\subsubsection {Results of the background estimation for $\nnbar\gamma$ and $\nnbar\gamma\gamma$}

\begin{figure}[hbtp]
	\begin{center}
		\includegraphics[width=0.49\textwidth]{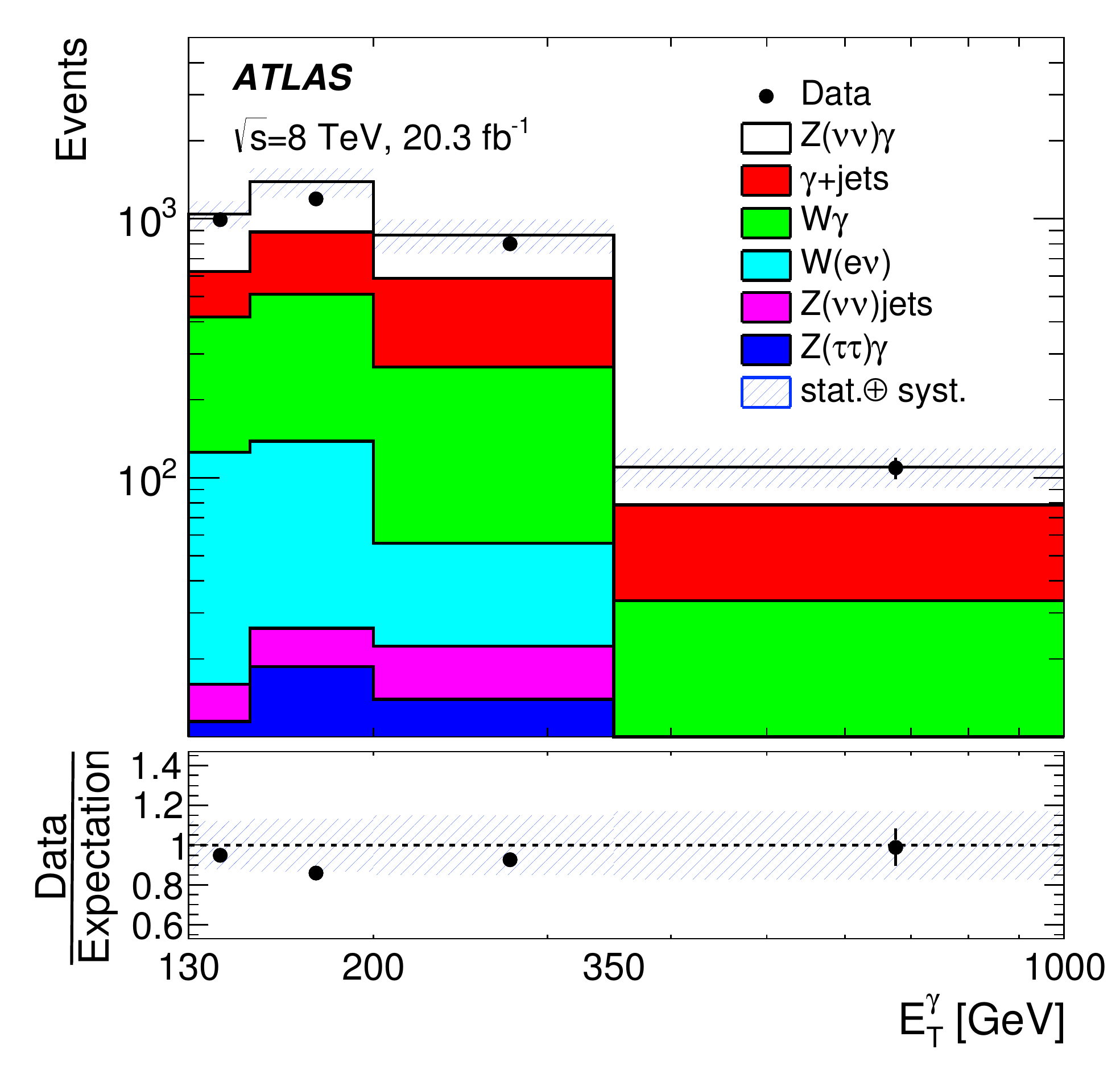}
		\includegraphics[width=0.49\textwidth]{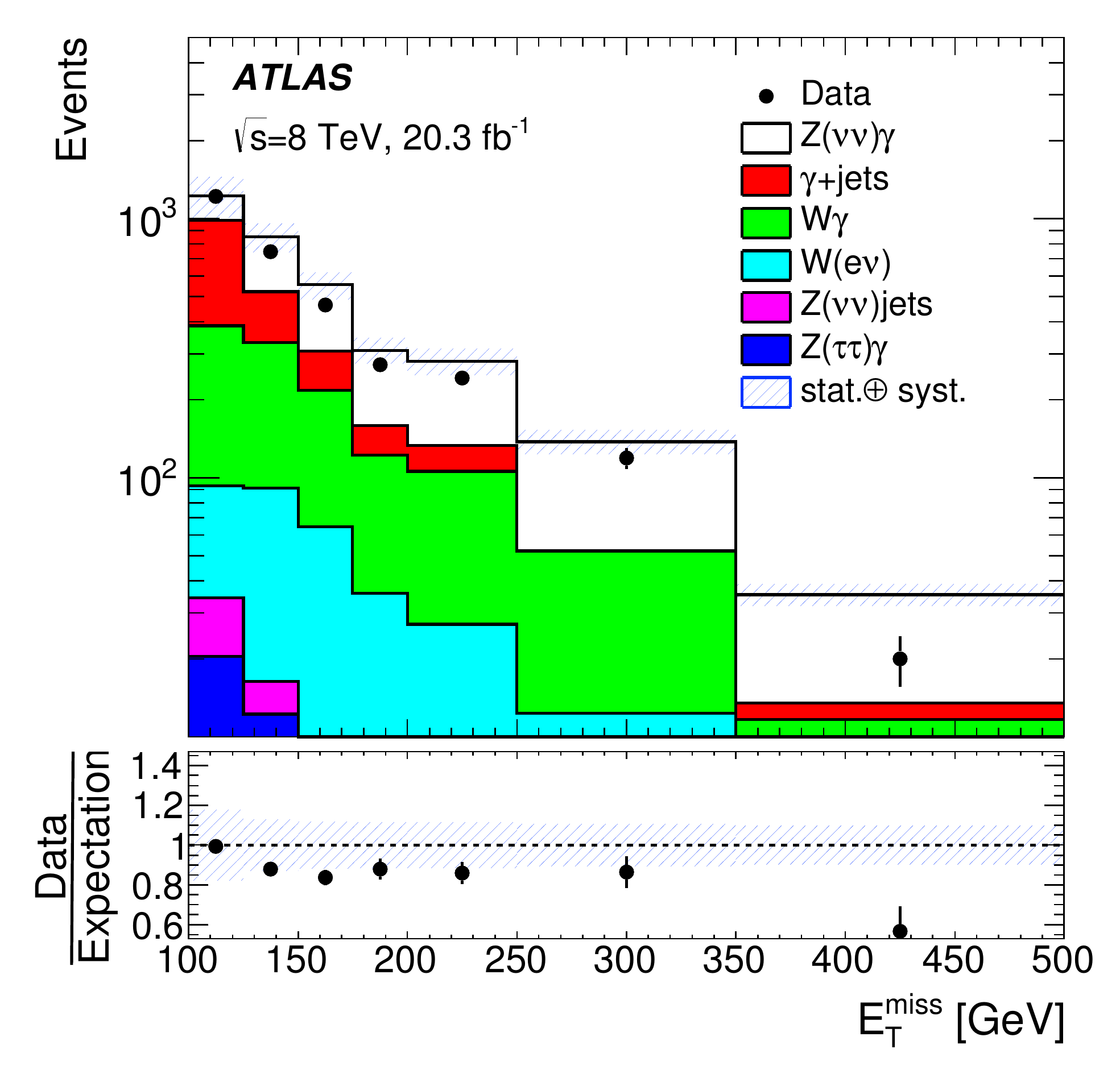}
	\end{center}
	\caption{The photon transverse energy $\ET$ (left) and missing transverse momentum $\met$ (right) distributions  from inclusive ($N_{\mathrm{jet}}\geq 0$) $\nu\bar{\nu}\gamma$ events. 
The numbers of candidates observed in data (points with error bars) are compared to the sum of the SM signal predicted from $\SHERPA$ and the various backgrounds discussed in Section~\ref{sec:bkgnngnngg}.
The uncertainty band on the sum of expected signal and backgrounds includes both the statistical and systematic uncertainties in the MC simulations and the data-driven background estimate added in quadrature.
The signal is normalized using the cross sections predicted by $\SHERPA$.
The theoretical uncertainties in the signal cross sections are evaluated bin-by-bin using MCFM, as described in Section~\ref{sec:theory_calc}.
The ratio of the numbers of candidates observed in data to the sum of expected signal and backgrounds is also shown.
	}
	\label{fig:ZnunuGPlots}
\end{figure}

\begin{figure}[hbtp]
	\begin{center}
			\includegraphics[width=0.49\textwidth]{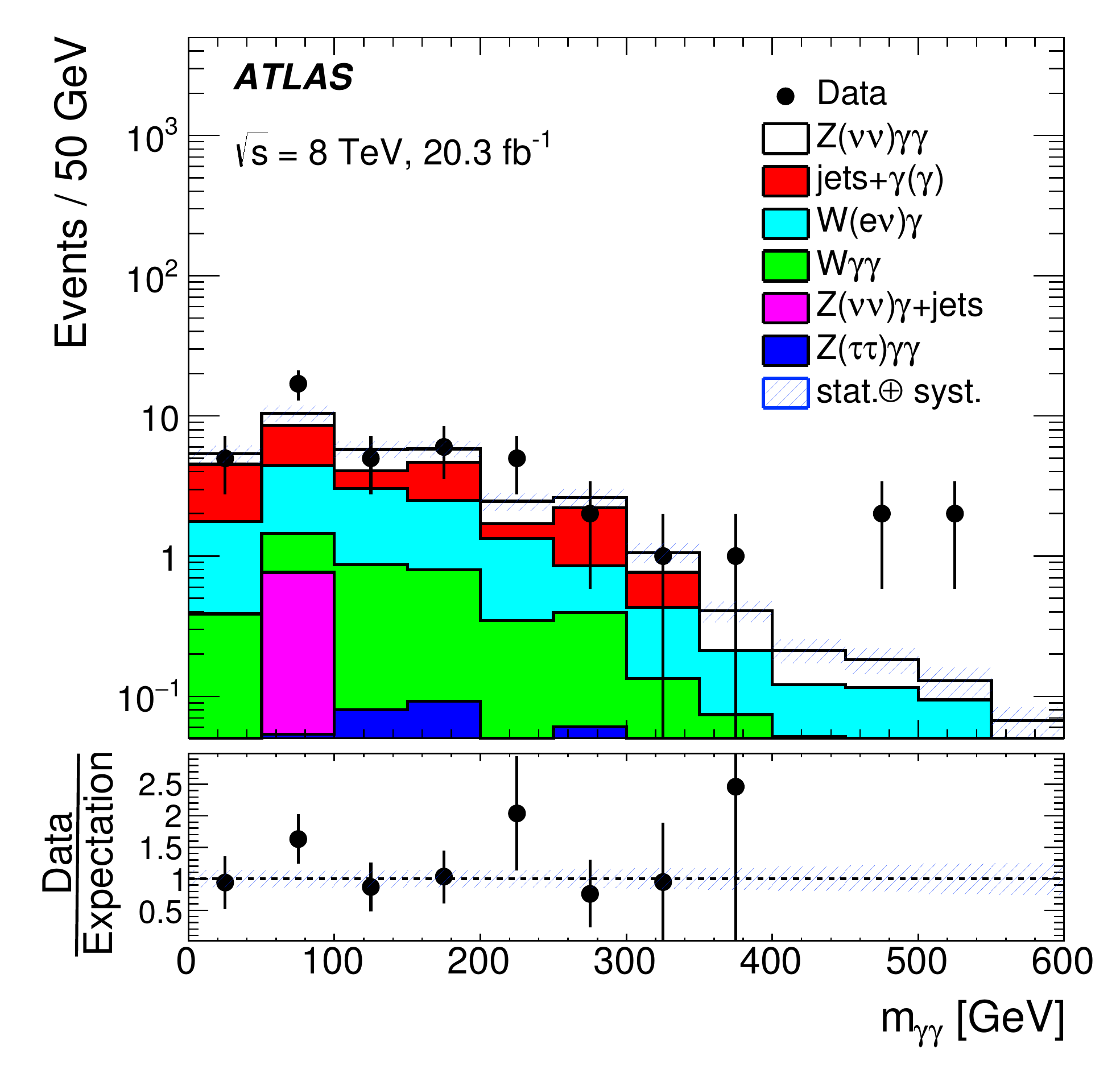}
			\includegraphics[width=0.49\textwidth]{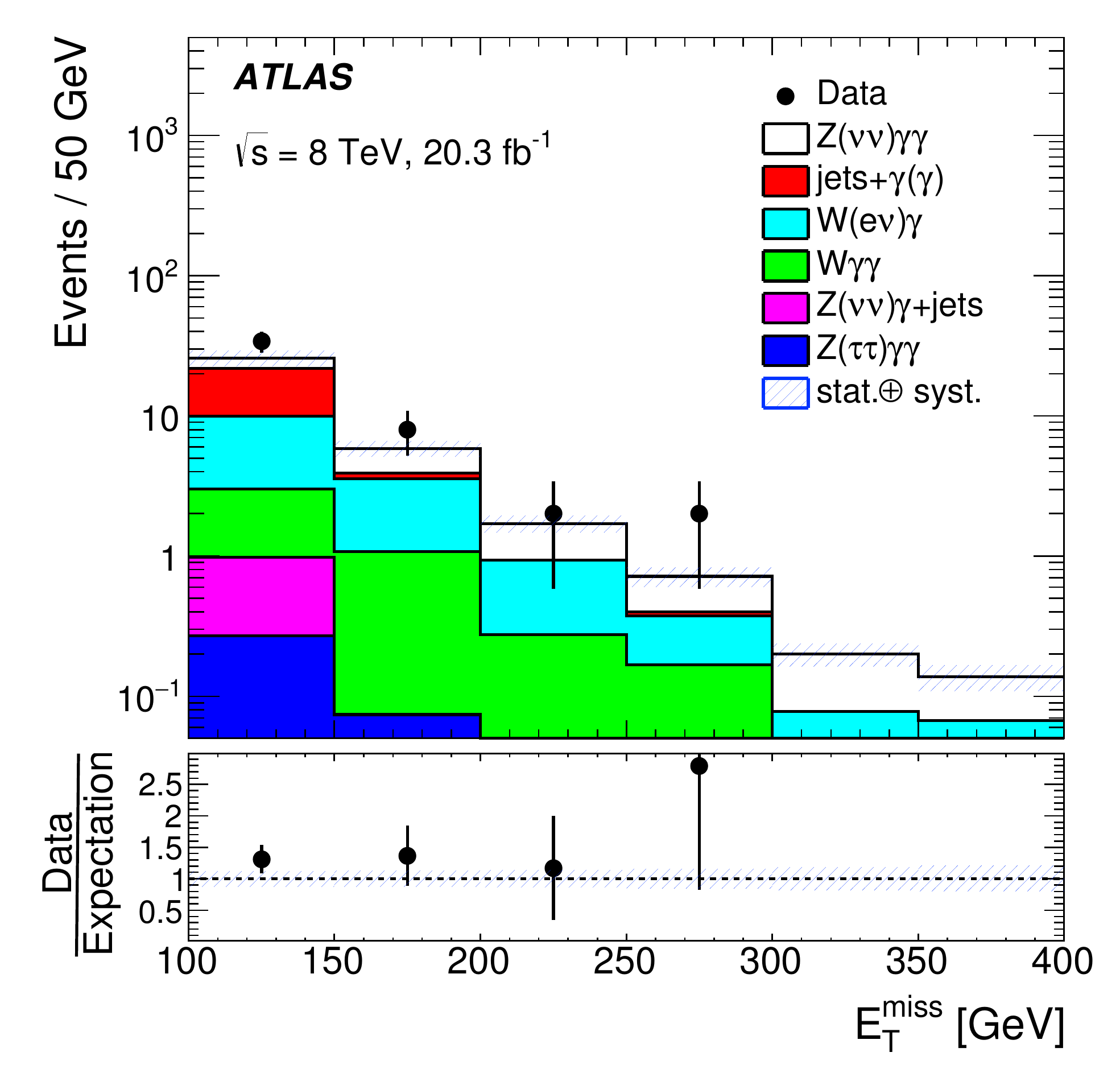}
	\end{center}
	\caption{The diphoton invariant mass $m_{\gamma\gamma}$ (left) and missing transverse momentum $\met$ (right) distributions from inclusive ($N_{\mathrm{jet}}\geq 0$) $\nu\bar{\nu}\gamma\gamma$ events. 
The numbers of candidates observed in data (points with error bars) are compared to the sum of the SM signal predicted from $\SHERPA$ and the various backgrounds discussed in Section~\ref{sec:bkgnngnngg}.
The uncertainty band on the sum of expected signal and backgrounds includes both the statistical and systematic uncertainties in the MC simulations and the data-driven background estimate added in quadrature.
The signal is normalized using the cross sections predicted by $\SHERPA$.
The theoretical uncertainties in the signal cross sections are evaluated using MCFM, as described in Section~\ref{sec:theory_calc}.
The ratio of the numbers of candidates observed in data to the sum of expected signal and backgrounds is also shown.
	}
	\label{fig:ZnunuGGPlots}
\end{figure}

A summary of the number of events observed in data and the background contributions in the $\nnbar\gamma(\gamma)$ channels is given in Tables~\ref{table:Znunug_results} and~\ref{table:Znunugg_results}. 
The photon transverse energy and the missing transverse momentum distributions from the selected events in the $\nnbar\gamma$ channel are shown in Figure \ref{fig:ZnunuGPlots}. 
The highest-$\ET$ photon is measured as $\ET^{\gamma}$=783 \GeV.
The diphoton invariant mass and the missing transverse momentum distributions from the selected events in the $\nnbar\gamma\gamma$ channel are shown in Figure \ref{fig:ZnunuGGPlots}.

\begin{table}[hbtp]
	\begin{center}
		\begin{tabular}{lcc}
			\hline \hline
			& \multicolumn{1}{c}{$N_{\mathrm{jets}}$ $\geq$ 0} & \multicolumn{1}{c}{$N_{\mathrm{jets}}$ = 0}\\
			$N^{\mathrm{obs}}_{\Zg}$ & $3085$ & $1039$  \\
                        \hline
			$N^{\gamma+\mathrm{jets}}_{\Zg}$ & $950\pm 30\pm 300$ & $9.2\pm 3.5\pm 0.7$  \\
			$N^{W(\ell\nu)\gamma}_{\Zg}$ & $900\pm 50\pm 300$ & $ 272\pm 14\pm 92$  \\
			$N^{W(e\nu)}_{\Zg}$ & $258\pm 38\pm 18$ & $147\pm 21\pm 10$ \\
			$N^{Z(\nnbar)+\mathrm{jets}}_{\Zg}$ & $22.9\pm0.5\pm6.1$ & $11.1\pm0.4\pm3.4$  \\
			$N^{Z(\tautau)\gamma}_{\Zg}$ & $46.2\pm 0.9\pm 3.2$ & $10.23\pm 0.43\pm 0.72$ \\
\hline
			$N^{\mathrm{bkg}}_{\Zg}$  & $2180\pm 70\pm 420$ & $450\pm 25\pm 93$ \\
			\hline
			$N^{\mathrm{sig}}_{\Zg}$ ($\SHERPA$) & $1221\pm 2\pm 65$ & $742\pm 2 \pm 44$ \\
			\hline \hline
		\end{tabular}
	\end{center}
	\caption{
Total number of events satisfying the $\nu\bar{\nu}\gamma$ selection requirements in data ($N^{\mathrm{obs}}_{\Zg}$), predicted number of signal events from $\SHERPA$ ($N^{\mathrm{sig}}_{\Zg}$), and the 
		expected number of background events for each of the sources and together ($N^{\mathrm{bkg}}_{\Zg}$) 
		with the inclusive ($N_{\mathrm{jets}}$ $\geq$ 0) and exclusive ($N_{\mathrm{jets}}$ =  0) selections.  
The first uncertainty is statistical and the second is the sum of all contributions to the systematic uncertainty. 
The statistical uncertainties arise from the numbers of events in the control regions and the simulation.
The systematic uncertainties in the signal include both the experimental uncertainties described in Section~\ref{sec:intXsec} and the theoretical uncertainties in the cross sections evaluated using MCFM, as described in Section~\ref{sec:theory_calc}.
}
	\label{table:Znunug_results}
\end{table}

\begin{table}[hbtp]
	\begin{center}
		\begin{tabular}{lcc}
			\hline \hline
			& \multicolumn{1}{c}{$N_{\mathrm{jets}}$ $\geq$ 0} & \multicolumn{1}{c}{$N_{\mathrm{jets}}$ = 0}\\
			$N^{\mathrm{obs}}_{\Zgg}$  & $46$ & $19$ \\
			\hline
			$N^{\mathrm{jets}+\gamma(\gamma)}_{\Zgg}$  & $12.2\pm6.7\pm1.8$ & $2.9\pm4.0\pm0.4$ \\
			$N^{W(\ell\nu)\gamma\gamma}_{\Zgg}$  & $3.6\pm0.1\pm3.6$ & $1.0\pm0.1\pm1.0$  \\
			$N^{W(e\nu)\gamma}_{\Zgg}$   & $10.4\pm0.5\pm2.1$ & $3.47\pm0.28\pm0.69$\\
			$N^{Z(\nnbar)\gamma+\mathrm{jets}}_{\Zgg}$  & $0.71\pm0.71\pm0.90$ & $0.71\pm0.71\pm0.75$ \\
			$N^{Z(\tautau)\gamma\gamma}_{\Zgg}$  & $0.381\pm0.055\pm0.027$ & $0.141\pm0.036\pm0.010$ \\
\hline
			$N^{\mathrm{bkg}}_{\Zgg}$  & $27.2\pm6.8\pm4.6$ & $8.3\pm4.1\pm1.5$ \\
			\hline
			$N^{\mathrm{sig}}_{\Zgg}$ ($\SHERPA$)  &  $7.54\pm0.07\pm0.34$ & $4.80\pm0.06\pm0.29$ \\
			\hline \hline
		\end{tabular}
	\end{center}
	\caption{Total number of events satisfying the $\nu\bar{\nu}\gamma\gamma$ selection requirements in data ($N^{\mathrm{obs}}_{\Zgg}$), predicted number of signal events from $\SHERPA$ ($N^{\mathrm{sig}}_{\Zgg}$), and the 
		expected number of background events for each of the sources and together ($N^{\mathrm{bkg}}_{\Zgg}$) 
		with the inclusive ($N_{\mathrm{jets}}$ $\geq$ 0) and exclusive ($N_{\mathrm{jets}}$ =  0) selections. 
The first uncertainty is statistical and the second is the sum of all contributions to the systematic uncertainty.  
The statistical uncertainties arise from the numbers of events in the control regions and the simulation.
The systematic uncertainties in the signal include both the experimental uncertainties described in Section~\ref{sec:intXsec} and the theoretical uncertainties in the cross sections evaluated using MCFM, as described in Section~\ref{sec:theory_calc}.
}
	\label{table:Znunugg_results}
\end{table}

\section {$Z\gamma$ and $Z\gamma\gamma$ cross sections}
\label{sec:crossection}

\subsection{Description of the cross-section measurements}
\label{sec:introXsec}

The number of signal events in each of the four production channels, \llg, $\nu\bar{\nu}\gamma$, \llgg, and $\nu\bar{\nu}\gamma\gamma$, 
is determined by subtracting the estimated backgrounds from the number of observed events. 
The signal yields are then corrected for detection efficiencies in the fiducial regions used for the measurements. 
The cross sections are calculated for slightly extended fiducial regions using SM predictions for the extrapolation.
These cross sections allow a combination of data obtained from the $Z$ boson to electron and muon decay channels and are more easily compared to predictions from theory.
The extended fiducial regions (see Table~\ref{table:ZggExtendedFiducial}) are defined at the particle level, as described below.
The methods used for the determination of the cross sections and their uncertainties are described in Section~\ref{sec:intXsec}.
The integrated and differential cross-section measurement results are presented in Sections~\ref{sec:xsectotal} and~\ref{sec:diffXsec}, respectively. 

"Particle level" refers to stable particles with a proper decay length $c\tau > 10$ mm which are produced from the hard scattering, 
including those that are the products of hadronization.
The fiducial regions are defined with the same object and event kinematic selection criteria as the reconstruction-level selections described in Section~\ref{sec:selection}. 
Compared with the fiducial regions,
the extended fiducial regions use a unified charged lepton pseudorapidity selection criterion $|\eta^{\ell}| < 2.47$ for $\leplep\gamma$ and $\leplep\gamma\gamma$ channels.
As for $\nnbar\gamma$ and $\nnbar\gamma\gamma$ channels, the extended fiducial regions remove the $\Delta \phi(\vec{p}_{\mathrm{T}}^{\mathrm{\;miss}},\gamma) > \pi/2$ and $\Delta \phi(\vec{p}_{\mathrm{T}}^{\mathrm{\;miss}},\gamma\gamma) > 5\pi/6$ requirements, respectively.
Final-state radiation is incorporated into the particle level definition of the leptons by including the contributions from the photons 
within a cone of $\Delta R = 0.1$ around the lepton direction.
The particle level jets are reconstructed using the anti-$k_t$ algorithm with a radius parameter of $R=0.4$, including all stable particles except for muons and neutrinos.
The photons at particle level are required to satisfy the isolation criterion of $\epsilon^{p}_{h} < 0.5$, 
where $\epsilon^{p}_{h}$ is the transverse energy carried by the closest particle-level jet in a cone of $\Delta R = 0.4$ around the photon direction, subtracting the photon $E_{\mathrm{T}}$ and then divided by the photon $E_{\mathrm{T}}$.  

\begin{table}[hbtp]
\begin{center}
\begin{tabular}{lcccc}
\hline
\hline
Cuts  & $\leplep\gamma$ & $\leplep\gamma\gamma$ & $\nnbar\gamma$ & $\nnbar\gamma\gamma$ \\
\hline
Lepton & $\pT^{\ell} > 25$ \GeV & $\pT^{\ell} > 25$ \GeV & - & -  \\
       & $|\eta^{\ell}| < 2.47$ & $|\eta^{\ell}| < 2.47$ & - & - \\
\hline
Boson &  $m_{\leplep} > 40$ \GeV & $m_{\leplep} > 40$ \GeV & $\pT^{\nnbar} > 100$ \GeV & $\pT^{\nnbar} > 110$ \GeV \\
\hline
Photon & $\ET^{\gamma} > 15$ \GeV & $\ET^{\gamma} > 15$ \GeV & $\ET^{\gamma} > 130$ \GeV & $\ET^{\gamma} > 22$ \GeV\\
       &  \multicolumn{4}{c}{$|\eta^\gamma| < 2.37$} \\
       &  $\Delta R(\ell,\gamma) > 0.7$ & $\Delta R(\ell,\gamma) > 0.4$ & - & -  \\
       & -	& $\Delta R(\gamma,\gamma) > 0.4$ & -	& $\Delta R(\gamma,\gamma) > 0.4$  \\
       &  \multicolumn{4}{c}{$\epsilon^p_h <$0.5} \\
\hline
Jet    &  \multicolumn{4}{c}{ $\pT^{\mathrm{jet}} >$ 30 \GeV, $|\eta^{\mathrm{jet}}| < 4.5$}\\
       &   $\Delta R(\mathrm{jet},\ell/\gamma) > 0.3$ & $\Delta R(\mathrm{jet},\ell/\gamma) > 0.3$ &  $\Delta R(\mathrm{jet},\gamma) > 0.3$ & $\Delta R(\mathrm{jet},\gamma) > 0.3$ \\
       &   \multicolumn{4}{c}{Inclusive : $N_{\mathrm{jet}} \geq 0$, Exclusive : $N_{\mathrm{jet}} = 0$}\\
\hline
\hline
\end{tabular}
\caption{Definition of the extended fiducial regions where the cross sections are measured. The variable $\pT^{\nnbar}$ is the transverse momentum of the $\Zboson$ boson decaying to a neutrino pair. 
The variable $\epsilon^{p}_{h}$ is the transverse energy carried by the closest particle level jet in a cone of $\Delta R = 0.4$ around the photon direction, excluding the photon and divided by the photon transverse energy.}
\label{table:ZggExtendedFiducial}
\end{center}
\end{table}

\subsection{Determination of extended fiducial cross sections}
\label{sec:intXsec}
The integrated cross sections for $Z\gamma$ and $Z\gamma\gamma$ production in the extended fiducial regions are calculated using :
\begin{linenomath}
\begin{equation}
\sigma_{\textrm{ext-fid}} = \frac{N - B}{A \cdot C \cdot \int{Ldt}},
\end{equation}
\end{linenomath}
where
$N$ is the number of candidate events observed, $B$ is the expected number of background events and
$\int{Ldt}$ is the integrated luminosity corresponding to the dataset analyzed.
The factors $C$ and $A$ correct for detection efficiency and acceptance, respectively:
\begin{itemize}
\item $C$ is defined as the number of reconstructed signal events satisfying all selection criteria
divided by the number of events that, at particle level, meet the acceptance criteria of the fiducial region.
\item $A$ is defined as the number of signal events within the fiducial region 
divided by the number of signal events within the extended fiducial region, which are both defined at particle level.
\end{itemize}
The corrections $A$ and $C$ are determined using the $Z\gamma$ and $Z\gamma\gamma$ signal events generated with $\SHERPA$. The numerical values are summarized in Table~\ref{table:ZggAC}.

\begin{table}[hbtp]
\begin{center}
\resizebox{\textwidth}{!}{
\begin{tabular}{lcccccc}
\hline
\hline
 & $\ee\gamma$ & $\mumu\gamma$ & $\nnbar\gamma $ & $\ee\gamma\gamma$ & $\mumu\gamma\gamma$ & $\nnbar\gamma\gamma $ \\
\hline
 & \multicolumn{6}{c} {$N_{\mathrm{jets}} \geq 0$ } \\
$C$ & $0.412 \pm 0.016$ &  $0.512\pm 0.017$ & $0.720\pm0.038$  & $0.329\pm0.016$ & $0.377\pm0.017$  & $0.516\pm0.022$ \\
$A$ & $0.9381 \pm 0.0012$ &  $0.9470\pm 0.0010$ & $0.9132\pm0.0055$  & $0.8841\pm0.0037$ & $0.8844\pm0.0041$  & $0.711\pm0.010$ \\
\hline
 & \multicolumn{6}{c} {$N_{\mathrm{jets}} = 0$ } \\
$C$ & $0.392\pm0.019$ &  $0.492\pm0.020$ & $0.718\pm0.042$ & $0.312\pm0.018$ & $0.365\pm0.019$ & $0.515\pm0.031$ \\
$A$ & $0.9380\pm0.0013$ &  $0.9469\pm0.0012$ & $0.9380\pm0.0010$ & $0.8852\pm0.0044$ & $0.8807\pm0.0050$ & $0.873\pm0.010$ \\
\hline
\hline
\end{tabular}
}
\caption{Summary of correction factors $C$ and acceptances $A$ for the $Z\gamma$ and $Z\gamma\gamma$ cross-section measurements.
The uncertainties include both the statistical and systematic uncertainties.}
\label{table:ZggAC}
\end{center}
\end{table}

Systematic uncertainties in the acceptances $A$ are evaluated by varying the PDFs and the renormalization and factorization scales.
The uncertainty in the acceptances due to the PDF is taken as the envelope of the internal uncertainties from three different PDF sets, namely, the CT10 PDF set, the MSTW2008NLO PDF set~\cite{Martin:2009iq}, and the NNPDF2.3 PDF set~\cite{Ball:2013htb}.
The internal uncertainty from each PDF set is estimated by comparing the acceptance using the PDF central set with the acceptance estimated using the PDF eigenvector sets.
The renormalization and factorization scale uncertainties are assessed by varying these two scales independently by a factor of two from their nominal values, and taking the envelope of the resulting variations.
The impact of PDF uncertainties varies from $0.04\%$ to $0.3\%$,
while the renormalization and factorization scale uncertainties cause variations  from $0.08\%$ to $1.5\%$.
The total uncertainties in the acceptance factors are summarized in Table~\ref{table:ZggAC} .

Systematic uncertainties affecting the correction factors $C$ can be grouped into two categories.
The first includes the uncertainties 
arising from the efficiencies of
the trigger, reconstruction, identification, and other selection requirements. The second category stems
from the uncertainties of energy and momentum scales and resolutions of the final-state objects and the simulation of pileup events.  
Table~\ref{table:Zggsyst} presents all the contributions to the uncertainties in $C$
determined using the methods described below. The total uncertainties in the correction factors are summarized in Table~\ref{table:ZggAC} .

The photon identification efficiencies are measured in data using a combination of three methods as described in Ref.~\cite{ATLAS-CONF-2012-123}.
The uncertainties induced by the photon identification efficiency are estimated to be $1.5\%$ and $0.5\%$ for the $\leplep\gamma$ and $\nnbar\gamma$ channels, respectively.
For the $\leplep\gamma\gamma$ and $\nnbar\gamma\gamma$ channels, after taking into account the correlations between the two photons, 
the resulting uncertainties are $2.1\%$ and $1.9\%$, respectively.  
The photon isolation efficiencies are determined from data by studying the electron isolation efficiencies using $Z\rightarrow \ee$ events. 
The estimated uncertainty increases from $0.5\%$ for photons with $\ET$ around 20 \GeV~to $8\%$ for photons with $\ET$ greater than 350 \GeV,
dominated by the limited size of the $Z\rightarrow \ee$ sample in data.

The reconstruction and identification efficiencies of electrons and muons are derived using a tag-and-probe method with $Z$ and $J/\psi$ events decaying into $\ee$ or $\mumu$ pairs~\cite{ATLAS-CONF-2014-032,PERF-2014-05}.
The uncertainties are evaluated to be $1.6\%$ for the electron channels, and $0.9\%$ for the muon channels. 
The uncertainties arising from the selection efficiencies of lepton isolation and impact parameters requirements are also measured with a tag-and-probe method using $Z$ events.
They are found to be $2.2\%$.
The uncertainties due to the modeling of trigger efficiencies are evaluated to be $1.9\%$ for the $\nnbar\gamma$ channel and no more than $0.5\%$ for the other channels~\cite{TRIG-2012-03,ATLAS-CONF-2012-048}. 
The uncertainty in the jet vertex fraction efficiency is estimated by varying the selection requirement to account for the difference between data and simulation.
For exclusive $N_{\mathrm{jets}}=0$ measurements, they are calculated to be no more than $0.6\%$ for all the channels.   

The energy scale and resolution and their uncertainties for electrons and photons are obtained using $Z\rightarrow \ee$ events~\cite{Aad:2012elid}. 
The systematic uncertainty due to the energy scale varies from $1.2\%$ to $2.7\%$ and that associated with the energy resolution is no more than $0.5\%$ for all the final states.
The muon momentum scale and resolution are studied using samples of $J/\psi$, $\Upsilon$, and $Z$ decays to muon pairs~\cite{PERF-2014-05}. 
The corresponding uncertainties are no more than $0.5\%$ in all the channels.  

The exclusive $N_{\mathrm{jets}}=0$ measurements are affected by the uncertainties in the jet energy scale and resolution, because these uncertainties change the distributions of the number of jets with $\pT > 30$ \GeV~and $|\eta|<4.5$. 
They are studied using MC simulation, as well as $\gamma+$jet, $Z+$jet, dijet, and multijet data events~\cite{ATLAS-CONF-2015-037}.
Their systematic effect varies from $0.8\%$ to $2.9\%$ for all channels.
The uncertainties in the energy and momentum scales and resolutions of reconstructed physics objects are propagated to the $\met$ calculation.
The uncertainties arising from the scale and resolution of the energy deposits that are
not associated with any reconstructed physics object, named the $\met$ soft-term~\cite{ATLAS-CONF-2013-082}, are no more than $0.5\%$ for the $\nnbar\gamma$ final state,
and vary from $0.4\%$ to $1.7\%$ for the $\nnbar\gamma\gamma$ final state.
As mentioned in Section~\ref{sec:signalMC}, the MC events are reweighted so that the pileup conditions in the simulation match the data. 
The pileup events are modeled by MC simulation.
The uncertainties associated with the modeling of the pileup events are estimated to be no more than $1.1\%$ for all the final states.

\begin{table}[hbtp]
\begin{center}
\resizebox{\textwidth}{!}{
\begin{tabular}{lcccccc}
\hline
\hline
 		& $\ee\gamma$ & $\mumu\gamma$ & $\nnbar\gamma $ & $\ee\gamma\gamma$ & $\mumu\gamma\gamma$ & $\nnbar\gamma\gamma $ \\
\hline
MC statistical uncertainty	& 0.3 (0.3)	& 0.2 (0.3)	& 0.1 (0.1)	& 1.9 (2.3)	& 1.8 (2.1)	& 0.6 (0.8)	\\
\hline
Efficiencies : \\
~~~~Trigger	& 0.2 (0.2) & 0.5 (0.5)	& 1.9 (1.9)	& 0.1 (0.1) & 0.5 (0.5)	& 0.2 (0.2)	\\
~~~~Photon identification	& 1.5 (1.5)	& 1.5 (1.5)	& 0.5 (0.5)	& 2.1 (2.1) 	& 2.1 (2.1)	& 1.9 (1.9)	\\
~~~~Photon isolation		& 0.5 (0.5)	& 0.5 (0.5)	& 4.5 (4.3)	& 1.2 (1.2)	& 1.2 (1.2)	& 2.8 (2.8)	\\
~~~~Lepton reconstruction and identification 	& 1.6 (1.6)	& 0.9 (0.9)	& $-$ ($-$)	& 1.6 (1.6)	& 0.9 (0.9)	& $-$ ($-$)		\\
~~~~Lepton isolation and impact parameter 	& 2.2 (2.2) 	& 2.2 (2.2)	& $-$ ($-$) 	& 2.2 (2.2)     & 2.2 (2.2)     & $-$ ($-$)	\\	
~~~~Jet vertex fraction		& $-$ (0.5)	& $-$ (0.6)	& $-$ (0.1)		& $-$ (0.5)	& $-$ (0.6)	& $-$ (0.2)	\\
		&		&		&		&		&		&		\\
Energy/momentum scale and resolution : \\
~~~~Electromagnetic energy scale	& 2.3 (2.5)	& 1.2 (1.3)	& 2.1 (2.4)	& 2.5 (2.7)	& 1.8 (1.9)	& 2.0 (2.8)	\\
~~~~Electromagnetic energy resolution & $<$0.05 ($<$0.05)	& $<$0.05 ($<$0.05)	& $<$0.05 (0.1)	& 0.2 (0.3)	& 0.3 (0.3)	& 0.4 (0.5)	\\
~~~~Muon momentum scale			& $-$ ($-$)		& 0.1 (0.2)	& $-$ ($-$)		& $-$ ($-$)		& 0.3 (0.2)	& $-$ ($-$)		\\
~~~~Muon momentum resolution		& $-$ ($-$)             & $<$0.05 ($<$0.05)     & $-$ ($-$)             & $-$ ($-$)             & 0.5 (0.5)     & $-$ ($-$)             \\
~~~~Jet energy scale			& $-$ (1.9)	& $-$ (1.9)	& $<$0.05 (2.2)	& $-$ (2.2)	& $-$ (1.8)	& 0.7 (2.9)	\\
~~~~Jet energy resolution		& $-$ (1.2)     & $-$ (1.4)     & $<$0.05 (1.0)     & $-$ (1.2)     & $-$ (0.8)     & 0.1 (1.9)     \\
~~~~$\met$ soft-term energy scale & $-$ ($-$) & $-$ ($-$) & 0.3 (0.5)	& $-$ ($-$)		& $-$ ($-$)		& 1.3 (1.7)	\\
~~~~$\met$ soft-term energy resolution & $-$ ($-$) & $-$ ($-$) & $<$0.05 ($<$0.05)       & $-$ ($-$)             & $-$ ($-$)             & 0.4 (0.7)     \\
                &               &               &               &               &               &               \\
Pileup simulation	& 0.8 (0.8)	& 0.6 (0.7)	& 0.2 (0.4)	& 0.8 (1.0)	& 1.1 (1.1)	& 0.6 (0.9)	\\
\hline
Total, without MC statistical uncertainty		& 4.0 (4.7)	& 3.2 (4.1)	& 5.3 (5.9)	& 4.5 (5.3)	& 4.1 (4.6)	& 4.3 (6.0)	\\
\hline
\hline

\end{tabular}
}
\caption{Relative systematic uncertainties, in $\%$, in the signal correction factor $C$ for each channel in the inclusive $N_{\mathrm{jets}} \geq 0$ (exclusive $N_{\mathrm{jets}}=0$) measurement.}
\label{table:Zggsyst}
\end{center}
\end{table}

\subsection{Integrated extended fiducial cross sections for $Z\gamma$ and $Z\gamma\gamma$ production}
\label{sec:xsectotal}

The measurements of the cross sections of each final state and the combined charged-lepton final states, along with their uncertainties, are based on the maximization of the profile-likelihood ratio:
\begin{linenomath}
\begin{equation}
	\Lambda(\sigma) = \frac{\mathcal{L}(\sigma, \hat{\hat{\boldsymbol{\theta}}}(\sigma))}{\mathcal{L}(\hat{\sigma}, \hat{\boldsymbol{\theta}})},
\end{equation}
\end{linenomath}
where $\mathcal{L}$ represents the likelihood function, $\sigma$ is the cross section
and $\boldsymbol{\theta}$ are the nuisance parameters corresponding to sources of the systematic uncertainties. 
The $\hat{\sigma}$ and $\hat{\boldsymbol{\theta}}$ terms denote the unconditional maximum-likelihood estimate of the parameters, i.e., where 
the likelihood is maximized for both $\sigma$ and $\boldsymbol{\theta}$. 
The $\hat{\hat{\boldsymbol{\theta}}}(\sigma)$ corresponds to the value of $\boldsymbol{\theta}$ that maximizes $\mathcal{L}$ for given parameter values of $\sigma$.
The likelihood function is defined as :
\begin{linenomath}
\begin{equation}
	\mathcal{L}(\sigma, \boldsymbol{\theta}) = \prod^{\mathrm{^{final}_{states}}}_{i} {\mathrm{Poisson}}(N_{i}~|~S_i(\sigma, \boldsymbol{\theta}) + B_i(\boldsymbol{\theta})) \
										\cdot {\mathrm{Gaussian}}( \boldsymbol{\theta_0}~|~\boldsymbol{\theta}). 
\end{equation}
\end{linenomath}
It corresponds to the product of the Poisson probability of observing $N_{i}$ events in each final state, given the expectation for the signal $S_i$ and background $B_i$,
and is multiplied by the Gaussian constraints on the systematic uncertainties $\boldsymbol{\theta}$ with central values $\boldsymbol{\theta_0}$ from auxiliary measurements as described in Section~\ref{sec:intXsec}.

The measured cross sections for the $Z\gamma$ and $Z\gamma\gamma$ processes in the extended fiducial regions defined in Table~\ref{table:ZggExtendedFiducial}
are summarized in Table~\ref{table:ZggCrossSections}.
The theoretical predictions in the table are described in Section~\ref{sec:comparison}.
The significance for the combination of $\ee\gamma\gamma$ and $\mumu\gamma\gamma$ processes is 6.3 (6.0) standard deviations for the inclusive (exclusive) selection.

The $Z\gamma$ inclusive (exclusive) cross sections in the extended fiducial regions are measured with a precision of $6\%$ ($6\%$) in the $\leplep\gamma$ final state and $50\%$ ($24\%$) in the $\nnbar\gamma$ final state.
The smaller uncertainty in the exclusive $\nnbar\gamma$ measurement results from the reduced background fraction as shown in Table~\ref{table:Znunug_results}.
The $Z\gamma\gamma$ inclusive (exclusive) cross sections in the extended fiducial regions are measured with a precision of $16\%$ ($19\%$) in the $\leplep\gamma\gamma$ final state and $70\%$ ($60\%$) in the $\nnbar\gamma\gamma$ final state.
The precision of the $Z\gamma$ cross-section measurements is driven by their systematic uncertainties.  
For the $Z\gamma\gamma$ cross sections, the precision of the measurements is dominated by the statistical uncertainty in the $\leplep\gamma\gamma$ final state, and is equally affected by statistical and systematic uncertainties in the $\nnbar\gamma\gamma$ final state.

The systematic uncertainties in the measured cross sections in Table~\ref{table:ZggCrossSections} arise from the uncertainties in the acceptances $A$ and correction factors $C$, as well as from the uncertainties in the estimates of backgrounds.
In the $\leplep\gamma$ and $\leplep\gamma\gamma$ final states the two sources have effects of comparable size on the measured cross sections, 
while in the $\nnbar\gamma$ and $\nnbar\gamma\gamma$ final states the uncertainties in the estimates of backgrounds dominate. 

Compared with the $Z\gamma$ measurements at $\sqrt{s} = 7$ \TeV~\cite{Aad:2013izg}, the systematic uncertainty is reduced in the $\leplep\gamma$ final state while it becomes larger in the $\nnbar\gamma$ final state. 
The reduced systematic uncertainty in the $\leplep\gamma$ final state mainly results from the reduced systematic uncertainty from photon identification efficiency, as well as the smaller statistical uncertainty in the data-driven estimate of the $Z$+jets background. 
The larger systematic uncertainty in the $\nnbar\gamma$ final state is largely a result of the increased photon $E_{\mathrm{T}}$ threshold requirement due to the increased single-photon trigger $E_{\mathrm{T}}$ threshold, which results in generally increased systematic uncertainties in the estimates of backgrounds. 

The measurements of the cross sections in the $\ee\gamma$ and $\mumu\gamma$ final states agree within one standard deviation. 
In order to assess the compatibility of the cross-section measurements in the $\ee\gamma\gamma$ and $\mumu\gamma\gamma$ final states, a profile-likelihood ratio is constructed, parameterized as a function of the difference in measured cross sections.
With this approach, the measurements are found to be compatible within 1.7 (1.8) standard deviations in the inclusive (exclusive) case.

\begin{table}[hbtp]
\begin{center}
\renewcommand{\arraystretch}{1.2}
\begin{tabular}{l|c|c|c}
\hline\hline
Channel & Measurement [fb]            & \textsc{MCFM} Prediction [fb] & NNLO Prediction [fb]\\
\hline\hline
\multicolumn{4}{c}{$N_{\mathrm{jets}} \geq 0$ }\\
\hline
$\ee\gamma$ & 1510 $\pm 15 (\mathrm{stat.}) ^{+91} _{-84} (\mathrm{syst.}) ^{+30} _{-28} (\mathrm{lumi.})$ & \multirow{3}{*}{1345$^{+66}_{-82}$} & \multirow{3}{*}{$1483^{+19}_{-37}$} \\
$\mumu\gamma$ & 1507 $\pm 13 (\mathrm{stat.}) ^{+78} _{-73} (\mathrm{syst.}) ^{+29} _{-28} (\mathrm{lumi.})$ &  & \\
\cline{1-2}
$\leplep\gamma$ & 1507 $\pm 10 (\mathrm{stat.}) ^{+78} _{-73} (\mathrm{syst.}) ^{+29} _{-28} (\mathrm{lumi.})$ &  & \\
\hline
$\nnbar\gamma$ &  68 $\pm 4 (\mathrm{stat.}) ^{+33} _{-32} (\mathrm{syst.}) \pm 1 (\mathrm{lumi.})$ & 68.2$ \pm 2.2$ & $81.4^{+2.4}_{-2.2}$\\
\hline\hline
\multicolumn{4}{c}{$N_{\mathrm{jets}} = 0$ }\\
\hline
$\ee\gamma$ & 1205 $ \pm 14 (\mathrm{stat.}) ^{+84} _{-75} (\mathrm{syst.}) \pm 23 (\mathrm{lumi.})$ &  \multirow{3}{*}{1191$^{+71}_{-89}$} & \multirow{3}{*}{$1230^{+10}_{-18}$} \\
$\mumu\gamma$ & 1188 $ \pm 12 (\mathrm{stat.}) ^{+68} _{-63} (\mathrm{syst.}) ^{+23} _{-22} (\mathrm{lumi.})$ &   &  \\
\cline{1-2}
$\leplep\gamma$ & 1189 $ \pm 9 (\mathrm{stat.}) ^{+69} _{-63} (\mathrm{syst.}) ^{+23} _{-22} (\mathrm{lumi.})$ &   &  \\
\hline
$\nnbar\gamma$ & 43 $ \pm 2 (\mathrm{stat.}) \pm 10 (\mathrm{syst.}) \pm 1 (\mathrm{lumi.})$ & 51.0$^{+2.1}_{-2.3}$ & $49.21^{+0.61}_{-0.52}$\\
\hline\hline
\multicolumn{4}{c}{}\\
\hline\hline
\multicolumn{4}{c}{$N_{\mathrm{jets}} \geq 0$ }\\
\hline
$\ee\gamma\gamma$ & 6.2 $^{+1.2} _{-1.1} (\mathrm{stat.}) \pm 0.4 (\mathrm{syst.}) \pm 0.1  (\mathrm{lumi.})$ & \multirow{3}{*}{3.70$^{+0.21}_{-0.11}$} & \\
$\mumu\gamma\gamma$ & 3.83 $^{+0.95} _{-0.85} (\mathrm{stat.}) ^{+0.48} _{-0.47} (\mathrm{syst.}) \pm 0.07 (\mathrm{lumi.})$ & & \\
\cline{1-2}
$\leplep\gamma\gamma$ & 5.07 $^{+0.73} _{-0.68} (\mathrm{stat.}) ^{+0.41} _{-0.38} (\mathrm{syst.}) \pm 0.10 (\mathrm{lumi.})$ &  \\
\hline
$\nnbar\gamma\gamma$ & 2.5 $^{+1.0} _{-0.9} (\mathrm{stat.}) \pm 1.1 (\mathrm{syst.}) \pm 0.1 (\mathrm{lumi.})$ & 0.737$^{+0.039}_{-0.032}$ \\
\hline\hline
\multicolumn{4}{c}{$N_{\mathrm{jets}} = 0$ }\\
\hline
$\ee\gamma\gamma$ & 4.6 $^{+1.0} _{-0.9} (\mathrm{stat.}) ^{+0.4} _{-0.3} (\mathrm{syst.}) \pm 0.1 (\mathrm{lumi.})$ & \multirow{3}{*}{2.91$^{+0.23}_{-0.12}$} &  \\
$\mumu\gamma\gamma$ & 2.38 $^{+0.77} _{-0.67} (\mathrm{stat.}) ^{+0.33} _{-0.32} (\mathrm{syst.}) ^{+0.05} _{-0.04} (\mathrm{lumi.})$ & & \\
\cline{1-2}
$\leplep\gamma\gamma$ & 3.48 $^{+0.61} _{-0.56} (\mathrm{stat.}) ^{+0.29} _{-0.25} (\mathrm{syst.}) \pm 0.07 (\mathrm{lumi.})$  & & \\
\hline
$\nnbar\gamma\gamma$ & 1.18 $^{+0.52} _{-0.44} (\mathrm{stat.}) ^{+0.48} _{-0.49} (\mathrm{syst.}) \pm 0.02 (\mathrm{lumi.})$ & 0.395$^{+0.049}_{-0.037}$ \\
\hline\hline
\end{tabular}
\caption{Measured cross sections for the $Z\gamma$ and $Z\gamma\gamma$ processes at $\sqrt{s} = 8$ \TeV~in the extended fiducial regions defined in Table~\ref{table:ZggExtendedFiducial}.
The SM predictions from the generator \textsc{MCFM} calculated at NLO, as well as the predictions at NNLO~\cite{Grazzini:2015nwa} (for $Z\gamma$ only), are also shown in the table with combined statistical and systematic uncertainties.
All \textsc{MCFM}  
~\cite{MCFM_2011} 
and NNLO predictions are corrected to particle level using parton-to-particle scale factors as described in Section~\ref{sec:theory_calc}.}
\label{table:ZggCrossSections}
\end{center}
\end{table}

\subsection{Differential extended fiducial cross section for $Z\gamma$ production}
\label{sec:diffXsec}

The measurements of differential cross sections allow the comparison of data results to theory predictions in terms of not only their overall normalizations, but also their shapes.
The measurements are performed for $Z\gamma$ production in several observables that are sensitive to higher-order perturbative QCD corrections.
These include the photon transverse energy $E_{\mathrm{T}}^{\gamma}$, the invariant mass of the $\llg$ three-body system, and the jet multiplicity $N_{\mathrm{jets}}$.   
The differential cross sections are defined in the extended fiducial region, and are extracted with an unfolding procedure to remove measurement 
inefficiencies and resolution effects from the observed distributions.
The procedure described in Ref.~\cite{Aad:2013izg} is followed, using an iterative Bayesian method~\cite{D'Agostini:1994zf}.
Events from simulated signal MC samples are used to generate a response matrix for each distribution. 
Each element of the response matrix is the 
conditional probability that an event is found in bin $i$ in the measurement given that it is in bin $j$ at the particle level.
In the first iteration, the prior distribution of the particle level prediction is given by the signal MC sample.
The response matrix and the measured distribution then modify the prior distribution, giving the posterior distribution at the particle level.
For each further iteration, the posterior distribution of the previous iteration is used as the new prior distribution.
Three iterations are found to be optimal, as too many iterations give rise to large statistical fluctuations, 
while too few can produce a result that is biased by the dependence on the initial prior distribution.

The statistical uncertainties of the unfolded distribution are estimated using pseudoexperiments,
generated by fluctuating each bin of the observed spectrum according to a Poisson distribution with 
the expected value equal to the observed yield.
The shape uncertainties from the number of signal MC events are also obtained by performing pseudoexperiments.
The sources of systematic uncertainty are discussed in Section~\ref{sec:intXsec},
with their impact on the unfolded distribution assessed by varying the response matrix
for each of the systematic uncertainty sources by one standard deviation and adding up the resulting changes in quadrature.
The results from the electron and muon channels are combined with equal weight, taking into account the correlations between the systematic uncertainties in the two channels. 

In addition to the systematic uncertainties described in Section~\ref{sec:intXsec}, the differences between the unfolded results with three iterations and the results with two or four iterations are taken as systematic uncertainties associated with the unfolding method.

\begin{figure}[hbtp]
\begin{center}
\includegraphics[width=0.49\textwidth]{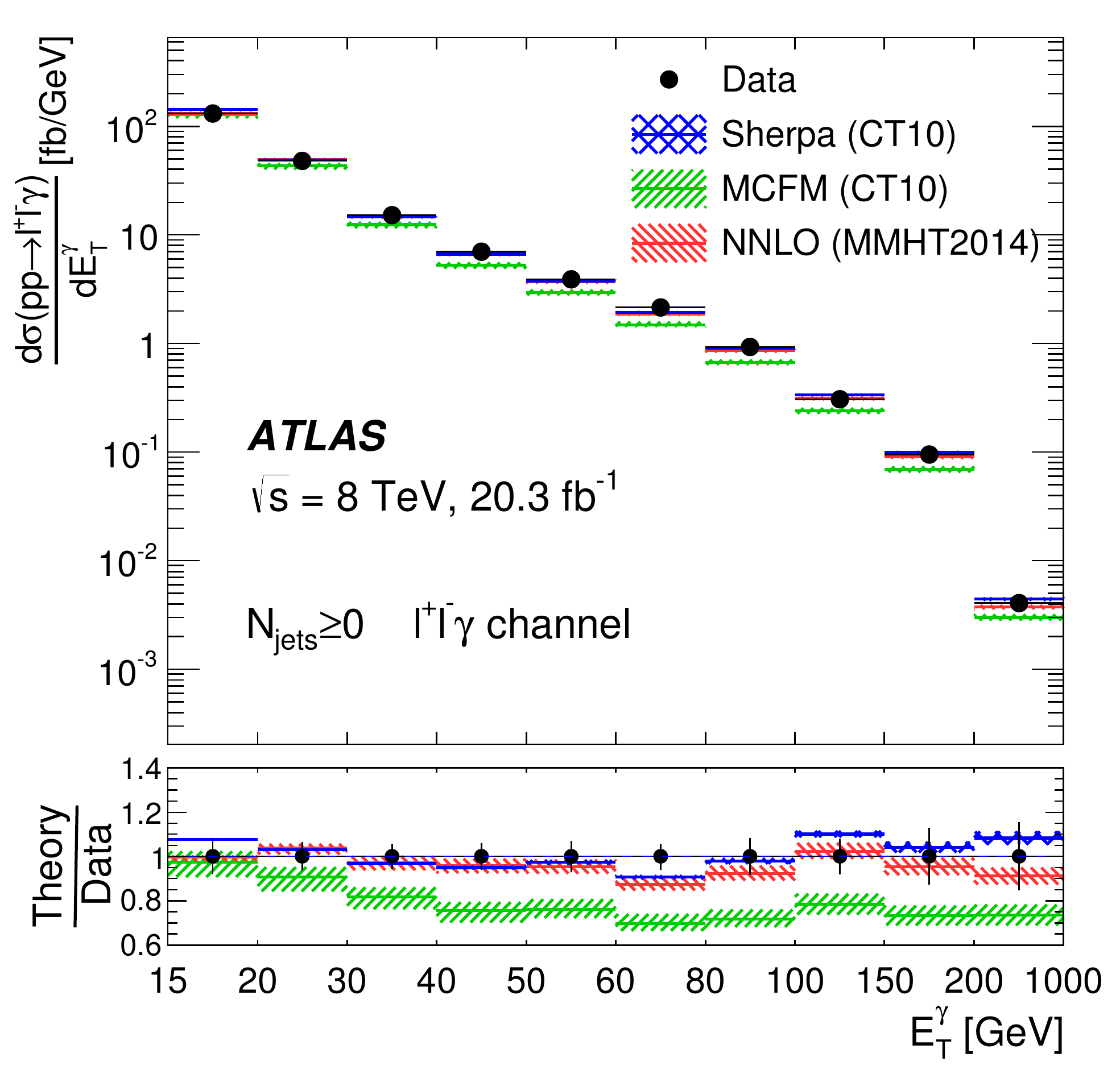}
\includegraphics[width=0.49\textwidth]{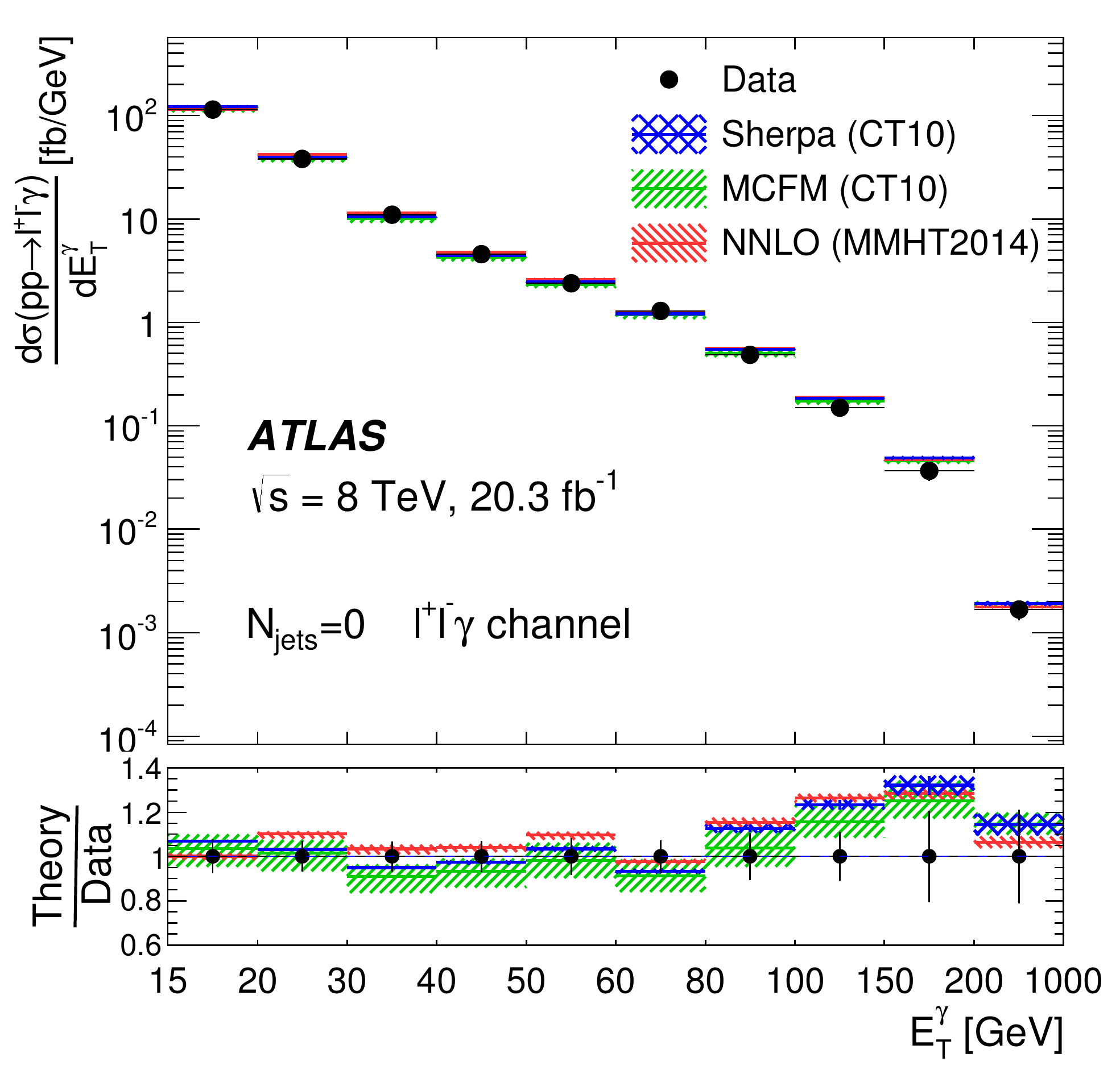}
\caption{The measured (points with error bars) and predicted differential cross sections as a function of $E_{\mathrm{T}}^{\gamma}$ for the $pp \rightarrow \leplep\gamma$ process in the inclusive $N_{\mathrm{jets}} \geq 0$ (left) 
and exclusive $N_{\mathrm{jets}} = 0$ (right) extended fiducial regions.
The error bars on the data points show the statistical and systematic uncertainties added in quadrature.
The MCFM and NNLO predictions are shown with shaded bands that indicate the theoretical uncertainties described in Section~\ref{sec:theory_calc}.
The \SHERPA predictions are shown with shaded bands indicating the statistical uncertainties from the size of the MC samples.
The lower plots show the ratios of the predictions to data (shaded bands).
The error bars on the points show the relative uncertainties of the data measurements themselves.
The bin size varies from 5~\GeV~to 800~\GeV.
}
\label{fig:UnfoldedPhotonEtllg}
\end{center}
\end{figure}

\begin{figure}[hbtp]
\begin{center}
\includegraphics[width=0.49\textwidth]{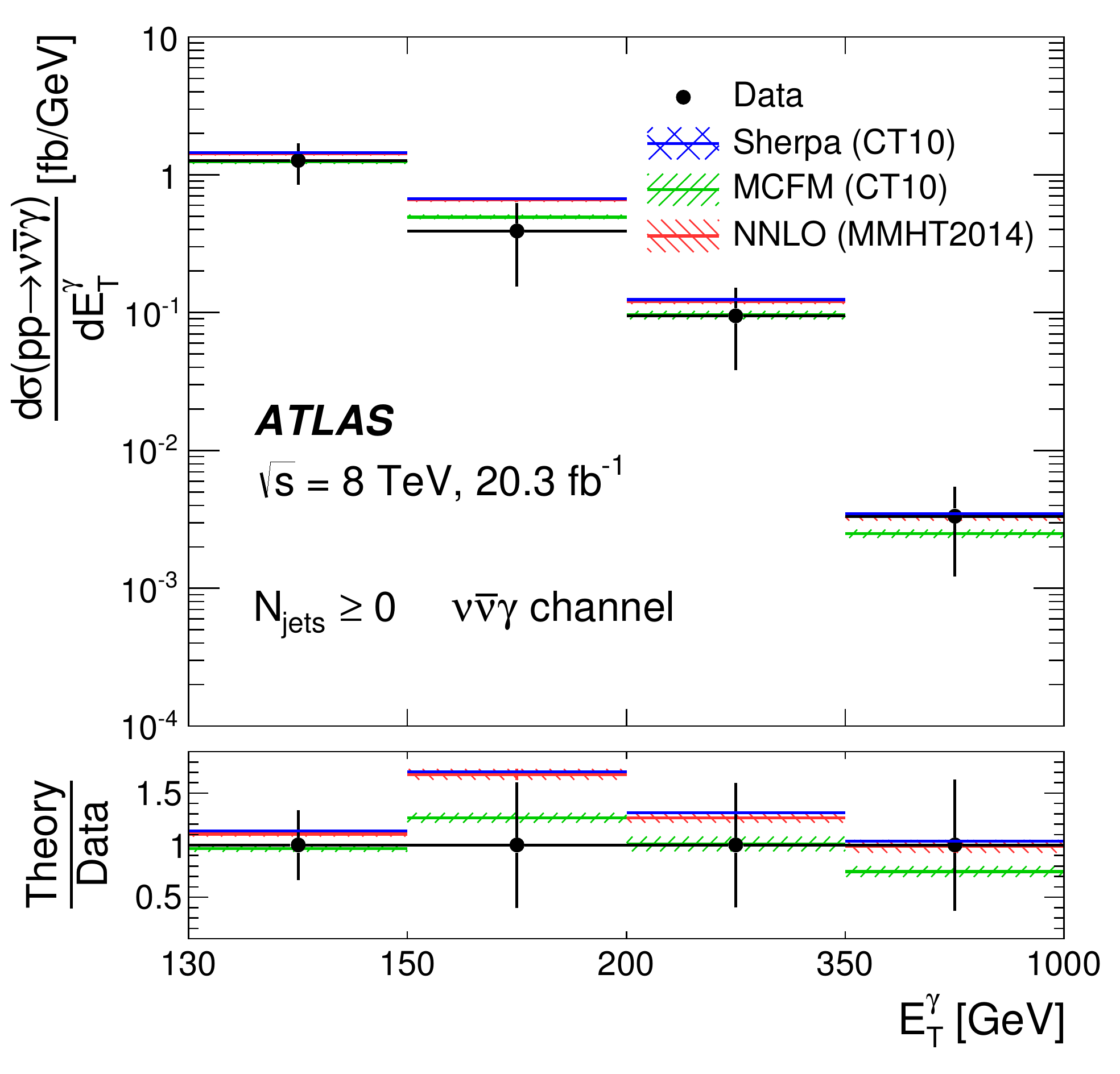}
\includegraphics[width=0.49\textwidth]{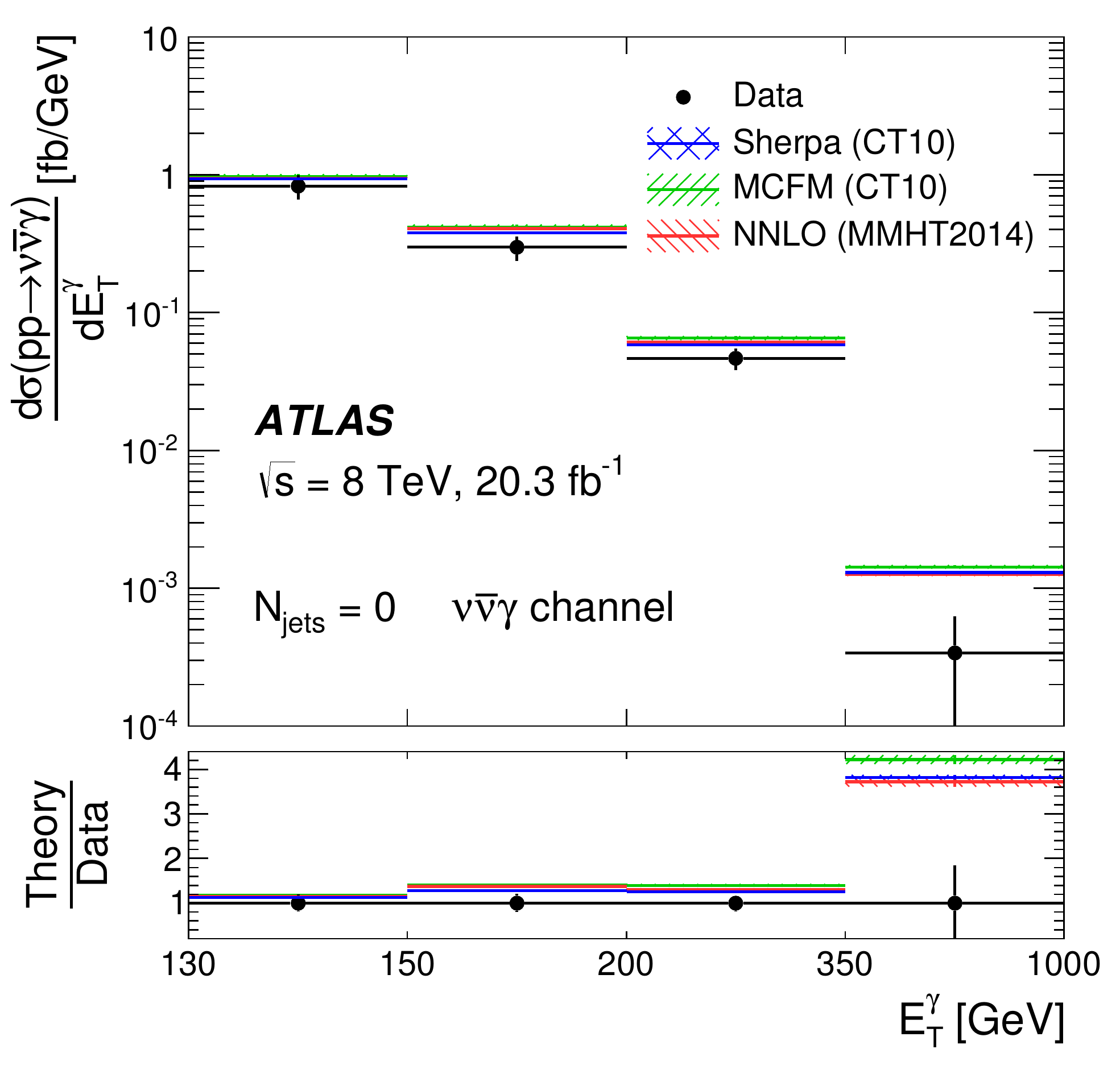}
\caption{The measured (points with error bars) and predicted differential cross sections as a function of $E_{\mathrm{T}}^{\gamma}$ for the $pp \rightarrow \nnbar\gamma$ process in the inclusive $N_{\mathrm{jets}} \geq 0$ (left)
and exclusive $N_{\mathrm{jets}} = 0$ (right) extended fiducial regions.
The error bars on the data points show the statistical and systematic uncertainties added in quadrature.
The MCFM and NNLO predictions are shown with shaded bands that indicate the theoretical uncertainties described in Section~\ref{sec:theory_calc}.
The \SHERPA predictions are shown with shaded bands indicating the statistical uncertainties from the size of the MC samples.
The lower plots show the ratios of the predictions to data (shaded bands).
The error bars on the points show the relative uncertainties of the data measurements themselves.
The bin size varies from 20~\GeV~to 650~\GeV.
}
\label{fig:UnfoldedPhotonEtnunug}
\end{center}
\end{figure}

\begin{figure}[hbtp]
\begin{center}
\includegraphics[width=0.49\textwidth]{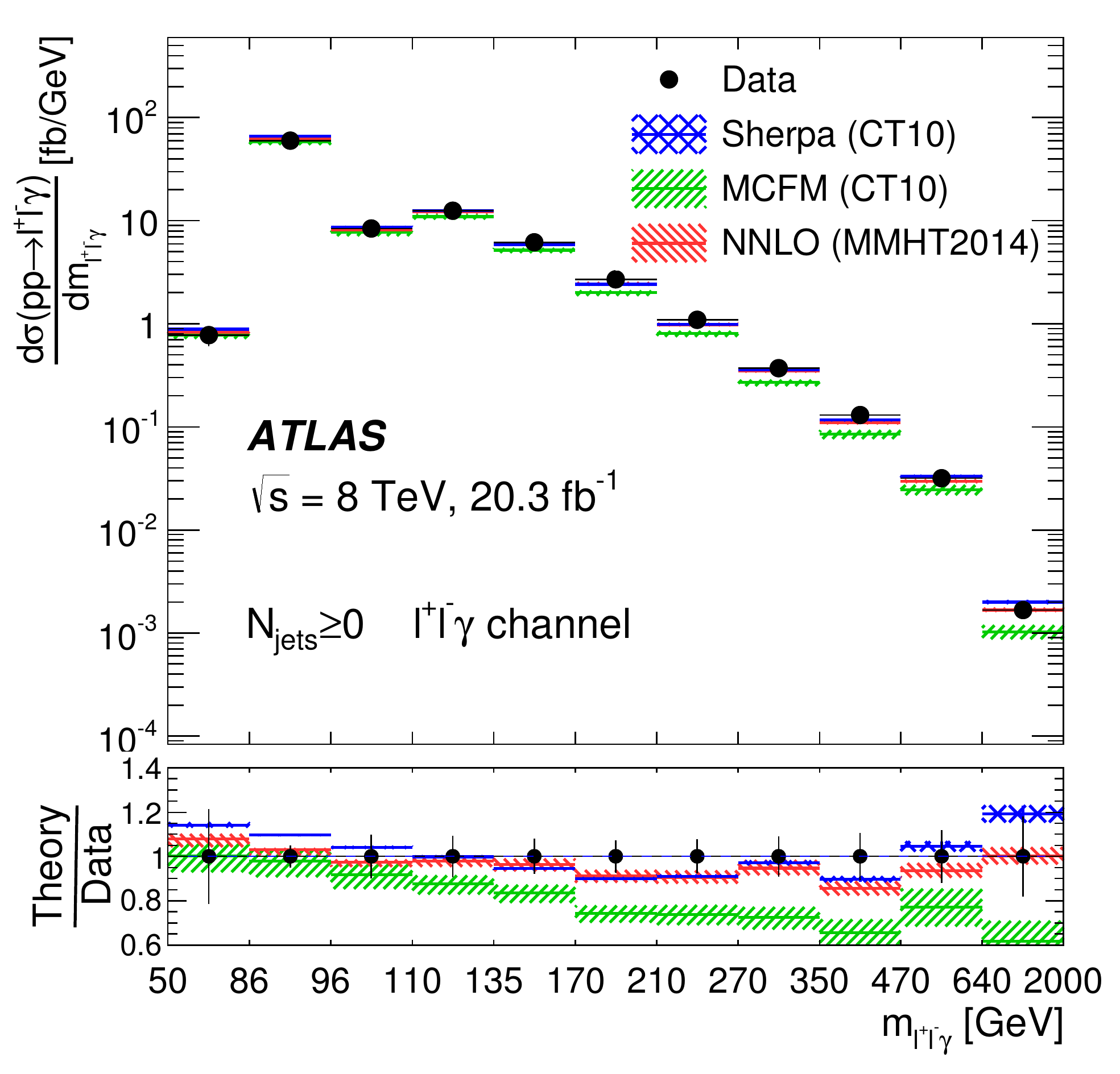}
\includegraphics[width=0.49\textwidth]{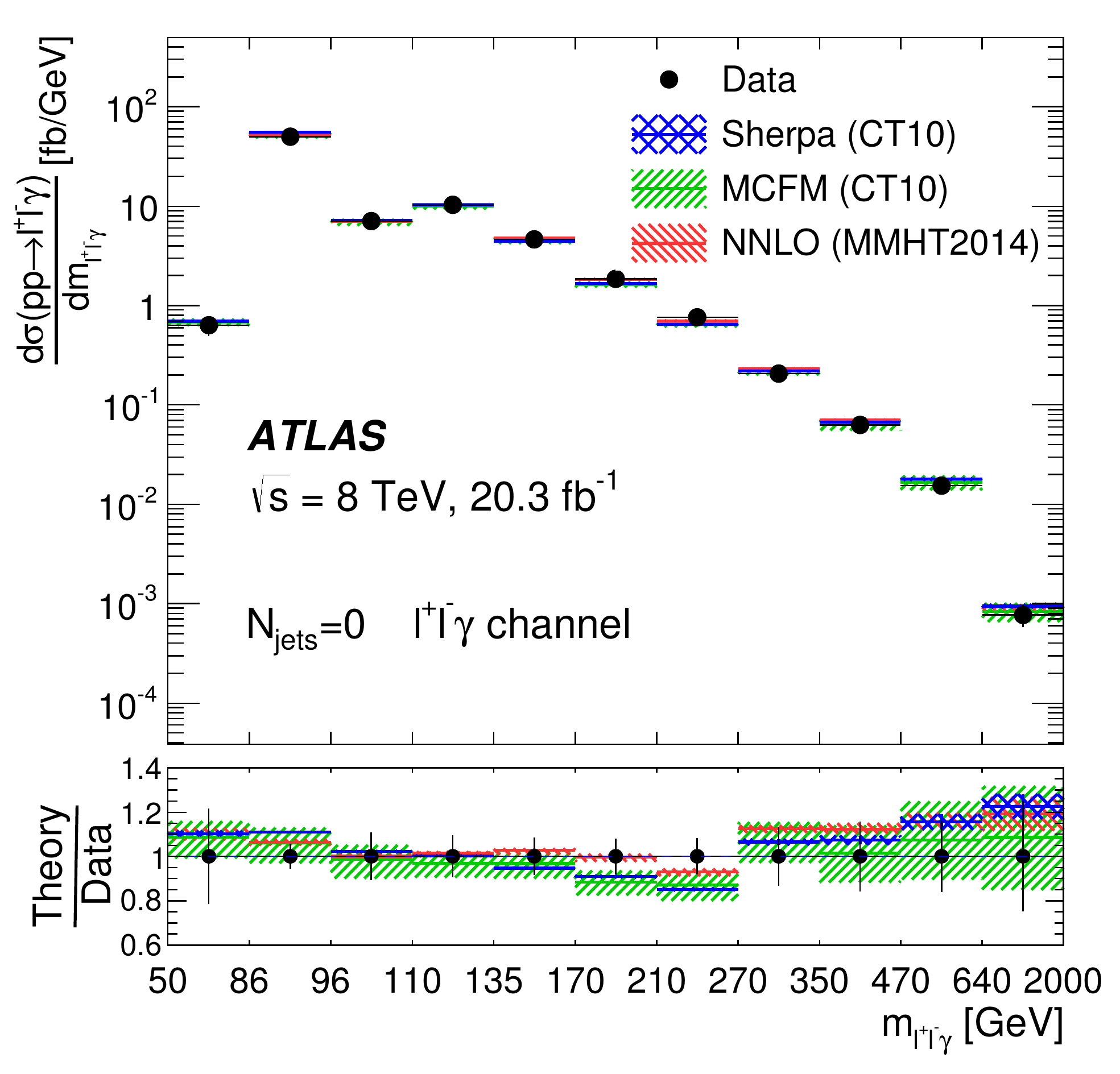}
\caption{The measured (points with error bars) and predicted differential cross sections as a function of $m_{\leplep\gamma}$ for the $pp \rightarrow \leplep\gamma$ process in the inclusive $N_{\mathrm{jets}} \geq 0$ (left) 
and exclusive $N_{\mathrm{jets}} = 0$ (right) extended fiducial regions.
The error bars on the data points show the statistical and systematic uncertainties added in quadrature.
The MCFM and NNLO predictions are shown with shaded bands that indicate the theoretical uncertainties described in Section~\ref{sec:theory_calc}.
The \SHERPA predictions are shown with shaded bands indicating the statistical uncertainties from the size of the MC samples.
The lower plots show the ratios of the predictions to data (shaded bands).
The error bars on the points show the relative uncertainties of the data measurements themselves.
The bin size varies from 10~\GeV~to 1360~\GeV.
}
\label{fig:UnfoldedMllg}
\end{center}
\end{figure}

\begin{figure}[hbtp]
\begin{center}
\includegraphics[width=0.49\textwidth]{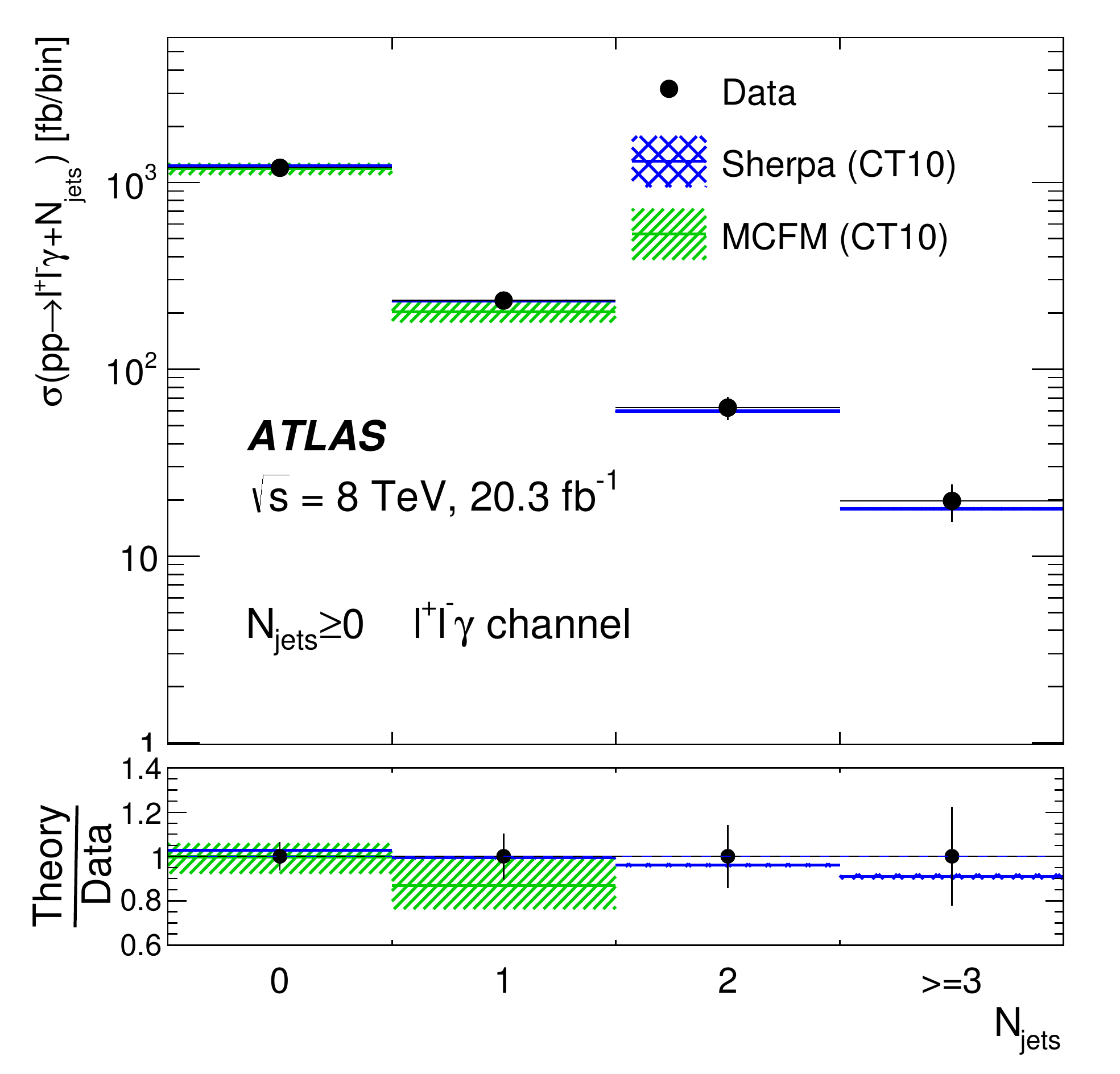}
\caption{The measured (points with error bars) and predicted cross sections as a function of $N_{\mathrm{jets}}$ for the $pp \rightarrow \leplep\gamma$ process in the extended fiducial region.
The error bars on the data points show the statistical and systematic uncertainties added in quadrature.
The MCFM prediction is shown with shaded bands that indicate the theoretical uncertainties described in Section~\ref{sec:theory_calc}.
The \SHERPA prediction is shown with shaded bands indicating the statistical uncertainties from the size of the MC samples.
The lower plot shows the ratios of the predictions to data (shaded bands).
The error bars on the points show the relative uncertainties of the data measurements themselves.
}
\label{fig:UnfoldedNJetsllg}
\end{center}
\end{figure}

The differential cross sections are presented as a function of $E_{\mathrm{T}}^{\gamma}$ in Figure~\ref{fig:UnfoldedPhotonEtllg} for the inclusive and exclusive measurements of the $\leplep\gamma$ channel 
and in Figure~\ref{fig:UnfoldedPhotonEtnunug} for the inclusive and exclusive measurements of the $\nnbar\gamma$ channel.
The differential cross sections are shown in Figure~\ref{fig:UnfoldedMllg} as a function of $m_{\leplep\gamma}$. 
Figure~\ref{fig:UnfoldedNJetsllg} shows the cross sections in the $\leplep\gamma$ channel measured in bins of jet multiplicity.
The predictions in the figures are described in Section~\ref{sec:comparison}.
As with the integrated cross sections shown in Table~\ref{table:ZggCrossSections}, the differential cross sections of the exclusive measurements in the $\nnbar\gamma$ channel have smaller uncertainties than the inclusive measurements.

\section {Comparison of measurements to Standard Model predictions}
\label{sec:comparison}

\subsection{Estimation of Standard Model expectations}
\label{sec:theory_calc}

The measurements of $\Zboson\gamma$ and $\Zboson\gamma\gamma$ production are compared to SM predictions using the parton shower Monte Carlo $\SHERPA$  1.4  and the NLO parton-level generator {\textsc{MCFM}}.\footnote{The MCFM predictions of $Z\gamma$ production include all the NLO QCD contributions of order $\alpha_{\mathrm{s}}$ and in addition the process $gg\rightarrow Z\gamma$, which is of order ${\alpha_{\mathrm{s}}}^2$. The contribution from gluon fusion is about $1\%$ ($2\%$) of the cross section in the inclusive extended fiducial region for the $\ell^{+}\ell^{-}\gamma$ ($\nu\bar{\nu}\gamma$) final state~\cite{Grazzini:2015nwa}.}
In addition, parton-level NNLO SM predictions for $\Zboson$$\gamma$ are compared to data using the calculations described in Ref.~\cite{Grazzini:2015nwa}. 
The theory predictions include off-shell $Z$ bosons and direct photons arising from initial-state radiation (from the quarks) and radiative $Z$-boson decay in the case of charged-lepton final states, and from fragmentation of final-state quarks and gluons into photons,
leading to the production channels $pp$ $\to\leplep\gamma(\gamma)+X$ and $pp$ $\to\nnbar\gamma(\gamma)+X$. In the $\SHERPA$ and {\textsc{MCFM}} generators, contributions from quark/gluon fragmentation into isolated photons are also included. The CT10 PDF set~\cite{Lai:2010vv} is used for the $\SHERPA$  and {\textsc{MCFM}} generation, and the MMHT2014 PDF set~\cite{Harland-Lang:2014zoa} is 
used for the NNLO predictions. The renormalization and factorization scales 
are set equal to $m_{\Zboson\gamma}$ ($m_{\Zboson\gamma\gamma}$) for the {\textsc{MCFM}} NLO generation of  
$\Zboson\gamma$ ($\Zboson\gamma\gamma$) events and to $\sqrt{m_\Zboson^2 + (\ET^{\gamma})^2}$ for the NNLO
$\Zboson\gamma$ predictions. The other electroweak parameters used are the default values~\cite{mcfm_man} from the authors of the generators.

The events
generated with $\SHERPA$ as described in Section~\ref{sec:signalMC} are also compared to
the measurements at particle level.
 For the NLO and NNLO parton-level predictions, parton-to-particle correction factors $C^{*}$(parton $\to$ particle) must be applied in order to obtain the particle level cross sections. 
These correction factors are computed as the ratios of the $pp\rightarrow Z\gamma(\gamma)$ cross sections predicted by $\SHERPA$ with hadronization and the underlying event disabled to the cross sections with them enabled.
The systematic uncertainties in the correction factors are evaluated by using an alternative parton-showering method~\cite{Carli:2010cg} within $\SHERPA$, and are found to be negligible compared to the statistical uncertainties. 
The particle level cross sections are 
 obtained by dividing the NLO and NNLO parton-level predictions by the $C^{*}(\mathrm{parton}\to \mathrm{particle})$ correction factors
 summarized in Table~\ref{tab:SF_p2p}. The corrections are a few percent for the inclusive cross sections and reach 
 about 10$\%$ for some exclusive channels. The correction factors in Table~\ref{tab:SF_p2p} apply to the predictions made for the $\Zboson\gamma$ and $\Zboson\gamma\gamma$ cross sections in the extended fiducial region described in Table~\ref{table:ZggExtendedFiducial}. 
 
The systematic uncertainties in the SM NLO cross sections are estimated
by varying the QCD scales by factors of 0.5 to 2.0 (independently for the
renormalization and factorization scales) and varying the 
CT10 PDFs by their uncertainties at 68\% confidence level. The uncertainties due to the contribution of photons from fragmentation of quarks or gluons are estimated by varying
the fraction of hadronic energy $\epsilon^{p}_{h}$ in the isolation cone from
0.25 to 0.75. For the NLO exclusive zero-jet cross sections
  the method suggested in Ref.~\cite{Stewart:2011cf} is used to estimate the
additional uncertainty due to the $N_{\mathrm{jet}} = 0$ requirement. The systematic uncertainties in the SM
NNLO cross sections are determined as described
in Ref.~\cite{Grazzini:2015nwa}.  In all cases the
uncertainties in the parton-to-particle correction factors are included.

\begin{table}[hbtp]
\begin{center}
\begin{tabular}{lcc}
\hline
\hline
                & $N_{\mathrm{jets}} \geq 0$ & $N_{\mathrm{jets}} = 0$   \\
\hline
$\leplep\gamma$       & $1.01708 \pm 0.00065$  & $0.96809 \pm 0.00078$ \\
$\nnbar\gamma$   & $0.9987 \pm 0.0025  $ & $0.9150 \pm 0.0030  $   \\
\hline
$\leplep\gamma\gamma$                & $1.0273 \pm 0.0039$ &  $0.9755 \pm 0.0047$ \\
$\nnbar\gamma\gamma$   & $1.0012 \pm 0.0076  $ & $0.873 \pm 0.010  $   \\
\hline
\hline
\end{tabular}
\caption{Parton-to-particle correction factors $C^{*}(\mathrm{parton}\to \mathrm{particle})$ obtained from the $\SHERPA$ MC samples.
For $\leplep\gamma$ and $\leplep\gamma\gamma$ channels the parton-to-particle level correction factors are the weighted average over both lepton flavors ($e$, $\mu$).
The uncertainties include both the statistical and systematic contributions.}
\label{tab:SF_p2p}
\end{center}
\end{table}

\subsection{Extended fiducial cross sections compared to SM predictions}

The measured extended fiducial cross sections for $pp$ $\to$ $\leplep$$\gamma$+$X$ and $pp$ $\to$ $\nnbar$$\gamma+X$ production are compared to SM predictions in Table~\ref{table:ZggCrossSections}. 
The estimates of the cross section at NLO and NNLO and their systematic uncertainties are obtained as described above. 
Predictions are made for both inclusive production (no restriction on the system recoil $X$) and exclusive production of events having no central ($|$$\eta$$|$ $<$ 4.5) jet with $\pT > 30$ \GeV. 
There is generally good agreement between the cross-section measurements for these  $\Zboson\gamma$ channels and the SM predictions; the NNLO calculation of the inclusive cross section for the $\Zboson$($\leplep$)$\gamma$ channel gives better agreement with the measurement than the NLO calculation.  

Requiring two photons with $\ET$ $>$ 15 $\GeV$ results in a $\leplep\gamma\gamma$ cross section a factor of approximately 400 times smaller than $\leplep\gamma$ production. 
The measurements for both the $\leplep\gamma\gamma$ and $\nnbar\gamma\gamma$  channels are compared to the NLO MCFM predictions in Table~\ref{table:ZggCrossSections}. The measurements in these channels are statistically limited, but the
data are consistent with the predicted SM cross sections. 
The measured cross sections and the MCFM predictions are compatible within 1.7 (0.9) standard deviations in the inclusive (exclusive) $\leplep\gamma\gamma$ channel, and within 1.2 standard deviations in the $\nnbar\gamma\gamma$ channel.

\subsection{Differential cross sections compared to SM predictions}

The background-subtracted, unfolded differential cross sections for the $\ET^{\gamma}$ spectra from
$pp\to \leplep \gamma+X$ and $pp \to \nnbar\gamma+X$ production are compared to SM
expectations in Figures~\ref{fig:UnfoldedPhotonEtllg} and~\ref{fig:UnfoldedPhotonEtnunug}. For inclusive
$pp \to \leplep \gamma+X$  the NLO calculation underestimates the production
of photons at high $\ET$, whereas the NNLO calculation and the $\SHERPA$ shower MC both agree with 
the data. For exclusive $pp \to \leplep \gamma+X$ production all three SM calculations are in
good agreement with the data, as are the SM predictions for the photon $\ET$ spectra from
$pp$ $\to$ $\nnbar$$\gamma+X$ production.

The differential spectra of the  $\leplep \gamma$ invariant mass from $pp \to \leplep \gamma+X$
are compared to data in Figure~\ref{fig:UnfoldedMllg}. For the exclusive channel all three SM predictions agree well with the
data. For the inclusive channel the NLO prediction underestimates the cross section at high 
$m_{\leplep\gamma}$, while the NNLO calculation is in good agreement with the data.

In Figure~\ref{fig:UnfoldedNJetsllg} the measured jet multiplicity spectrum from $\leplep \gamma$ events is compared to NLO {\textsc{MCFM}} predictions
for zero and one jet, and to $\SHERPA$ for zero to three jets. These SM predictions are in
agreement with the data.

\section {Limits on triple and quartic gauge-boson couplings}
\label{sec:agc}

\subsection{Anomalous triple gauge-boson couplings $\Zboson$$\Zboson$$\gamma$ and $\Zboson$$\gamma$$\gamma$}

Within the Standard Model, vector-boson self-interactions are completely fixed by the model's $S\kern -0.15em U(2)_L \times U(1)_Y$ gauge structure~\cite{PhysRevD.47.4889}. Their observation is thus a crucial test of the model. Any deviation from the SM prediction is called an anomalous coupling.
Anomalous triple gauge-boson couplings for $\Zboson\gamma$ production can be parameterized by four CP-violating ($h_1^V$, $h_2^V$) and 
four CP-conserving ($h_3^V$, $h_4^V$) complex parameters (where $V=Z$, $\gamma$).
All of these parameters are zero at tree level in the SM. Since the CP-conserving couplings $h_{3,4}^V$ and the CP-violating couplings $h_{1,2}^V$ do not interfere and their 
sensitivities to aTGCs are nearly identical~\cite{PhysRevD.47.4889}, 
the limits from this study are expressed in terms of the CP-conserving parameters $h_{3,4}^V$.

The yields of $\Zboson\gamma$ events with high $\ET^{\gamma}$ with the exclusive zero-jet selection are used to set the limits.
The exclusive selection is used since it significantly reduces the SM contribution at high $\ET^{\gamma}$ and therefore optimizes the sensitivity to anomalous couplings.
The contributions from aTGCs increase with 
the $\ET$ of the photon, and the search is optimized to have the highest sensitivity by using the extended fiducial cross sections for $\Zboson
\gamma$ production with $\ET^{\gamma}$ greater than 250 \GeV~for $\leplep\gamma$ and greater than 400 \GeV~for $\nnbar\gamma$. The neutrino channel has the highest sensitivity to aTGCs.
The measured cross sections and the SM predictions in these high-$\ET^\gamma$ phase-space regions (aTGC regions) are shown in Table~\ref{table:aTGC_reg_SM_cs}.

\begin{table}[hbtp]
\begin{center}
\begin{spacing}{1.25}
\begin{tabular}{l c c}
\hline\hline
Channel  & Measurement [fb]  & Prediction [fb] \\
\hline
$\leplep\gamma$ ($E_{\mathrm{T}}^\gamma>$ 250 \GeV) & $0.42^{+0.16}_{-0.13}(\mathrm{stat.})^{+0.07}_{-0.04}(\mathrm{syst.})$ & $0.660\pm0.015(\mathrm{stat.})\pm0.018(\mathrm{syst.})$  \\
$\nnbar\gamma$ ($E_{\mathrm{T}}^\gamma>$ 400 \GeV)& $0.06^{+0.15}_{-0.10}(\mathrm{stat.})^{+0.04}_{-0.04}(\mathrm{syst.})$  & $0.466\pm0.021(\mathrm{stat.})\pm0.020(\mathrm{syst.})$  \\
\hline\hline
\end{tabular}
\end{spacing}
\end{center}
\vspace*{-8mm}
\caption{Theoretical \textsc{MCFM} SM and observed cross sections in chosen aTGC regions (with the exclusive selection) for the channels studied. 
The $\ET^{\gamma}$ threshold is $250$ \GeV~for the electron and muon channels and is $400$ \GeV~for the neutrino channel. 
The first uncertainty is statistical, the second is systematic.}
\label{table:aTGC_reg_SM_cs}
\end{table}

\begin{figure}[hbtp]
\begin{center}
\includegraphics[width=0.45\textwidth]{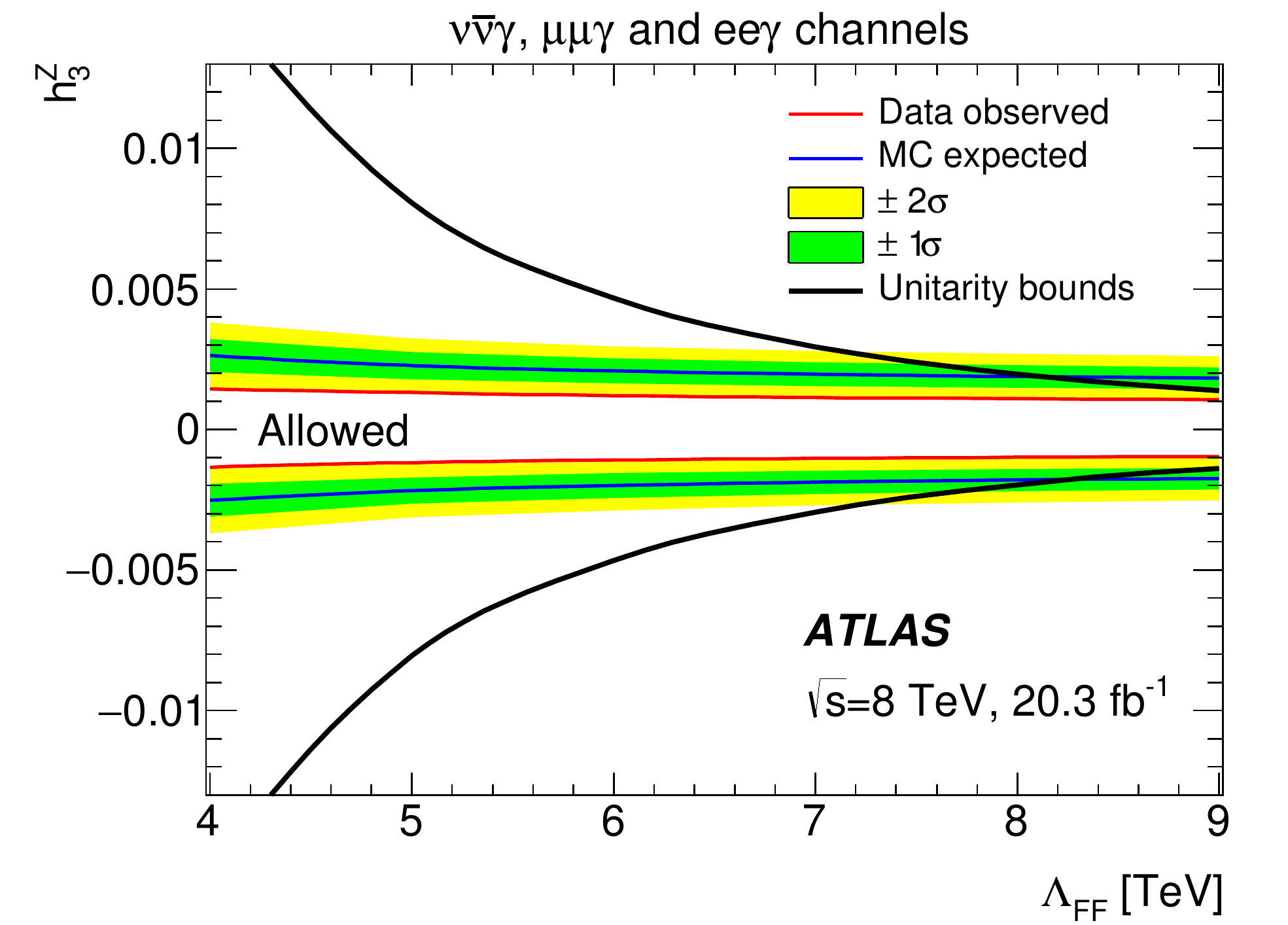}
\includegraphics[width=0.45\textwidth]{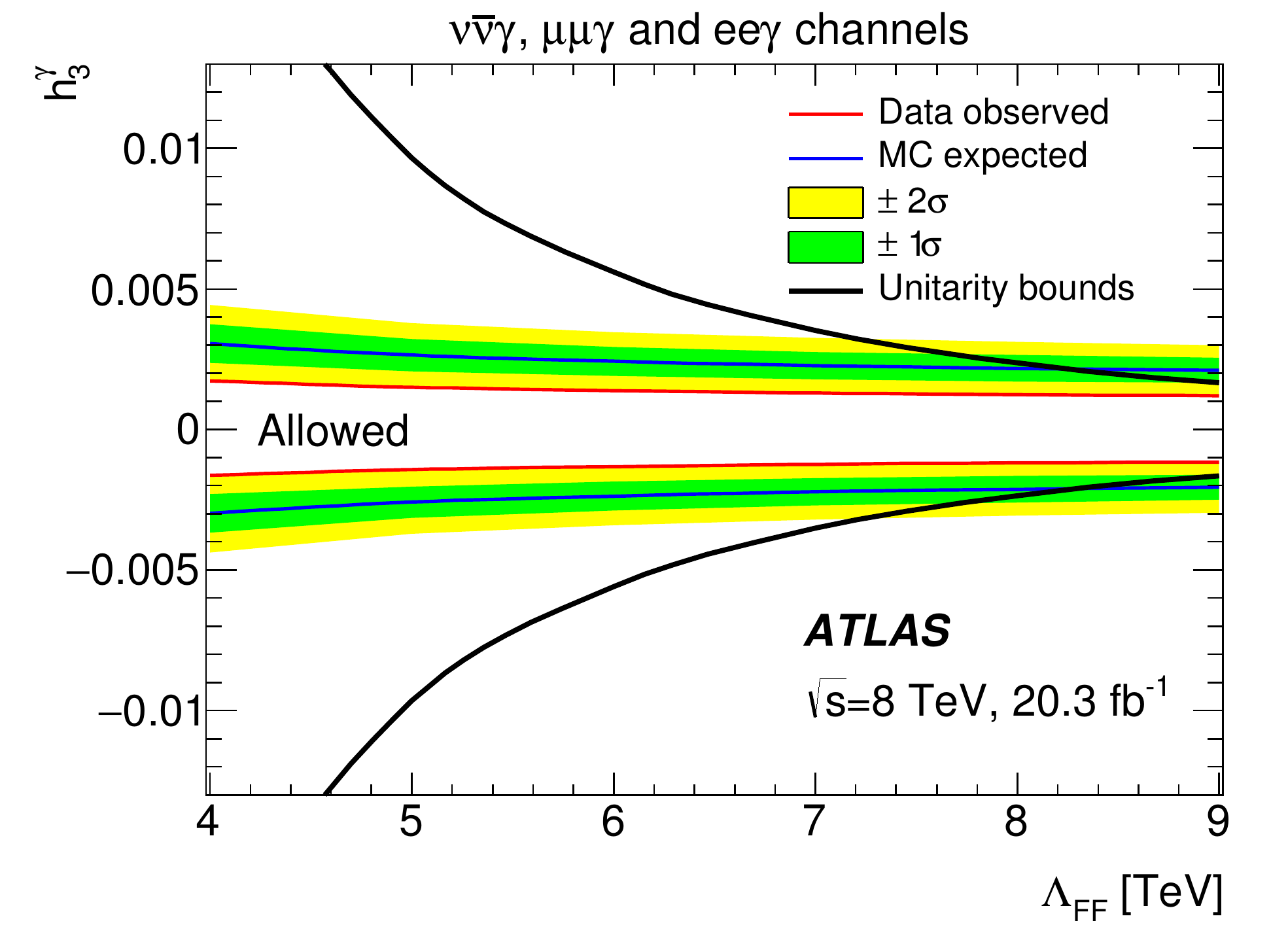}
\caption{Dependencies of the observed limits, expected limits and unitarity bounds on the form factor energy scale $\Lambda_{\mathrm{FF}}$ for $h_3^\Zboson$ (left) and $h_3^{\gamma}$ (right). $\Lambda_{\mathrm{FF}}\le8$ \TeV~can be chosen to obtain the unitarized limits. The green and yellow bands show areas of variation for the expected limits by 1$\sigma$ and 2$\sigma$, respectively.}
\label{fig:h3_Lambda_opt}
\end{center}
\end{figure}

\begin{figure}[hbtp]
\begin{center}
\includegraphics[width=0.45\textwidth]{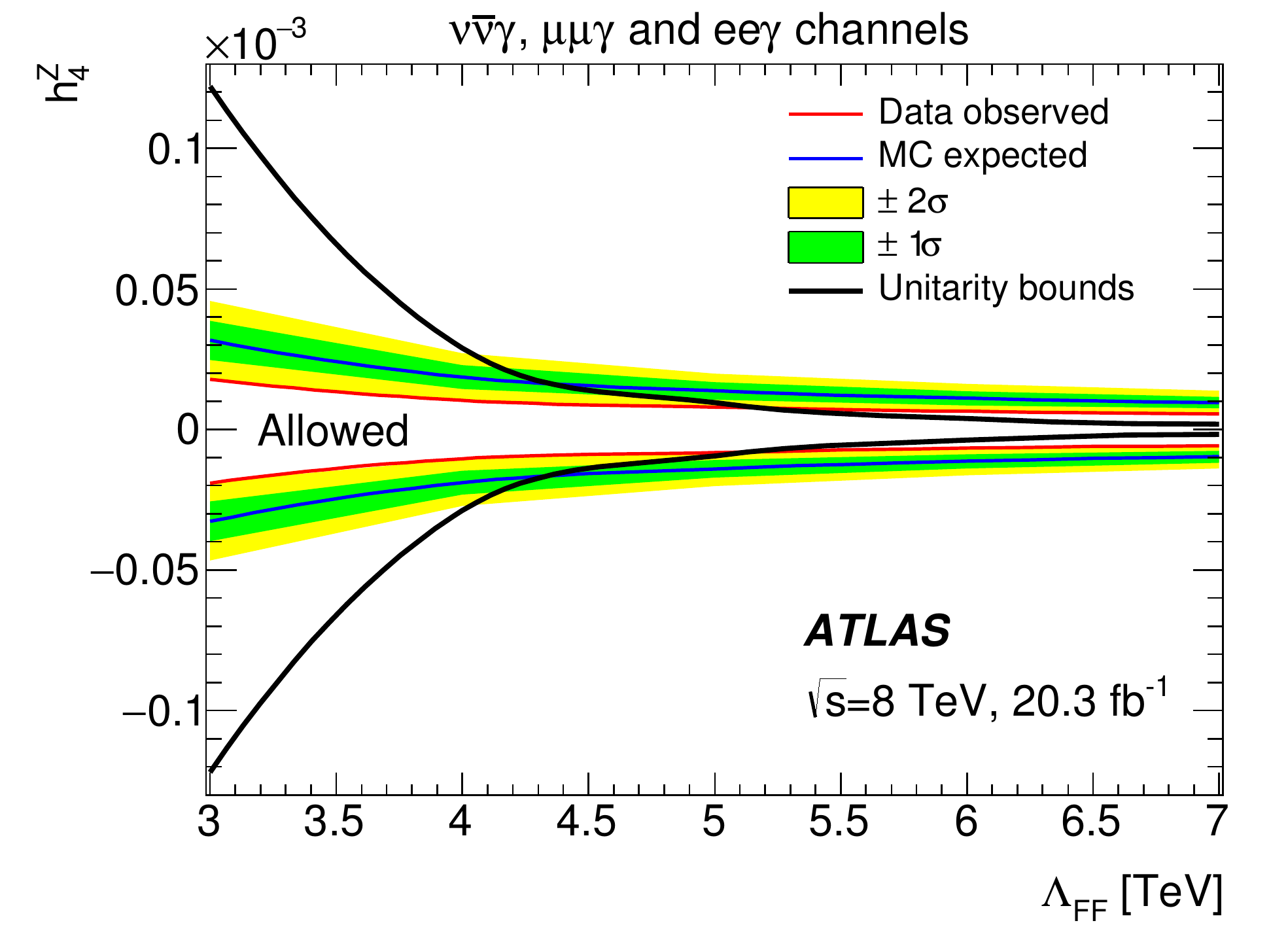}
\includegraphics[width=0.45\textwidth]{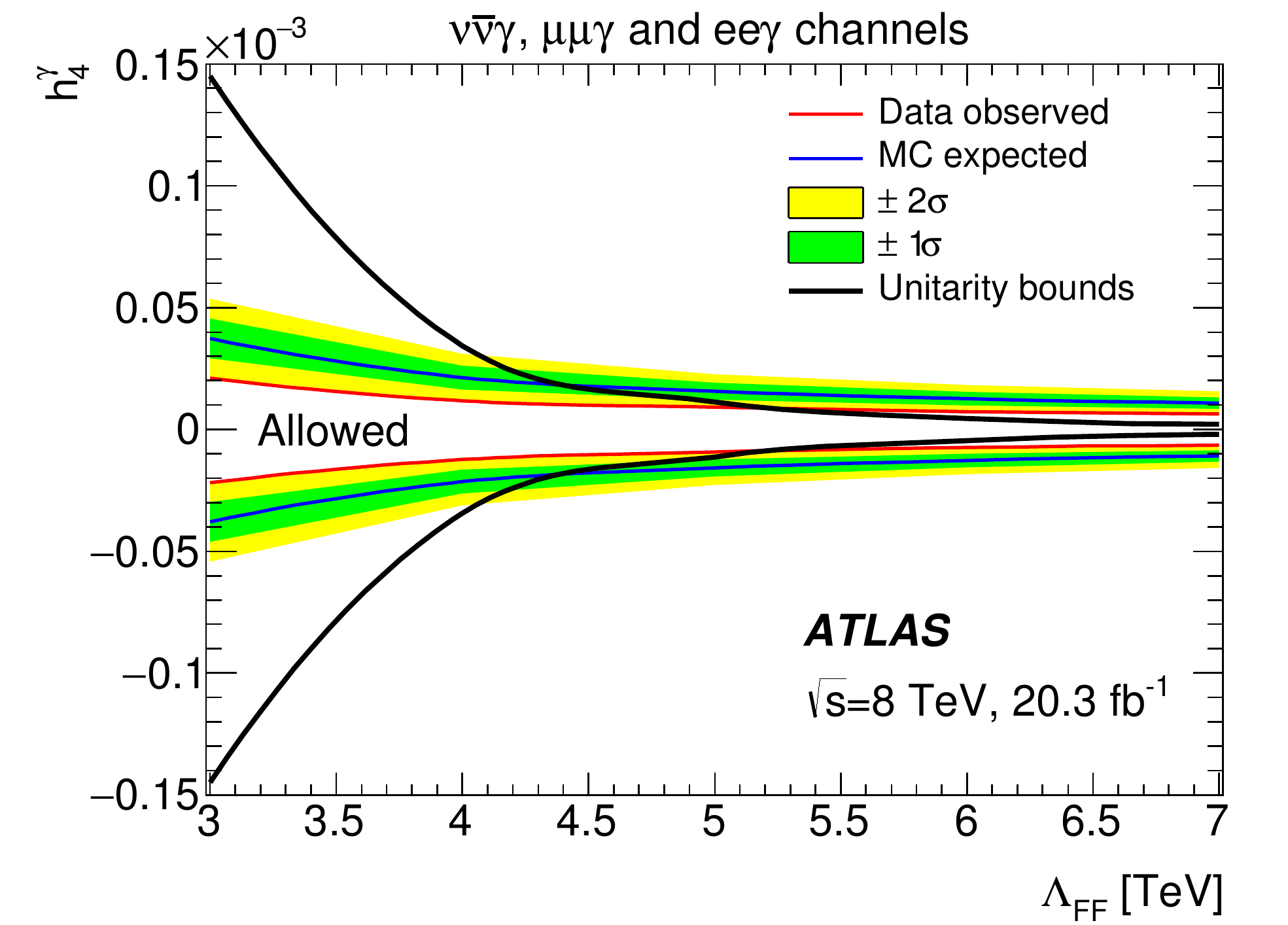}
\caption{Dependencies of the observed limits, expected limits and unitarity bounds on the form factor energy scale $\Lambda_{\mathrm{FF}}$ for $h_4^\Zboson$ (left) and $h_4^{\gamma}$ (right). $\Lambda_{\mathrm{FF}}\le4$ \TeV~can be chosen to obtain the unitarized limits. The green and yellow bands show areas of variation for the expected limits by 1$\sigma$ and 2$\sigma$, respectively.}
\label{fig:h4_Lambda_opt}
\end{center}
\end{figure}

Form factors (FF) are introduced to avoid unitarity violation at very high parton center-of-mass energy $\sqrt{\hat{s}}$: 
$h_3^V(\hat{s})=h_{3}^V/(1+\hat{s}/\Lambda_{\mathrm{FF}}^2)^n$ and $h_4^V(\hat{s})=h_{4}^V/(1+\hat{s}/\Lambda_{\mathrm{FF}}^2)^n$,
with the form factor exponent $n$ set to three for $h_3^V$ and four for $h_4^V$ to preserve unitarity~\cite{EuPhJ.16.105}, where $\Lambda_{\mathrm{FF}}$ is the approximate energy scale at which contributions from physics beyond the SM would become directly observable.
The dependencies of the unitarity bounds on the aTGC parameters from the scale $\Lambda_{\mathrm{FF}}$ calculated as in Ref.~\cite{PLB.201.3} are shown in Figures~\ref{fig:h3_Lambda_opt} and ~\ref{fig:h4_Lambda_opt}, where the observed and expected limits are derived as discussed below. 
A form factor with $\Lambda_{\mathrm{FF}}$ = 4 \TeV~is chosen as the lowest scale to
preserve unitarity for all the studied parameters. The limits on aTGCs are
also given without a form factor ($\Lambda_{\mathrm{FF}}=\infty$) as a benchmark, although
unitarity is not preserved in this case.

The cross-section predictions with aTGCs ($\sigma_{\Zboson\gamma}^{\mathrm{aTGC}}$) are obtained from the \textsc{MCFM} generator.
The number of expected $\Zboson\gamma$ events in the exclusive aTGC region ($N^{\mathrm{aTGC}}_{\Zboson\gamma}(h_3^V,h_4^V)$, where $V = Z$ or $\gamma$) is obtained using

\begin{equation}
N^{\mathrm{aTGC}}_{\Zboson\gamma}(h_3^V,h_4^V) = \sigma_{\Zboson\gamma}^{\mathrm{aTGC}}(h_3^V,h_4^V) \times C_{\Zboson\gamma} \times A_{\Zboson\gamma} \times \frac{1}{C^{*}(\mathrm{parton}\to \mathrm{particle})} \times \int 
\mathcal{L} dt.
\label{eqn:NaTGC}
\end{equation}

The anomalous couplings influence the kinematic properties of the $\Zboson\gamma$ events and thus the corrections for event reconstruction ($C_{\Zboson
\gamma}$). The maximum variations of $C_{\Zboson\gamma}$ due to nonzero aTGC parameters within the measured aTGC limits are quoted as additional systematic uncertainties. Since the influence of the anomalous couplings on the acceptance corrections ($A_{\Zboson\gamma}$) and parton-to-particle ($C^{*}(\mathrm{parton}\to \mathrm{particle})$) corrections is an order of magnitude smaller than on $C_{\Zboson\gamma}$, it is neglected.

The limits on a given aTGC parameter are extracted from a frequentist profile-likelihood test, as explained in Section~\ref{sec:xsectotal}. The profile likelihood combines the observed number of exclusive $\Zboson\gamma$ candidate events for the $\ET^{\gamma}$ threshold mentioned above, the expected signal as a function of aTGC as described in Equation~\ref{eqn:NaTGC}, and the estimated number of background 
events separately for each channel. A point in the aTGC space is accepted (rejected) at the 95\% confidence level (C.L.) if fewer (more) than 95\% of the randomly generated 
pseudoexperiments exhibit larger profile-likelihood-ratio values than that observed in data. A pseudoexperiment in this context is a set of randomly generated numbers of events, which follow the Poisson distribution with the mean equal to the sum of the number of expected signal events and the estimated number of background events. The systematic uncertainties are included in 
the likelihood function as nuisance parameters with correlated Gaussian constraints, and all nuisance parameters are fluctuated in each 
pseudoexperiment.

\begin{figure}[hbtp]
\begin{center}
\includegraphics[width=0.45\textwidth]{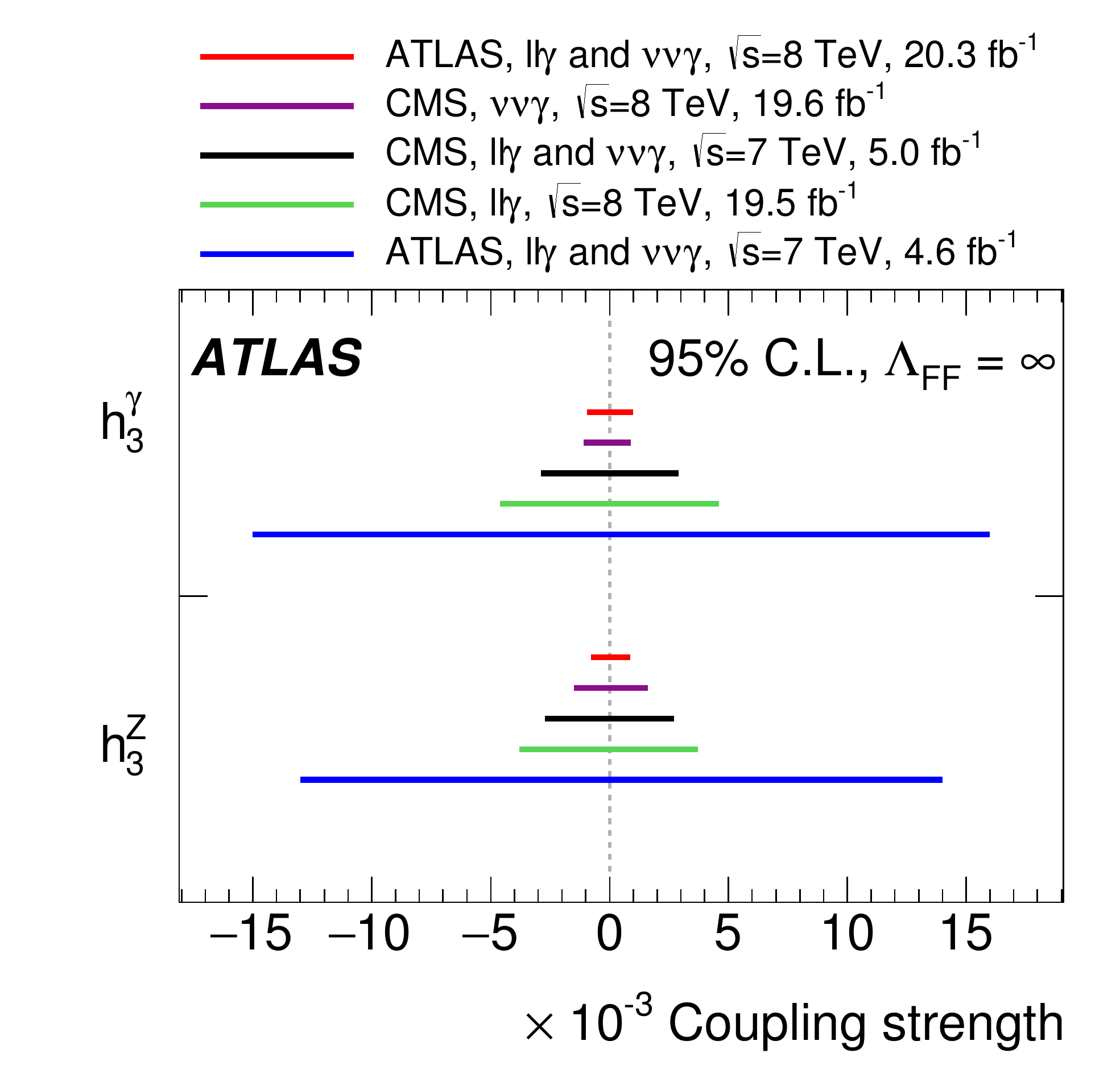}
\includegraphics[width=0.45\textwidth]{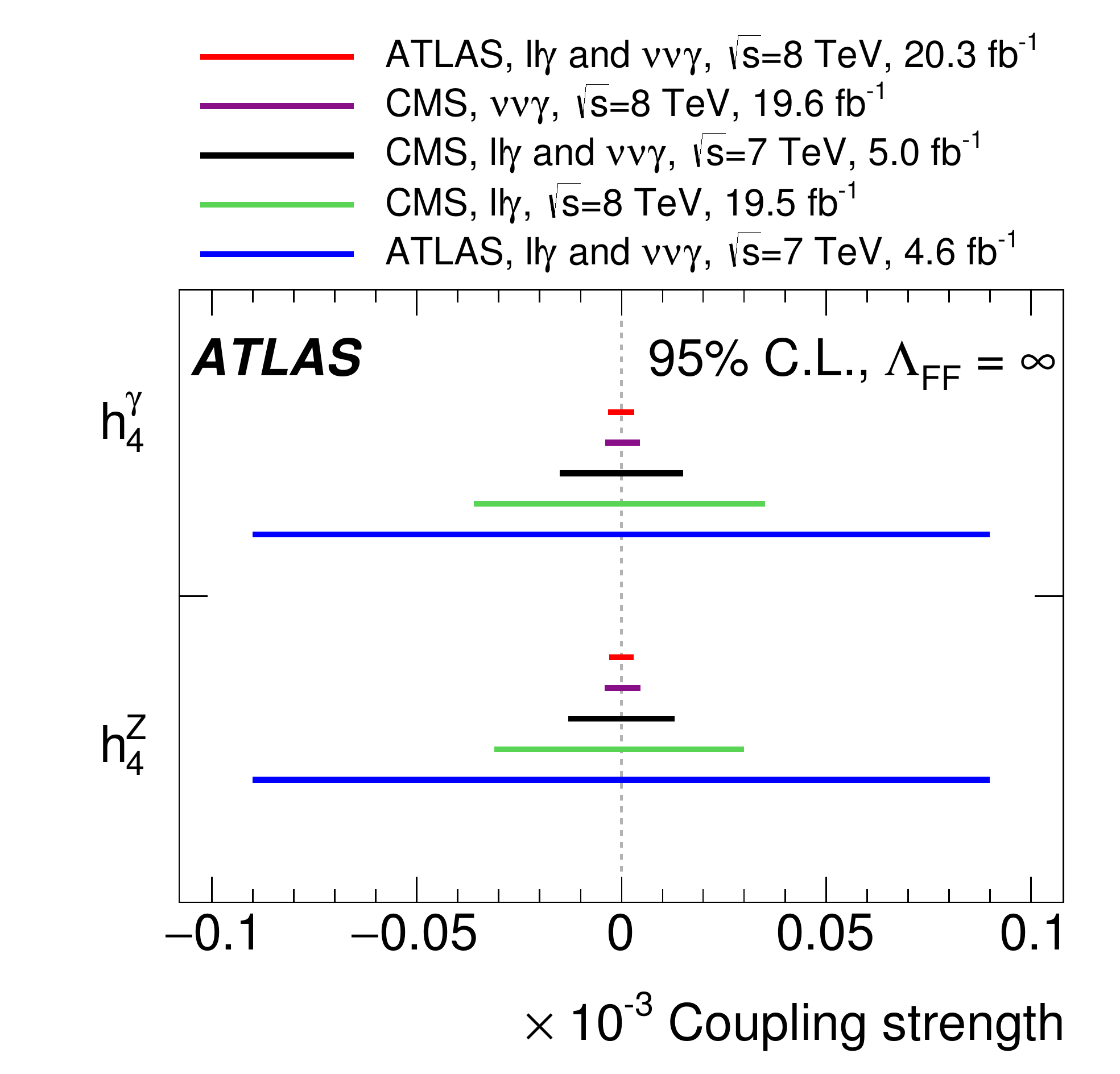}
\caption{The 95\% C.L. nonunitarized intervals ($\Lambda_{\mathrm{FF}}=\infty$) for anomalous couplings from current and previous ATLAS results and CMS results for the neutral aTGC $h_3^{\gamma}$, $h_3^\Zboson$ (left) and $h_4^{\gamma}$, $h_4^\Zboson$ (right) as obtained from $\Zboson\gamma$ events.}
\label{fig:h3_limit_comp}
\end{center}
\end{figure}

The allowed ranges for the anomalous couplings are shown in Table~\ref{table:observedExpected1DLimitsATGC} for $\Zboson\Zboson\gamma$ ($h_3^Z$ and $h_4^Z$) and $\Zboson\gamma\gamma$ ($h_3^\gamma$ and $h_4^\gamma$) vertices. These results are compared in Figure~\ref{fig:h3_limit_comp} with the previous ATLAS results~\cite{Aad:2013izg} and results from the CMS experiment~\cite{CMS:2013zg,Chatrchyan:2013nda,Khachatryan:2015kea,bib:CMSZnunug15}.

\begin{table}[hbtp]
\begin{center}
\begin{tabular}{ccc}
\hline\hline
Process & \multicolumn{2}{c}{$pp \rightarrow \leplep\gamma$ and $pp \rightarrow \nnbar\gamma$}\\
\hline
$\Lambda_{\mathrm{FF}}$ & \multicolumn{2}{c}{$\infty$}\\
 & Observed 95\% C.L. & Expected 95\% C.L. \\
$h_3^\gamma$ &  $[-9.5, 9.9] \times 10^{-4}$ & $[-1.8, 1.8] \times 10^{-3} $\\
$h_3^\Zboson$ & $[-7.8, 8.6] \times 10^{-4}$ & $[-1.5, 1.5] \times 10^{-3} $\\
$h_4^\gamma$ &  $[-3.2, 3.2] \times 10^{-6}$ & $[-6.0, 5.9] \times 10^{-6} $\\
$h_4^\Zboson$ & $[-3.0, 2.9] \times 10^{-6}$ & $[-5.5, 5.4] \times 10^{-6} $\\
\hline
$\Lambda_{\mathrm{FF}}$ & \multicolumn{2}{c}{4 \TeV}\\
 & Observed 95\% C.L. & Expected 95\% C.L. \\
$h_3^\gamma$ &  $[-1.6, 1.7] \times 10^{-3}$ & $[-3.0, 3.1] \times 10^{-3} $\\
$h_3^\Zboson$ & $[-1.3, 1.4] \times 10^{-3}$ & $[-2.5, 2.6] \times 10^{-3} $\\
$h_4^\gamma$ &  $[-1.2, 1.1] \times 10^{-5}$ & $[-2.2, 2.1] \times 10^{-5} $\\
$h_4^\Zboson$ & $[-1.0, 1.0] \times 10^{-5}$ & $[-1.9, 1.9] \times 10^{-5} $\\
\hline\hline
\end{tabular}
\end{center}
\caption{Observed and expected one-dimensional limits on $h_3^V$ and $h_4^V$, assuming that any excess in data over background predictions is due solely to $h_3^V$ or $h_4^V$ and that only one of them is nonzero.}
\label{table:observedExpected1DLimitsATGC}
\end{table}

The 95\% C.L. limits on each aTGC parameter are obtained with the other aTGC parameters set to their SM values using a one-dimensional profile-likelihood fit. The dependence of these observed and expected limits versus $\Lambda_{\mathrm{FF}}$ is shown in Figures~\ref{fig:h4_Lambda_opt} and ~\ref{fig:h3_Lambda_opt}. 
The obtained observed limits are almost a factor of two better than the expected limits, which is due to a downward fluctuation in the region of high $\ET^{\gamma}$ for the $\nnbar\gamma$ channel. All anomalous couplings considered are found to be compatible with the SM value zero. The observed limits on $h_3^{\gamma}, h_3^{Z}$ are at the level of $0.8$--$1.7 \times 10^{-3}$ and those on $h_4^{\gamma}, h_4^{Z}$ are at the level of $0.3$--$1.2\times 10^{-5}$ as shown in Table~\ref{table:observedExpected1DLimitsATGC}. These limits are the most stringent to date. 

\begin{figure}[hbtp]
\begin{center}
\includegraphics[width=0.45\textwidth]{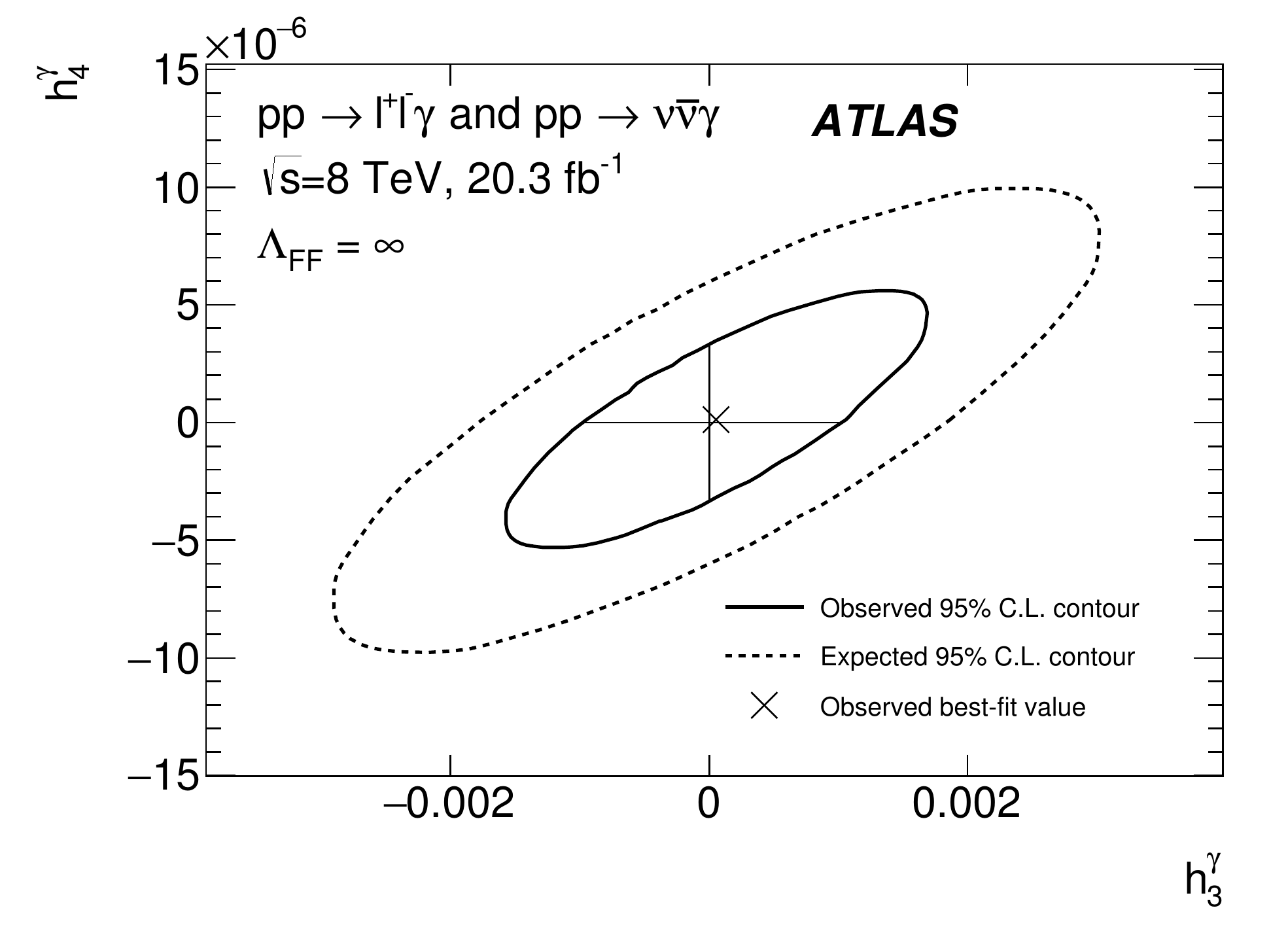}
\includegraphics[width=0.45\textwidth]{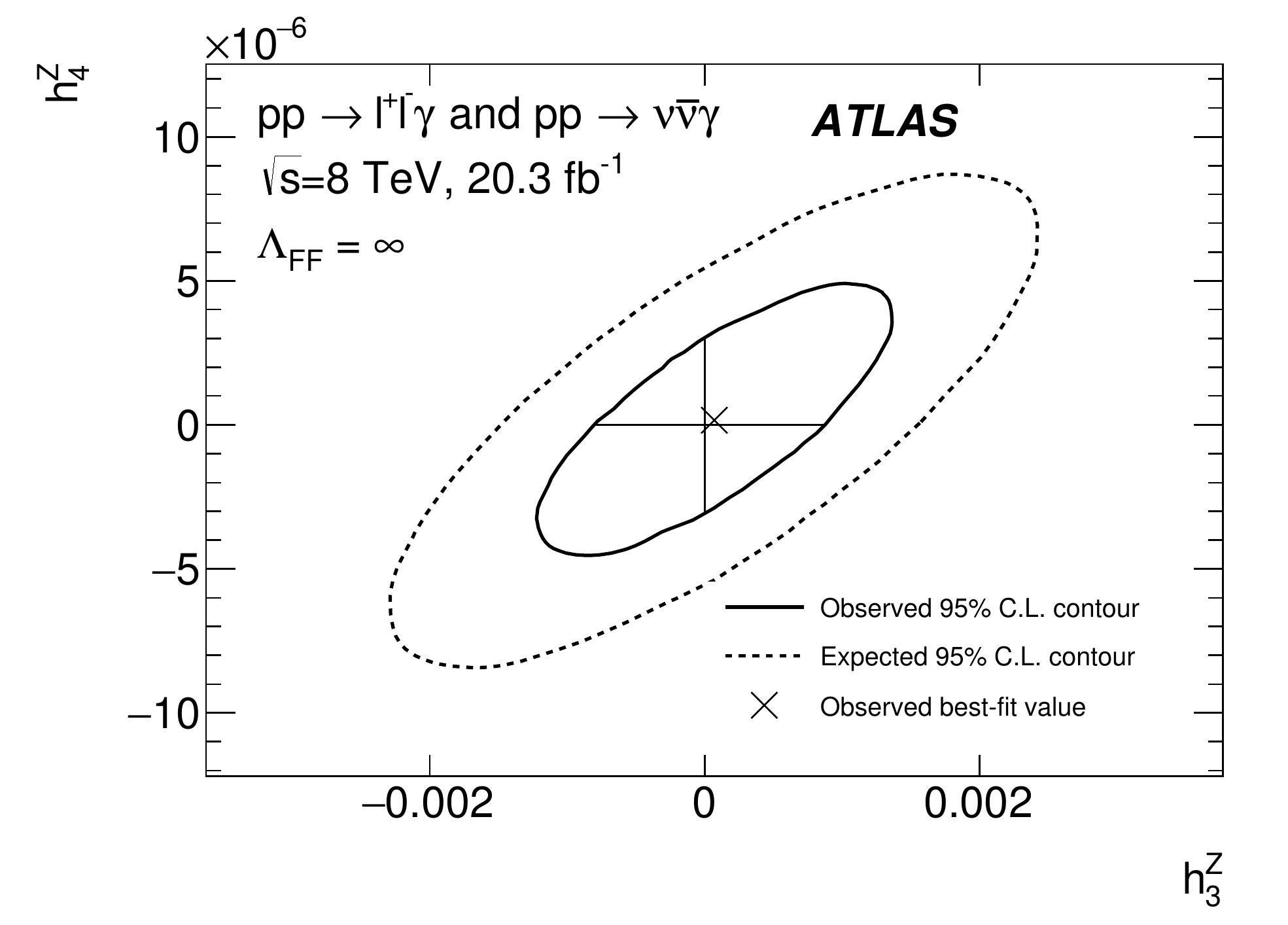}
\caption{Observed (solid ellipse) and expected (dashed ellipse) 95\% C.L. contours shown in the two-parameter planes for pairs of anomalous couplings $h_3^{\gamma}$ and $h_4^{\gamma}$ (left), $h_3^\Zboson$ and $h_4^\Zboson$ (right), corresponding to an infinite cutoff scale. The horizontal and vertical lines inside each contour correspond to the limits found in the one-parameter fit procedure, and the ellipses indicate the correlations between the one-parameter fits. The cross inside each contour corresponds to the observed best-fit value.}
\label{fig:TwoDimensionalLimits_inf}
\end{center}
\end{figure}

\begin{figure}[hbtp]
\begin{center}
\includegraphics[width=0.45\textwidth]{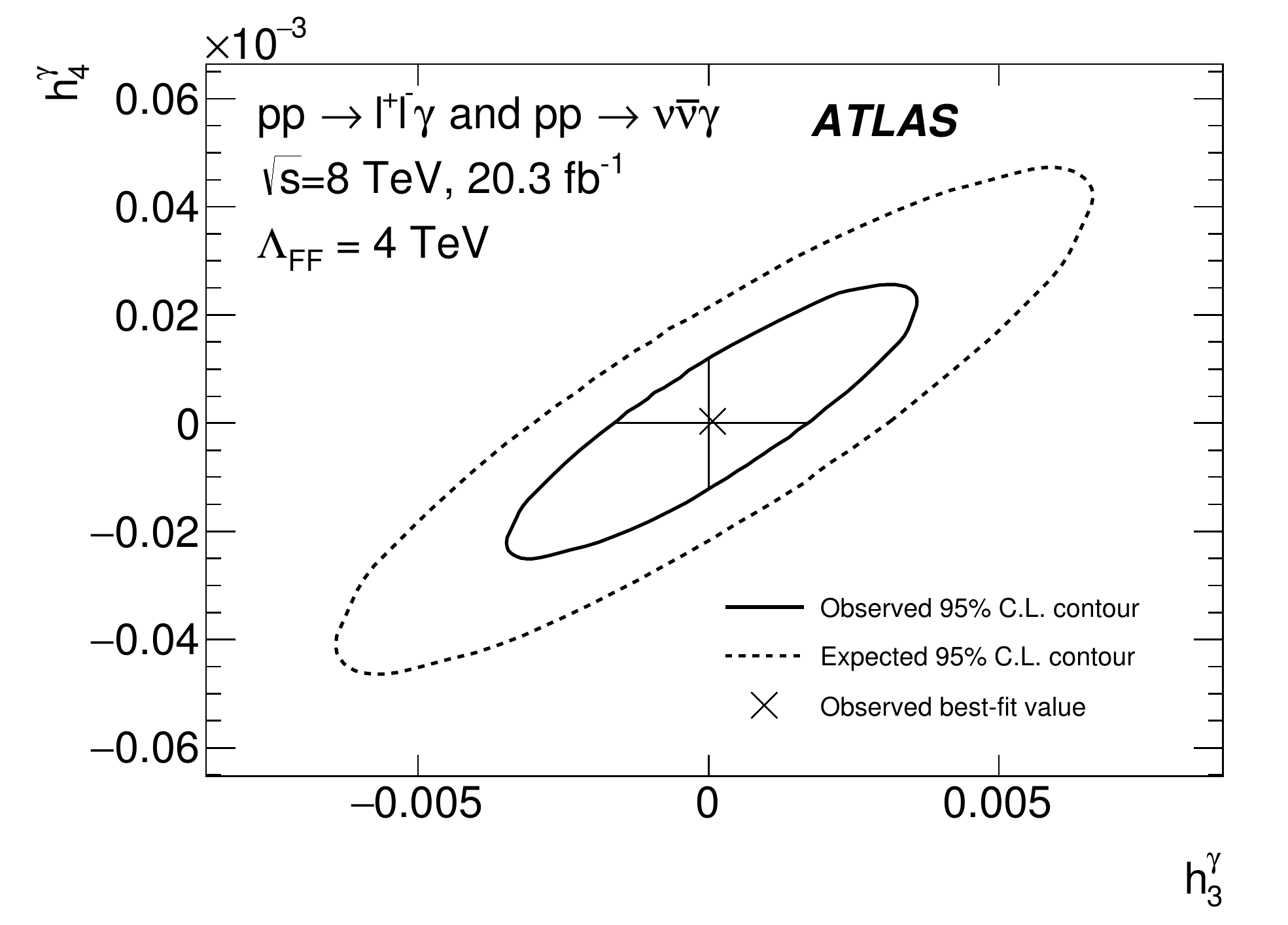}
\includegraphics[width=0.45\textwidth]{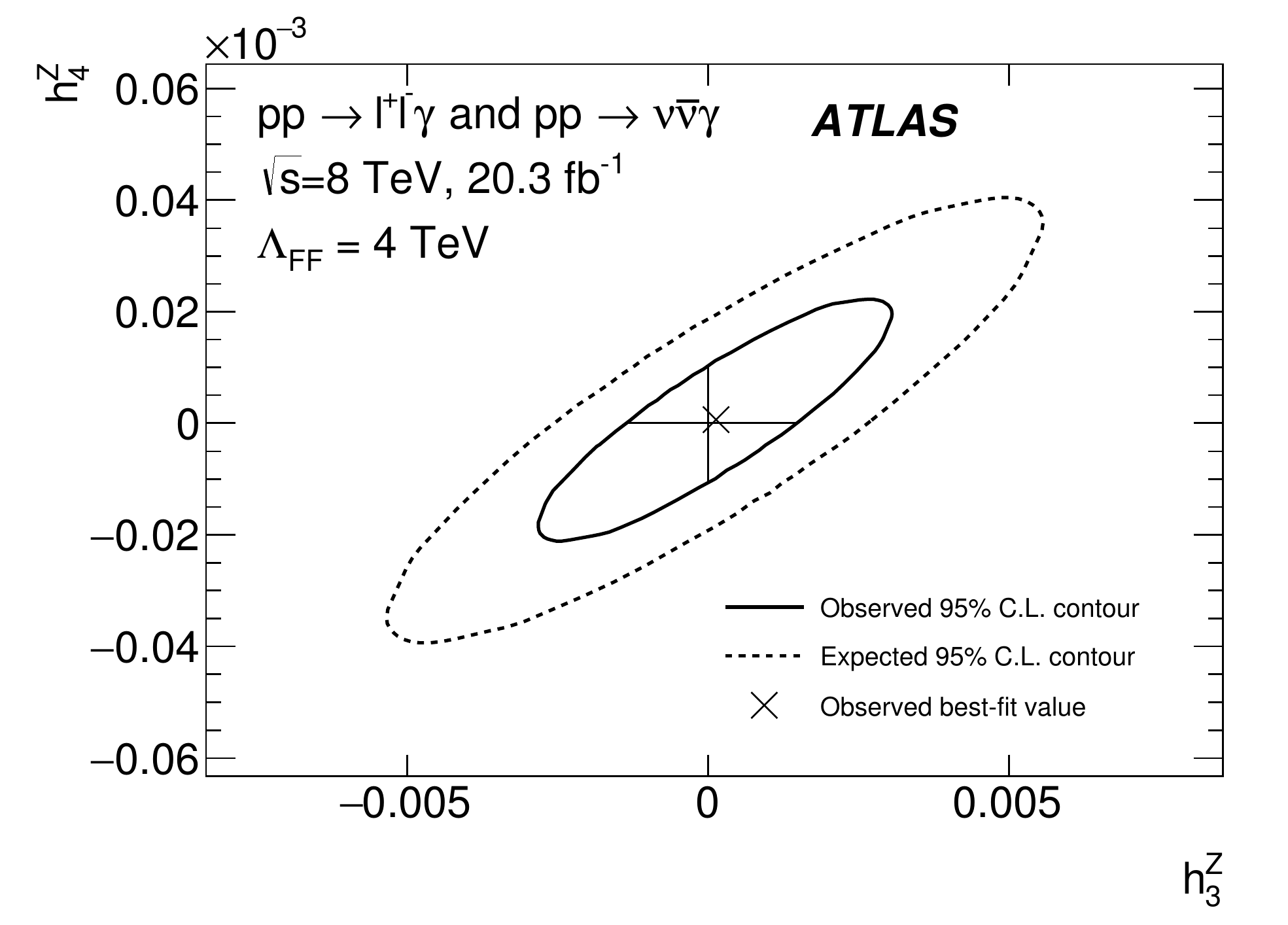}
\caption{Observed (solid ellipse) and expected (dashed ellipse) 95\% C.L. contours shown in the two-parameter planes for pairs of anomalous couplings $h_3^{\gamma}$ and $h_4^{\gamma}$ (left), $h_3^\Zboson$ and $h_4^\Zboson$ (right), corresponding to a $\Lambda_{\mathrm{FF}}=4$ \TeV~cutoff scale. The horizontal and vertical lines inside each contour correspond to the limits found in the one-parameter fit procedure, and the ellipses indicate the correlations between the one-parameter fits. The cross inside each contour corresponds to the observed best-fit value.}
\label{fig:TwoDimensionalLimits_4TeV}
\end{center}
\end{figure}

The limits on all possible combinations of each pair of aTGC are also evaluated by the same method. The 95\% C.L. regions in two-parameter aTGC space are shown as contours on the $(h_3^{\gamma}, h_4^{\gamma})$ and $(h_3^{\Zboson}, h_4^{\Zboson})$ planes in Figures~\ref{fig:TwoDimensionalLimits_inf} and~\ref{fig:TwoDimensionalLimits_4TeV}, since only these pairs are expected to interfere~\cite{PhysRevD.47.4889}. 

Since all sensitivity of the measurement of aTGCs is contained in a single measurement of the $\Zboson\gamma$ cross section in the high-$\ET^\gamma$ 
regions, the likelihood ratio used to obtain the two-parameter limits has one effective degree of freedom. 
Therefore the results obtained for the aTGC frequentist limits found in the one-parameter fit are identical to the corresponding limits obtained from the two-parameter 
fits at the points where the other aTGC is zero.

\subsection{Anomalous quartic gauge-boson couplings $Z$$Z$$\gamma\gamma$ and $Z$$\gamma$$\gamma\gamma$}

Triboson $Z\gamma\gamma$ production in the SM has no contributions from the quartic gauge-boson couplings $ZZ\gamma\gamma$ and $Z\gamma\gamma\gamma$.
However, physics beyond the SM could induce these anomalous neutral QGCs, enhancing the cross section for $Z\gamma\gamma$ production and modifying the kinematic distribution of the final-state $Z$ boson and photons.
The effect of such new couplings can be modeled using an effective field theory (EFT)~\cite{bib:eft} that includes higher-dimensional operators~\cite{bib-eboli-dim8}. 

The event generator \textsc{Vbfnlo} is used to produce the $Z\gamma\gamma$ events with the aQGCs introduced using EFT dimension-8 operators with coefficients $f_{T0}/\Lambda^{4}$, $f_{T5}/\Lambda^{4}$, $f_{T9}/\Lambda^{4}$, $f_{M2}/\Lambda^{4}$, and $f_{M3}/\Lambda^{4}$ in the linear Higgs-doublet representation~\cite{bib-eboli-dim8}
for the aQGC parameterization~\cite{bib:snowmass_ewk}. In this formalism, the parity conserving effective Lagrangian, which induces pure quartic couplings of the weak gauge bosons,
is introduced by employing the linear representation for the higher order operators and assuming that the recently observed Higgs boson belongs to a $S\kern -0.15em U(2)_{L}$ doublet~\cite{bib-eboli-dim8}.
Dimension-8 operators are the lowest-dimension operators that lead to quartic gauge-boson couplings without exhibiting triple gauge-boson vertices.
The $f_{T,x}$ operators contain only the field strength tensor while the $f_{M,x}$ operators contain both the Higgs double derivatives and the field strength.
A weak boson field is either from the covariant derivative of the Higgs doublet field or from the field strength tensor.
In the SM, all these aQGC operator coefficients are equal to zero.
The parameters $f_{T0}/\Lambda^{4}$ and  $f_{T5}/\Lambda^{4}$ are most sensitive to production of aQGC effects, $f_{T9}$ can only be probed via neutral QGCs
such as $Z\gamma\gamma$ while $f_{M2}/\Lambda^{4}$ and $f_{M3}/\Lambda^{4}$ are chosen since they can be related to dimension-6 operators constrained by LEP experiments and CMS~\cite{bib:snowmass_ewk},
which allows further comparisons and future aQGC combinations across different experiments.
The corresponding coefficients $a_0$ and $a_c$ in the LEP formalism can be translated in the context of EFT dimension-8 operators (for $ZZ\gamma\gamma$/$Z\gamma\gamma\gamma$ vertices) according to the formalism transformation equation as follows~\cite{bib:snowmass_ewk}: 

\begin{eqnarray}
\frac{f_{M2}}{\Lambda^4} = - \frac{a_0}{\Lambda^2} \frac{s_w^2}{2 v^2 c_w^2},
\\
\frac{f_{M3}}{\Lambda^4} = \frac{a_c}{\Lambda^2} \frac{s_w^2}{2 v^2 c_w^2}.
\end{eqnarray}

Form factors are introduced to restore unitarity at a very high parton center-of-mass energy $\sqrt{\hat{s}}$: 
$f_i(\hat{s})=f_{i}/(1+\hat{s}/\Lambda_{\mathrm{FF}}^2)^n$.
The parameter $\Lambda_{\mathrm{FF}}$ is chosen to preserve unitarity up to $\sqrt{\hat{s}} = 8$ \TeV~with the FF exponent $n$ set to two.

In order to have better sensitivities to aQGCs, the measured $Z\gamma\gamma$ exclusive (zero-jet) fiducial cross section is used with the additional requirement
$m_{\gamma\gamma} >$ 300 (200) \GeV~for $\nnbar\gamma\gamma$ ($\leplep\gamma\gamma$) channel.  
The SM backgrounds in these aQGC-optimized regions are estimated using the same methods as described in Section~\ref{sec:backgrounds} for the $Z\gamma\gamma$ cross-section measurements. 
Theory predictions for the SM signal and data observations in these aQGC extended fiducial regions are shown in Table~\ref{table:aQGC_reg_SM_cs}.

\begin{table}[hbtp]
\begin{center}
\begin{spacing}{1.25}
\begin{tabular}{l c c c}
\hline\hline
Channel & Measurement [fb] & Prediction [fb]  \\
\hline
$\leplep\gamma\gamma$ ($m_{\gamma\gamma}>$ 200 \GeV) & $0.12^{+0.11}_{-0.07}(\mathrm{stat.})^{+0.03}_{-0.01}(\mathrm{syst.})$ & $0.0674\pm0.0013(\mathrm{stat.})\pm0.0053(\mathrm{syst.})$ \\
$\nnbar\gamma\gamma$ ($m_{\gamma\gamma}>$ 300 \GeV) & $0.16^{+0.17}_{-0.11}(\mathrm{stat.})^{+0.04}_{-0.01}(\mathrm{syst.})$ & $0.0499\pm0.0008(\mathrm{stat.})\pm0.0062(\mathrm{syst.})$  \\
\hline\hline
\end{tabular}
\end{spacing}
\end{center}
\vspace*{-8mm}
\caption{Theoretical \textsc{Vbfnlo} SM and observed cross sections in chosen aQGC regions (with the exclusive selection) for the channels studied. 
The $m_{\gamma\gamma}$ threshold is $200$ \GeV~for the electron and muon channels and is $300$ \GeV~for the neutrino channel.
The first uncertainty is statistical, the second is systematic.}
\label{table:aQGC_reg_SM_cs}
\end{table}

\begin{figure}[hbtp]
\begin{center}
\includegraphics[width=0.49\textwidth]{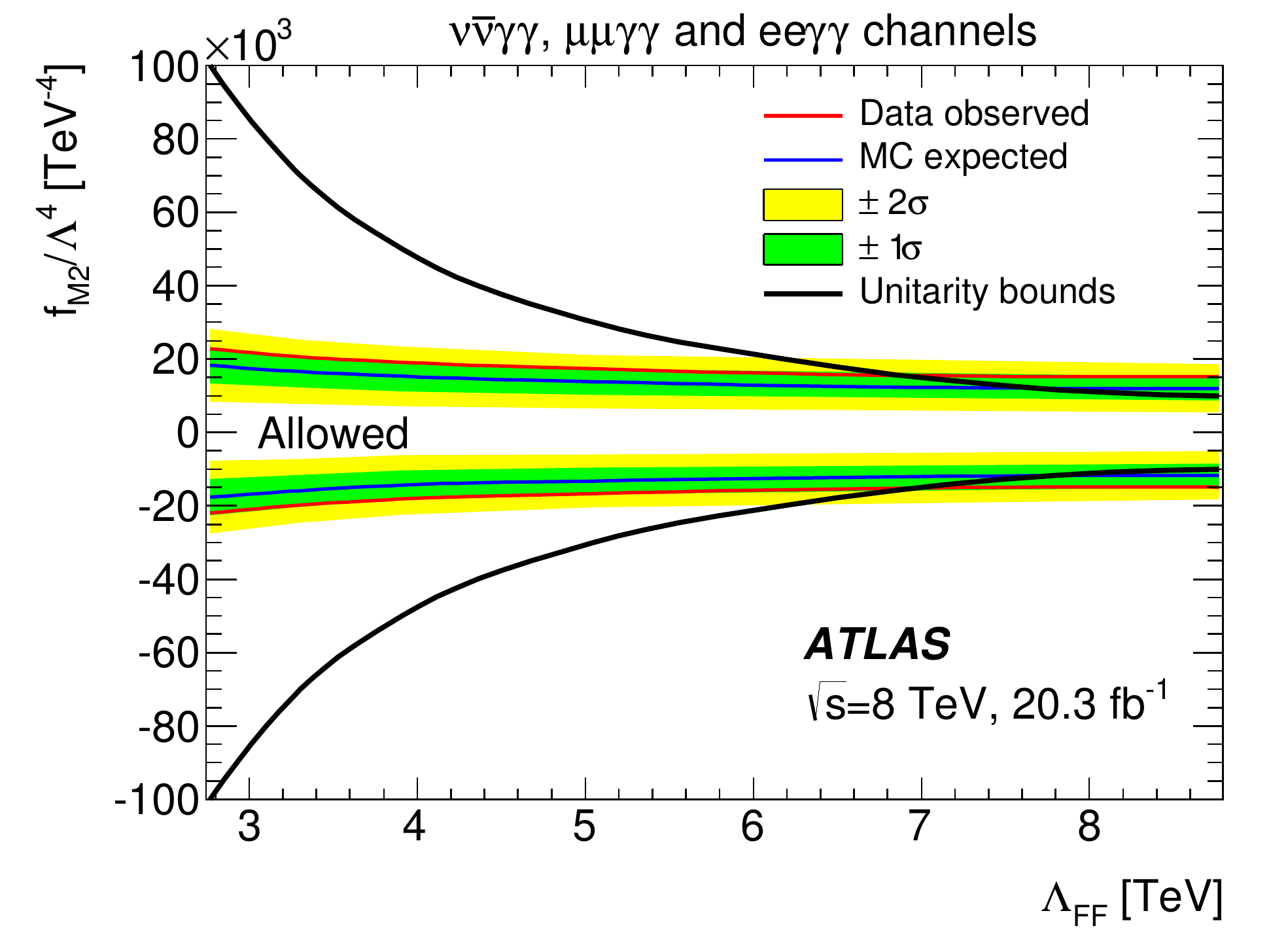}
\includegraphics[width=0.49\textwidth]{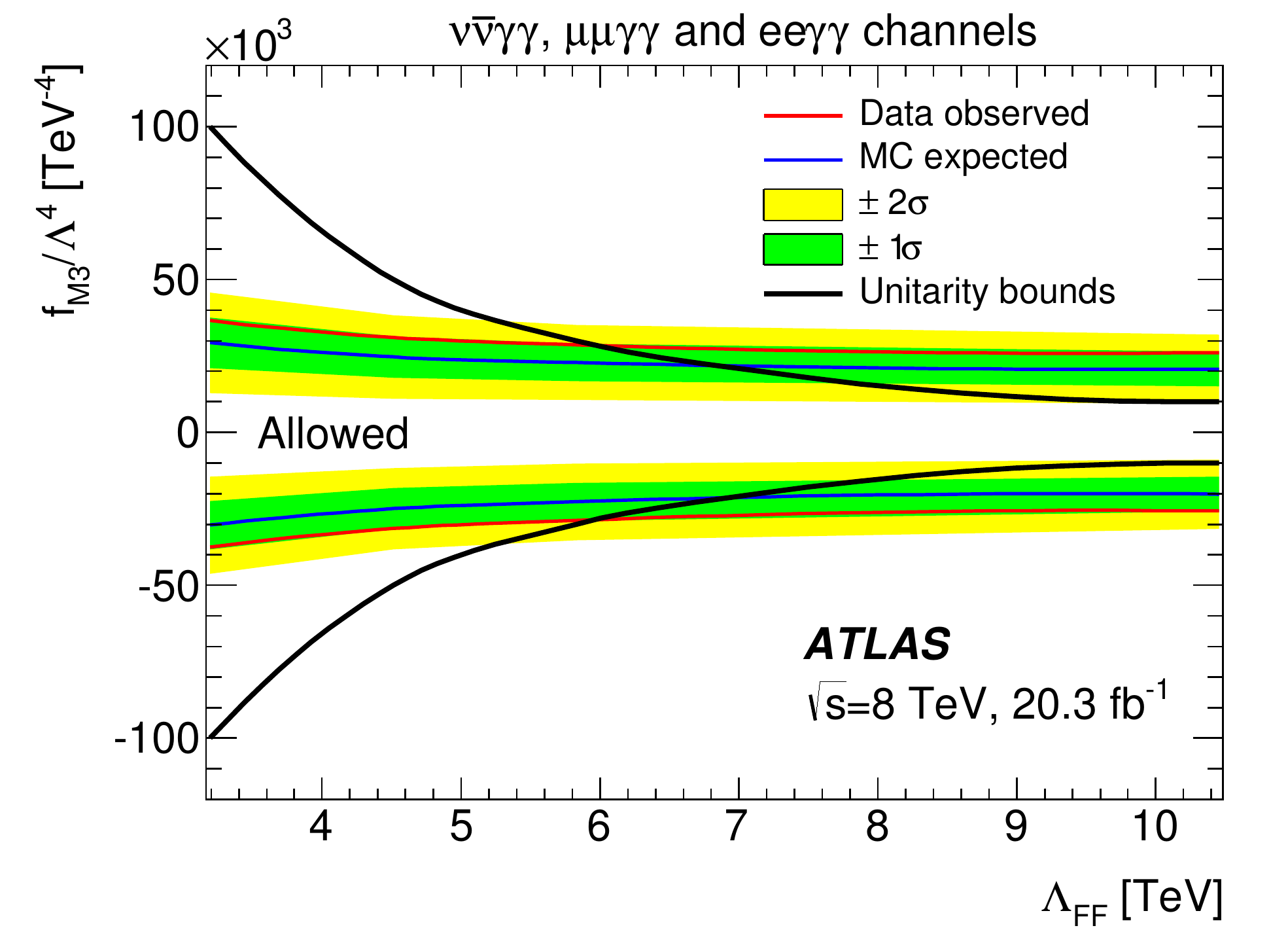}\\
\includegraphics[width=0.49\textwidth]{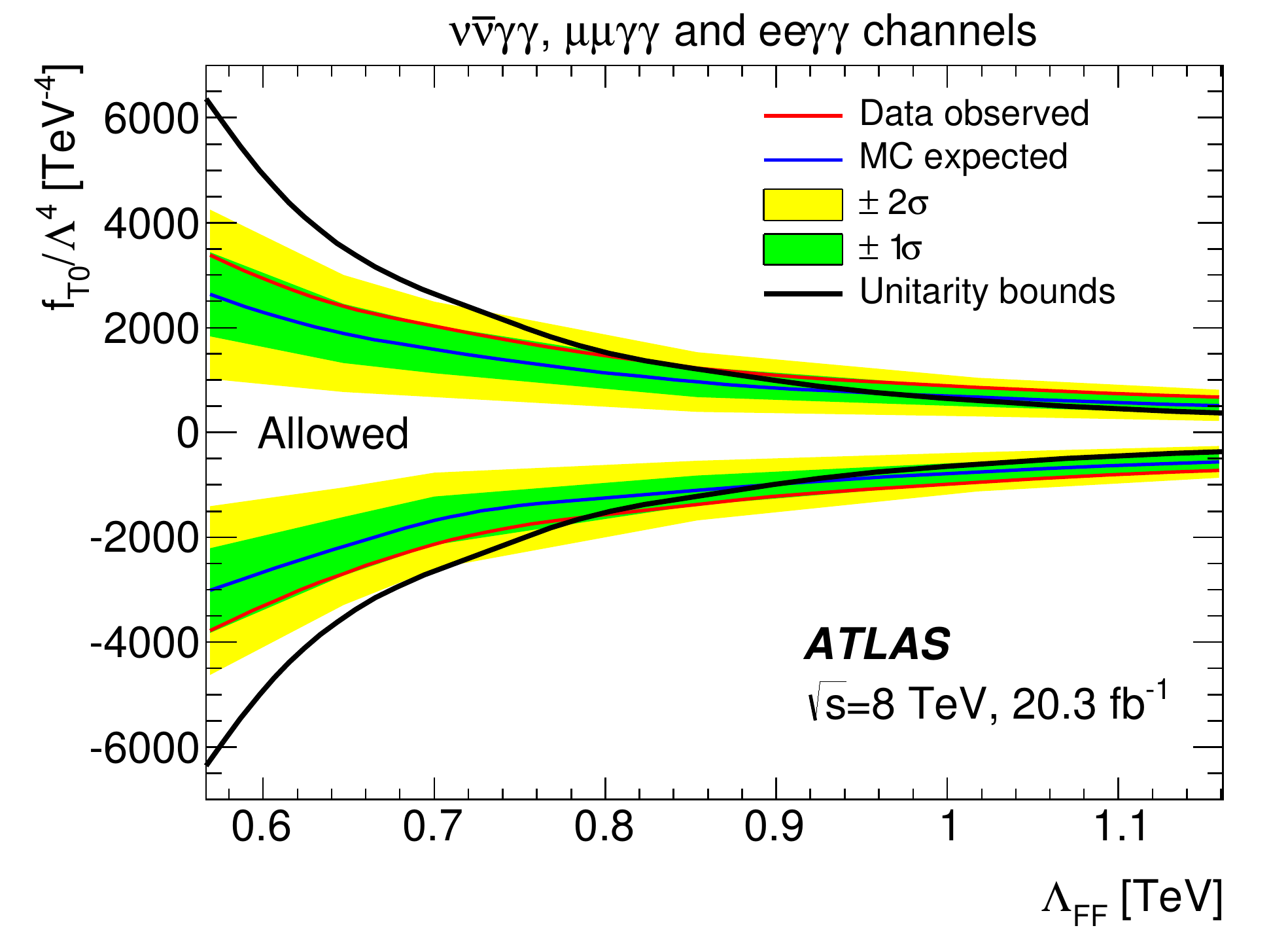}
\includegraphics[width=0.49\textwidth]{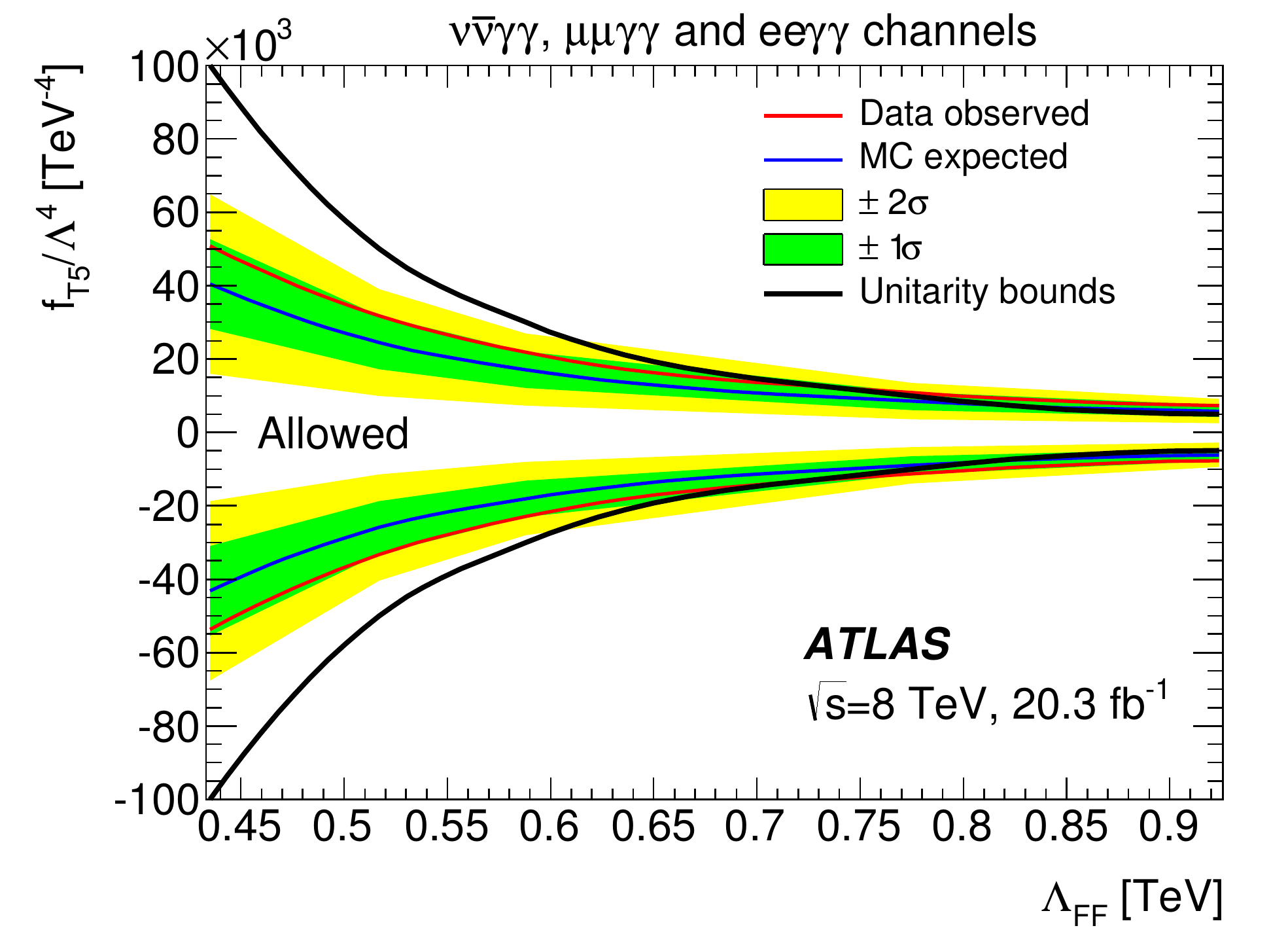}\\
\includegraphics[width=0.49\textwidth]{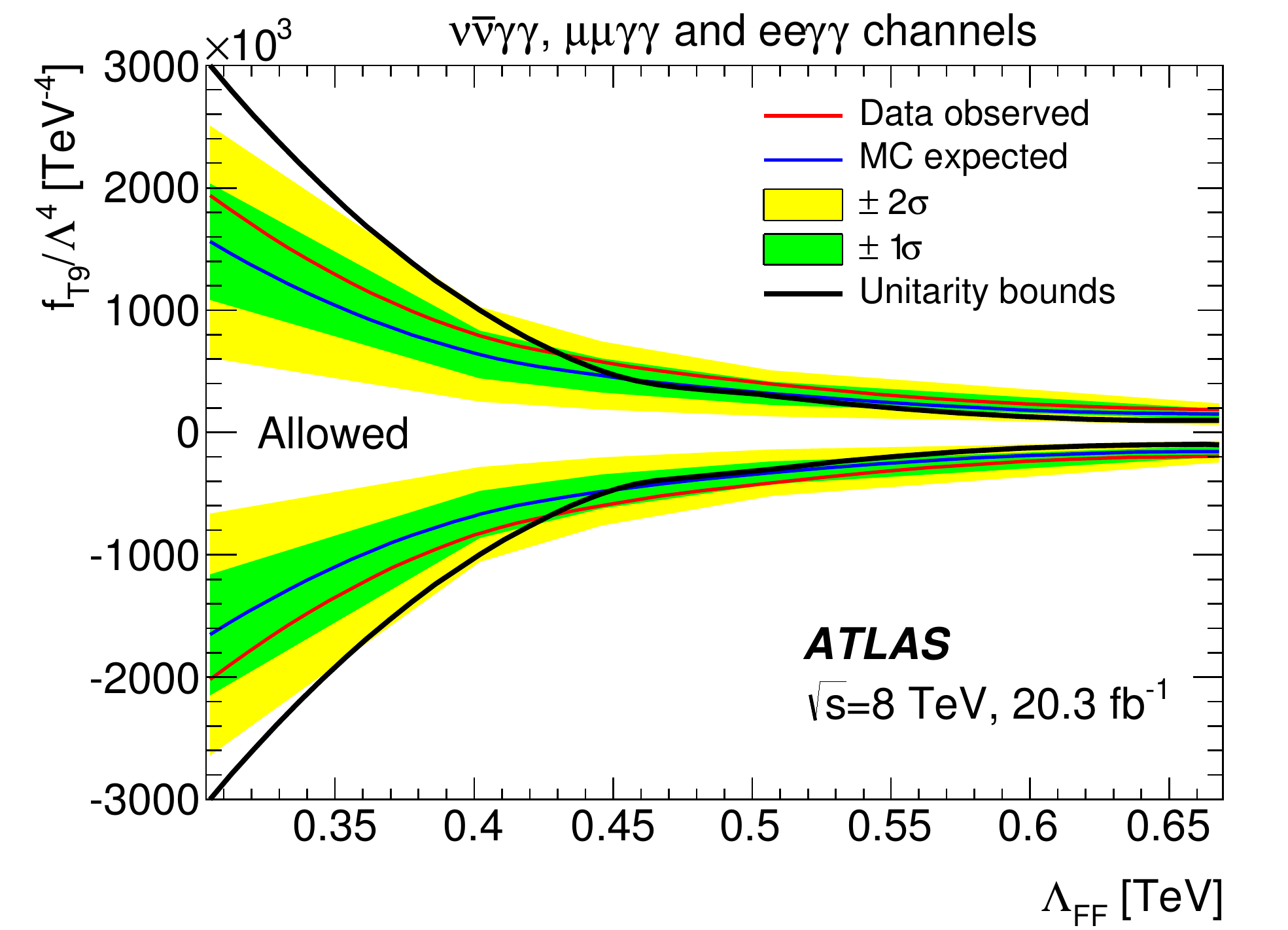}
\caption{Dependencies of the observed limits, expected limits and unitarity bounds on the form factor energy scale $\Lambda_{\mathrm{FF}}$ for $f_{M2}$ (top left), $f_{M3}$ (top right), $f_{T0}$ (center left), $f_{T5}$ (center right), $f_{T9}$ (bottom). The green and yellow bands show areas of variation for the expected limits by 1$\sigma$ and 2$\sigma$, respectively.}
\label{fig:aQGCvsLambdaFF}
\end{center}
\end{figure}

\begin{figure}[hbtp]
\begin{center}
\includegraphics[width=0.49\textwidth]{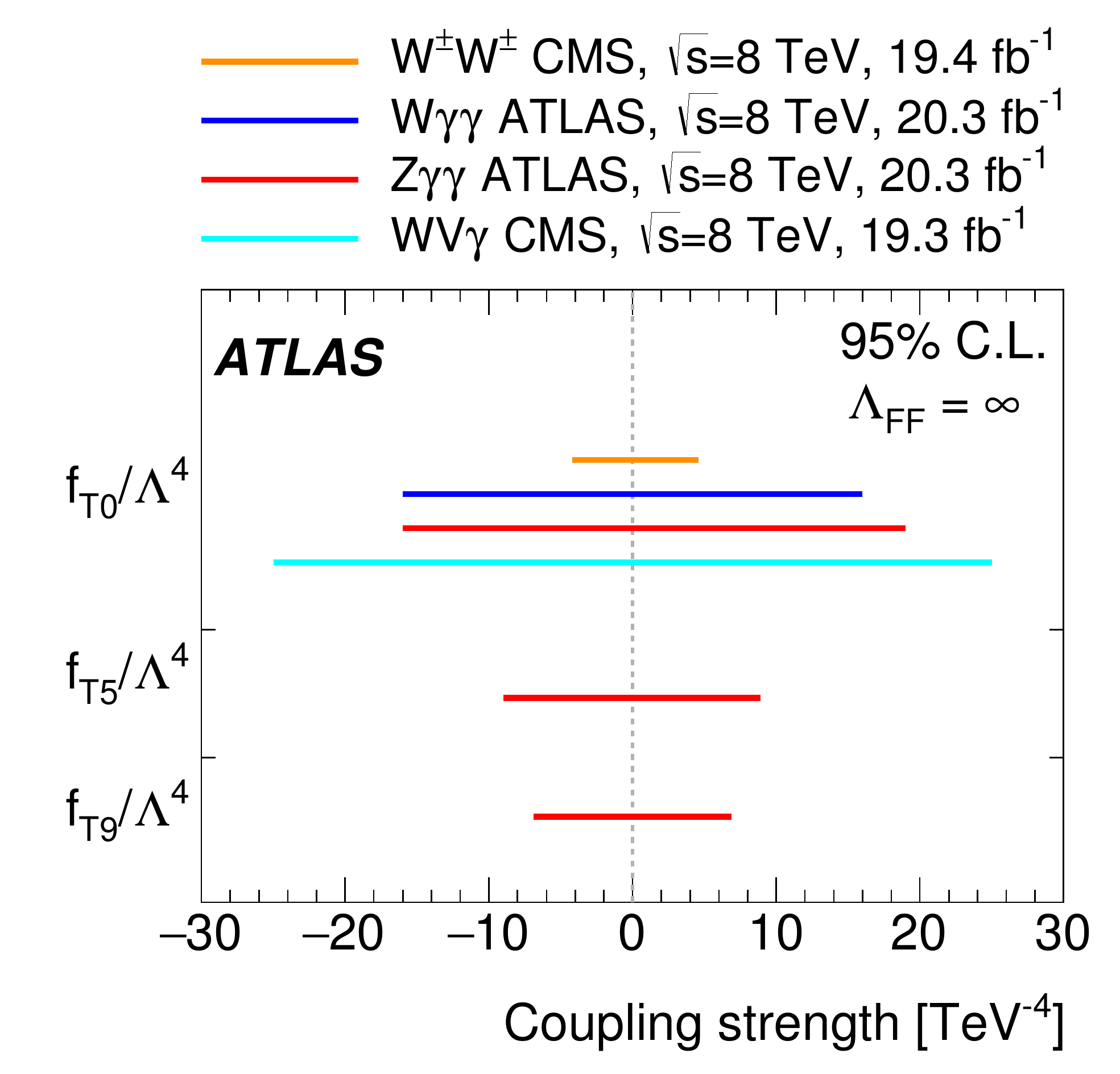}
\includegraphics[width=0.49\textwidth]{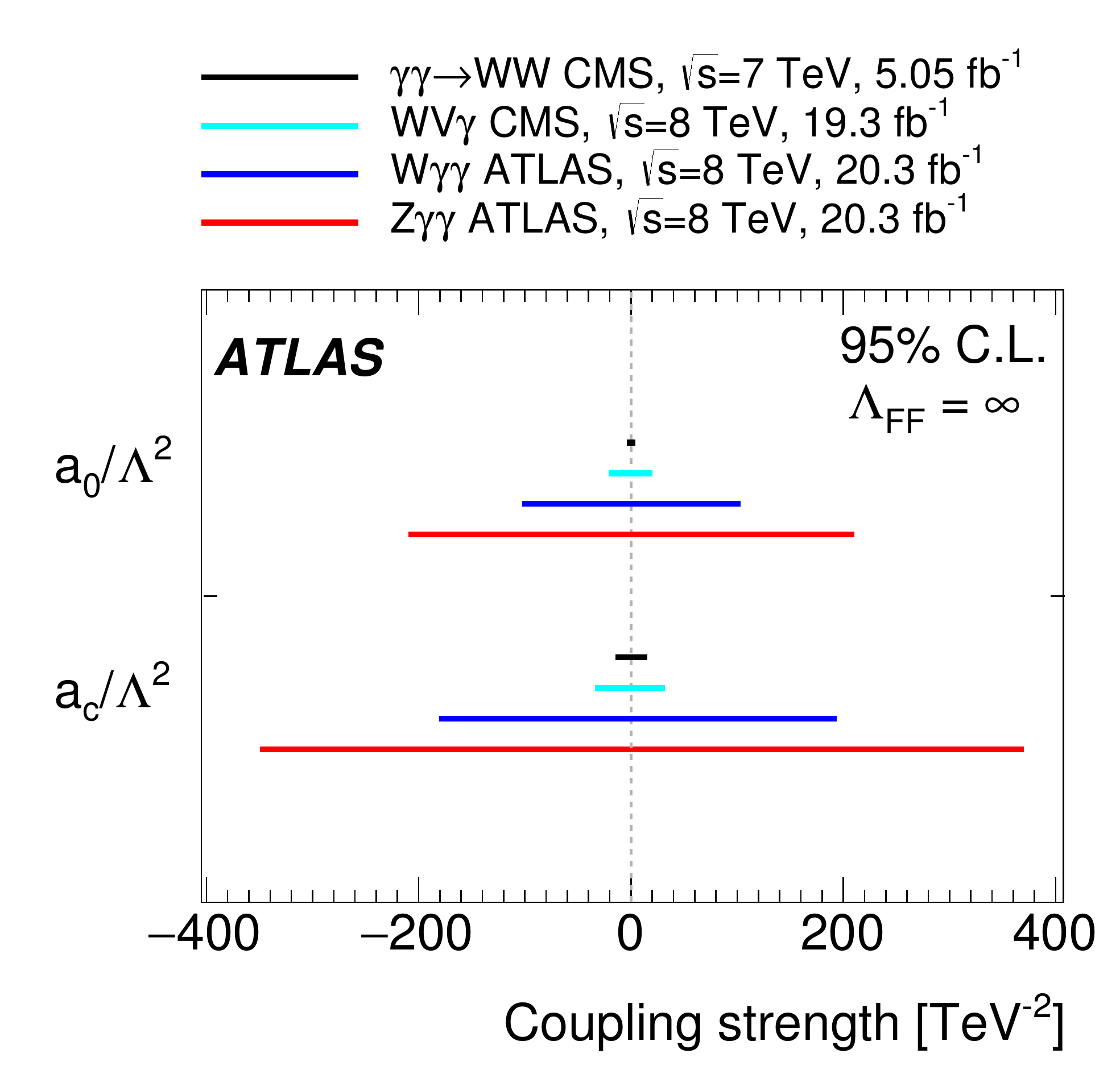}
\caption{Comparison of the observed limits for $f_{T0}/\Lambda^{4}$, $f_{T5}/\Lambda^{4}$, $f_{T9}/\Lambda^{4}$ (on the left) and LEP parameters~\cite{Belanger:1992ac} $a_0/\Lambda^{2}$ and $a_c/\Lambda^{2}$ (on the right) without FF unitarization. 
The limits are presented in the formalism as implemented in {\textsc{MadGraph5\_aMC@NLO}}~\cite{Degrande:2013rea} (left) and in the LEP formalism~\cite{bib:snowmass_ewk} (right).}
\label{fig:ununitarizedComp}
\end{center}
\end{figure}

The reconstruction efficiency $C_{\Zboson\gamma\gamma}$ is calculated from simulation samples with nonzero aQGCs using the events generated at LO by \textsc{Vbfnlo} and parton-showered by \textsc{Pythia8}. The deviation of
the reconstruction efficiency from that for SM production using $\SHERPA$ is taken as an additional uncertainty of 20\%~(60\%) 
for the  $\nnbar$ channel
($\leplep$ channels). The differences in $A_{\Zboson\gamma\gamma}$ and $C^*$(parton $\rightarrow$ particle) between aQGC and SM simulation samples are at the percent level and were neglected. The expected and observed 95\% C.L. limits of each dimension-8 operator coefficient are derived from one-dimensional profile-likelihood fits
as described in the aTGC study. The $\Lambda_{\mathrm{FF}}$-dependent observed/expected limits are obtained using the signal cross-section parameterization produced at LO by \textsc{Vbfnlo} and shown in Figure~\ref{fig:aQGCvsLambdaFF}.
The unitarity bounds versus $\Lambda_{\mathrm{FF}}$ are also plotted in the figure with the FF exponent $n$ equal to two.
Table~\ref{tab:observedExpected1DLimitsAQGC} shows the expected and observed 95\% C.L. limits with no unitarization restriction along with those respecting unitarity bounds at the maximum allowed 
value of $\Lambda_{\mathrm{FF}}$ according to the \textsc{Vbfnlo} estimation.
The limits without unitarization are compared to the limits from the most recent CMS
results~\cite{bib:CMSWWaa,bib:CMSWVa,bib:CMSWWss} and ATLAS results~\cite{STDM-2013-05} in Figure~\ref{fig:ununitarizedComp}.
The limits are presented in the formalism as implemented in \textsc{Vbfnlo}~\cite{Degrande:2013rea}, except for the ones in Figure~\ref{fig:ununitarizedComp}, 
which are presented in the formalism as implemented in {\textsc{MadGraph5\_aMC@NLO}}~\cite{Degrande:2013rea} (left plot) and in the LEP formalism~\cite{bib:snowmass_ewk} (right plot) in order to be compared to other results.

\begin{table}[hbtp]
	\begin{center}
	\begin{tabular}{c|c|c|c|c}
	\hline\hline
	$n$ & $\Lambda_{\mathrm{FF}}$ [\TeV] & Limits 95\% C.L. & Observed [\TeV$^{-4}$] & Expected [\TeV$^{-4}$]  \\ \hline
	\multirow{5}{*}{$0$}& \multirow{5}{*}{$\infty$}	
	& $f_{M2}/\Lambda^4$ & $[-1.6, 1.6]\times10^4$  & $[-1.2, 1.2]\times10^4$ \\
	&& $f_{M3}/\Lambda^4$ & $[-2.9, 2.7]\times10^4$  & $[-2.2, 2.2]\times10^4$ \\
	&& $f_{T0}/\Lambda^4$ & $[-0.86, 1.03]\times10^2$  & $[-0.65, 0.82]\times10^2$ \\
	&& $f_{T5}/\Lambda^4$ & $[-0.69, 0.68]\times10^3$  & $[-0.52, 0.52]\times10^3$ \\
	&&$f_{T9}/\Lambda^4$ & $[-0.74, 0.74]\times10^4$  & $[-0.58, 0.59]\times10^4$ \\
 \hline
	\multirow{5}{*}{$2$} & 5.5 & $f_{M2}/\Lambda^4$ & $[-1.8, 1.9]\times10^4$  & $[-1.4, 1.5]\times10^4$ \\
	& 5.0 & $f_{M3}/\Lambda^4$ & $[-3.4, 
3.3]\times10^4$  & $[-2.6, 2.6]\times10^4$\\
	& 0.7 & $f_{T0}/\Lambda^4$ & $[-2.3, 2.1]\times10^3$  & $[-1.9, 1.6]\times10^3$ \\
	& 0.6 & $f_{T5}/\Lambda^4$ & $[-2.3, 2.2]\times10^4$  & $[-1.8, 1.8]\times10^4$ \\
        & 0.4 & $f_{T9}/\Lambda^4$ & $[-0.89, 
0.86]\times10^6$  & $[-0.71, 0.68]\times10^6$ \\	 \hline\hline
\end{tabular}
\end{center}
\caption{Observed and expected one-dimensional limits on aQGC parameters. Form factor exponent $n=0$ corresponds to infinite scale limits without any form factor.}
\label{tab:observedExpected1DLimitsAQGC}
\end{table}

\FloatBarrier
\section{Summary}
\label{sec:summary}

The production cross section of $Z$ bosons in association with isolated high-energy photons is measured using 20.3 $\ifb$ of $pp$ collisions at $\sqrt{s} = 8$ \TeV~collected with the ATLAS detector at the LHC.
The analyses use the decays $Z$$\to$$\nnbar$ and $Z$/$\gamma^*$$\to$ $\ee$ or $\mumu$ with 
$m_{\leplep} > 40$ \GeV. The $\Zboson$/$\gamma^*$ decays to charged leptons are triggered on using electrons or muons with large transverse momentum. 
The production channels studied are $pp\to\ell^{+}\ell^{-}\gamma + X$ 
and $pp\to\ell^{+}\ell^{-}\gamma\gamma + X$ where the photons are required to have $\ET$ $>$ 15 \GeV. The events with $\Zboson$ decays to neutrinos are selected using high-$\ET$ photon triggers.   
The production channels studied are $pp\to\nu\bar{\nu}\gamma + X$ with photon $\ET$ $>$ 130 \GeV~and $pp\to\nu\bar{\nu}\gamma\gamma + X$ where the photons have $\ET$ $>$ 22 \GeV. 
In all production channels the photons are required to be isolated and to satisfy tight identification criteria. The dominant backgrounds arise from jets faking photons and these are evaluated using data-driven techniques.

The cross sections and kinematic distributions for channels with $\Zboson$/$\gamma^*$ decays to electrons and muons are combined assuming lepton universality and presented for a single charged-lepton flavor in fiducial regions 
defined by the lepton and photon acceptance.
For the channels with $\Zboson$ decays to neutrinos, the cross sections and kinematics are quoted for the sum of the three neutrino flavors. This leads to studies of the following four production channels:
\begin{itemize}
\item $pp\to\ell^{+}\ell^{-}\gamma + X$,
\item $pp\to\ell^{+}\ell^{-}\gamma\gamma + X$,
\item $pp\to\nu\bar{\nu}\gamma + X$,
\item $pp\to\nu\bar{\nu}\gamma\gamma + X$.
\end{itemize}

The cross sections are measured in a fiducial region, for both the inclusive case, with no requirements on the recoil system $X$, and the exclusive case in which there are no jets with $\pT > 30$ \GeV~within $|\eta|< 4.5$.

The data are compared to SM predictions using a parton shower Monte Carlo  ($\SHERPA$) and parton-level perturbative calculations carried out at  NLO ({\textsc{MCFM}}) and NNLO, corrected by parton-to-particle scale factors.

There is good agreement between the measurements and the SM predictions. $\SHERPA$ reproduces the kinematic spectra, including the jet multiplicity spectrum, in the single-photon production channels $\ell^{+}\ell^{-}\gamma + X$ and $\nu\bar{\nu}\gamma + X$. The NLO and NNLO matrix element generators are used to predict the photon $\ET$ and $m_{\leplep\gamma}$ differential spectra in these single-photon channels, and the magnitude of the cross sections. There is good agreement between data and the SM predictions, with the NNLO calculations needed to account for the production of the high-$\ET$ photons where the NLO calculation significantly underestimates the data. In the two-photon production channels
$\ell^{+}\ell^{-}\gamma\gamma + X$ and  $\nu\bar{\nu}\gamma\gamma + X$ the cross sections are compared to the NLO predictions. 
The measurements in these channels are statistically limited, but the data and SM predictions agree within the uncertainties.

Having found no significant deviations from SM predictions, the data are used to set limits on anomalous couplings of photons and $\Zboson$ bosons. These could result from $\Zboson$/$\gamma^*$ $s$-channel production 
coupled to a final-state $\Zboson$ boson and one photon (anomalous triple gauge-boson couplings, or aTGCs), or a final-state $\Zboson$ boson and two photons (anomalous quartic gauge-boson couplings, or aQGCs). The limits on the 
aTGCs are determined using a modified SM Lagrangian with operators proportional to parameters conventionally denoted as  $h_3^V$ and $h_4^V$ ($V = Z$ or $\gamma$). The contributions from aQGCs 
are introduced using an effective field theory concentrating on those operators most sensitive to the $Z\gamma\gamma$ final state. 
Limits are derived for these aTGC and aQGC parameters.

\section*{Acknowledgments}

We thank CERN for the very successful operation of the LHC, as well as the
support staff from our institutions without whom ATLAS could not be
operated efficiently.

We acknowledge the support of ANPCyT, Argentina; YerPhI, Armenia; ARC, Australia; BMWFW and FWF, Austria; ANAS, Azerbaijan; SSTC, Belarus; CNPq and FAPESP, Brazil; NSERC, NRC and CFI, Canada; CERN; CONICYT, Chile; CAS, MOST and NSFC, China; COLCIENCIAS, Colombia; MSMT CR, MPO CR and VSC CR, Czech Republic; DNRF and DNSRC, Denmark; IN2P3-CNRS, CEA-DSM/IRFU, France; GNSF, Georgia; BMBF, HGF, and MPG, Germany; GSRT, Greece; RGC, Hong Kong SAR, China; ISF, I-CORE and Benoziyo Center, Israel; INFN, Italy; MEXT and JSPS, Japan; CNRST, Morocco; FOM and NWO, Netherlands; RCN, Norway; MNiSW and NCN, Poland; FCT, Portugal; MNE/IFA, Romania; MES of Russia and NRC KI, Russian Federation; JINR; MESTD, Serbia; MSSR, Slovakia; ARRS and MIZ\v{S}, Slovenia; DST/NRF, South Africa; MINECO, Spain; SRC and Wallenberg Foundation, Sweden; SERI, SNSF and Cantons of Bern and Geneva, Switzerland; MOST, Taiwan; TAEK, Turkey; STFC, United Kingdom; DOE and NSF, United States of America. In addition, individual groups and members have received support from BCKDF, the Canada Council, CANARIE, CRC, Compute Canada, FQRNT, and the Ontario Innovation Trust, Canada; EPLANET, ERC, FP7, Horizon 2020 and Marie Sk{\l}odowska-Curie Actions, European Union; Investissements d'Avenir Labex and Idex, ANR, R{\'e}gion Auvergne and Fondation Partager le Savoir, France; DFG and AvH Foundation, Germany; Herakleitos, Thales and Aristeia programmes co-financed by EU-ESF and the Greek NSRF; BSF, GIF and Minerva, Israel; BRF, Norway; Generalitat de Catalunya, Generalitat Valenciana, Spain; the Royal Society and Leverhulme Trust, United Kingdom.

The crucial computing support from all WLCG partners is acknowledged
gratefully, in particular from CERN and the ATLAS Tier-1 facilities at
TRIUMF (Canada), NDGF (Denmark, Norway, Sweden), CC-IN2P3 (France),
KIT/GridKA (Germany), INFN-CNAF (Italy), NL-T1 (Netherlands), PIC (Spain),
ASGC (Taiwan), RAL (UK) and BNL (USA) and in the Tier-2 facilities
worldwide.


\printbibliography



\newpage 
\begin{flushleft}
{\Large The ATLAS Collaboration}

\bigskip

G.~Aad$^\textrm{\scriptsize 87}$,
B.~Abbott$^\textrm{\scriptsize 114}$,
J.~Abdallah$^\textrm{\scriptsize 65}$,
O.~Abdinov$^\textrm{\scriptsize 12}$,
B.~Abeloos$^\textrm{\scriptsize 118}$,
R.~Aben$^\textrm{\scriptsize 108}$,
M.~Abolins$^\textrm{\scriptsize 92}$,
O.S.~AbouZeid$^\textrm{\scriptsize 138}$,
N.L.~Abraham$^\textrm{\scriptsize 150}$,
H.~Abramowicz$^\textrm{\scriptsize 154}$
% ...
\bigskip
\\
% ...
$^{12}$ Institute of Physics, Azerbaijan Academy of Sciences, Baku, Azerbaijan\\
% ...
$^{65}$ University of Iowa, Iowa City IA, United States of America\\
% ...
$^{87}$ CPPM, Aix-Marseille Universit{\'e} and CNRS/IN2P3, Marseille, France\\
% ...
$^{92}$ Department of Physics and Astronomy, Michigan State University, East Lansing MI, United States of America\\
% ...
$^{108}$ Nikhef National Institute for Subatomic Physics and University of Amsterdam, Amsterdam, Netherlands\\
% ...
$^{114}$ Homer L. Dodge Department of Physics and Astronomy, University of Oklahoma, Norman OK, United States of America\\
% ...
$^{118}$ LAL, Univ. Paris-Sud, CNRS/IN2P3, Universit{\'e} Paris-Saclay, Orsay, France\\
% ...
$^{138}$ Santa Cruz Institute for Particle Physics, University of California Santa Cruz, Santa Cruz CA, United States of America\\
% ...
$^{150}$ Department of Physics and Astronomy, University of Sussex, Brighton, United Kingdom\\
% ...
$^{154}$ Raymond and Beverly Sackler School of Physics and Astronomy, Tel Aviv University, Tel Aviv, Israel\\
% ...
\end{flushleft}


\end{document}